\documentclass[11pt,a4paper]{article}

\usepackage[text={450pt,650pt},headheight={14.5pt},centering]{geometry}

\usepackage{mathptmx}

\usepackage{microtype}

\usepackage{braket}
\usepackage{mathtools}
\usepackage{amsmath}
\usepackage{amssymb}
\usepackage{dsfont}

\usepackage{graphicx}

\usepackage[font=small]{caption}
\usepackage{booktabs}
\usepackage{subfig}

\usepackage{tikz}
\usetikzlibrary{calc}

\newlength\vertdist
\newlength\hordist
\newlength{\nodewidth}
\tikzstyle arrow up=[->,out=45,in=315]
\tikzstyle arrow down=[->,out=225,in=135]
\tikzstyle state=[inner sep=1pt]
\tikzstyle dynkin node=[shape=circle,draw,thick,inner sep=0pt,minimum size=5mm]

\newlength\sitedist
\newlength\labeldist
\tikzstyle odd site=[shape=rectangle,draw,thick,inner sep=0pt,outer sep=0pt,minimum size=4mm]
\tikzstyle even site=[shape=rectangle,draw,fill=black!25!white ,thick,inner sep=0pt,outer sep=0pt,minimum size=4mm]
\tikzstyle null site=[shape=circle,draw,fill=black,thick,inner sep=0pt,outer sep=0pt,minimum size=1.75mm]
\tikzstyle chain line=[]

\usepackage{collref}
\usepackage[sort,compress]{cite}
\usepackage{hyperref}

\numberwithin{equation}{section}

\renewcommand{\a}{\alpha}
\newcommand{\ad}{a^\dag}

\newcommand{\earel}[1]{\mathrel{}&\hspace{-2\arraycolsep}#1\hspace{-2\arraycolsep}&\mathrel{}}
\newcommand{\eq}{\earel{=}}

\DeclareMathOperator{\tr}{tr}
\DeclareMathOperator{\Res}{Res}

\newcommand{\abs}[1]{{\left| #1 \right|}}

\newcommand{\AdS}{\text{AdS}}
\newcommand{\CFT}{\text{CFT}}
\newcommand{\Sphere}{\text{S}}
\newcommand{\CP}{\text{CP}}
\newcommand{\Torus}{\text{T}}

\newcommand{\struc}{f}

\newcommand{\alg}[1]{\mathfrak{#1}}
\newcommand{\gen}[1]{{#1}}
\newcommand{\comm}[2]{[#1,#2]}
\newcommand{\acomm}[2]{\{#1,#2\}}
\newcommand{\rep}[1]{\mathbf{#1}}

\newcommand{\adj}[3]{\left(\text{ad}_{#1}\right)^{#2}(#3)}
\newcommand{\adjb}[2]{\text{ad}_{#1}(#2)}
\newcommand{\adjbb}[1]{\text{ad}_{#1}}

\newcommand{\Ya}[1]{\mathcal{Y}(#1)}
\newcommand{\genY}[1]{\widehat{\mathfrak{#1}}}
\newcommand{\copro}{\Delta}

\begin{document}
\thispagestyle{empty}

\begingroup\parindent0pt
\begin{flushright}
{\small DMUS-MP-12/09}\\
{\small ITP-UU-12/39}\\
{\small SPIN-12/36}
\end{flushright}
\vspace*{5em}
\begingroup\LARGE
On the massless modes of the $\AdS_3/\CFT_2$ integrable systems
\par\endgroup
\vspace{2em}
\begingroup\large
Olof Ohlsson Sax${}^1$, Bogdan Stefa\'nski, jr.${}^2$ and Alessandro Torrielli${}^3$
\par\endgroup
\vspace{4em}
\begingroup\itshape
1. Institute for Theoretical Physics and Spinoza Institute, Utrecht University, 3508 TD Utrecht, The Netherlands

2. Centre for Mathematical Science, City University of London, Northampton Square, EC1V 0HB London, UK

3. Department of Mathematics, University of Surrey, Guildford, GU2 7XH, UK\par\endgroup
\vspace{1em}
\begingroup\ttfamily
1. O.E.OlssonSax@uu.nl

2. Bogdan.Stefanski.1@city.ac.uk

3. a.torrielli@surrey.ac.uk
\par\endgroup
\vspace{4em}
\endgroup

\paragraph{Abstract.}
We make a proposal for incorporating massless modes into the spin-chain of the $AdS_3/CFT_2$ integrable system. We do this by considering the $\alpha\rightarrow 0$ limit of the alternating $\alg{d}(2,1;\alpha)^2$ spin-chain constructed in~arXiv:1106.2558. In the process we encounter integrable spin-chains with non-irreducible representations at some of their sites. We investigate their properties and construct their R-matrices in terms of Yangians.

\newpage

\tableofcontents{}

\section{Introduction, review and summary}

The AdS/CFT correspondence~\cite{Maldacena:1997re,Witten:1998qj,Gubser:1998bc} provides a remarkable non-perturbative duality between quantum gauge and gravity theories. For certain classes of dual theories integrability has given a detailed, calculable description of how (in the planar limit) the correspondence works. For a review and a more complete list of references see~\cite{Beisert:2010jr,Arutyunov:2009ga}. Within the AdS/CFT correspondence, integrable structures were first identified in the case of the maximally supersymmetric duality between type IIB superstring theory on $AdS_5\times S^5$ and ${\cal N}=4$ super-Yang-Mills (SYM) $SU(N)$ gauge theory~\cite{Minahan:2002ve,Bena:2003wd}. However, we now know that integrability gives a handle on several other classes of dual pairs. These include other $AdS_5/CFT_4$ duals such as orbifolds, orientifolds and deformations (for a review and more complete list of references see~\cite{Zoubos:2010kh}) of the ${\cal N}=4$ dual pair.\footnote{Integrability in the context of AdS/CFT of other four-dimensional conformal gauge theories~\cite{Gaiotto:2009gz,ReidEdwards:2010qs,Colgain:2011hb,Aharony:2012tz} has been investigated in~\cite{Gadde:2009dj,Gadde:2010zi,Gadde:2010ku,Liendo:2011xb,Liendo:2011wc}.} Remarkably, integrability has also been instrumental in understanding the $AdS_4/CFT_3$ duality of the ABJM Chern-Simons theory~\cite{Bagger:2007jr,Aharony:2008ug} as initiated in~\cite{Minahan:2008hf,Gaiotto:2008cg,Gromov:2008qe} and Type IIA string theory on $AdS_4\times CP^3$~\cite{Arutyunov:2008if,Stefanski:2008ik,Gomis:2008jt}.\footnote{For a review and a much more extensive list of references see~\cite{Klose:2010ki}.} 

\subsection{The \texorpdfstring{$AdS_3/CFT_2$}{AdS3/CFT2} dualities}

More recently, the integrability approach to the $AdS_3/CFT_2$ correspondence~\cite{Maldacena:1997re,Maldacena:1998bw,Seiberg:1999xz}  has been developed in~\cite{Babichenko:2009dk}. Related recent advances in this area include~\cite{David:2008yk,David:2010yg,OhlssonSax:2011ms,Rughoonauth:2012qd,Sundin:2012gc,Cagnazzo:2012se}.  The gravity side of these dualities is given by Type IIB superstring theory on $AdS_3\times S^3\times M$ where $M=T^4$ or $M=S^3\times S^1$.~\footnote{Another background of this type is $M=K3$. However, for our analysis it can be viewed as an orbifold of $T^4$.} For brevity, we will henceforth refer to these two backgrounds, and their corresponding dualities, as $T^4$ and $S^1$, respectively. In order to satisfy the supergravity equations of motion, the radii of the $AdS_3$ and $S^3$ factors are related to one another
\begin{eqnarray}
T^4\qquad\qquad &:& \frac{1}{R_{AdS}^2}=\frac{1}{R_{S}^2} \\
S^1\qquad\qquad &:& \frac{1}{R_{AdS}^2}=\frac{1}{R_{+}^2}+\frac{1}{R_{-}^2}\,.
\end{eqnarray}
Above, $R_\pm$ are the radii of the two three-spheres in the $S^1$ geometry, and because of the above relation it is convenient to define $\alpha$ and $\phi$ as
\begin{equation}
\alpha\equiv \sin^2\phi\equiv\frac{R_{AdS}^2}{R_{+}^2}\,.
\end{equation}
As $\alpha\rightarrow 0$ the $S^1$ theory becomes the (decompactified) $T^4$ theory.\footnote{The limit $\alpha\rightarrow 1$ is identical to the $\alpha\rightarrow 0$ limit: the role of the two $S^3$ factors is interchanged. Throughout the text we wil only refer to the $\alpha\rightarrow 0$ limit and all our results will apply equally to $\alpha\rightarrow 1$.}
The radii and other moduli of the $S^1$ and $T^4$ factors can be chosen freely. Both backgrounds also require fluxes to support the geometries, which can be suitable combinations of R-R and NS-NS fluxes. Throughout this paper we will consider backgrounds with only R-R fluxes.\footnote{Type IIB S-duality relates these backgrounds to more general combinations of fluxes. A detailed analysis of the $T^4$ dual pair's moduli space, including S-duality is given in~\cite{Larsen:1999uk}.} Our motivation for this comes from the observation that in higher-dimensional $AdS$ R-R backgrounds, integrability has been very successful at understanding 
the (stringy) quantum gravity at small string coupling. We expect this to continue to be the case for $AdS_3/CFT_2$ as well. In particular, we will investigate the dual pairs as functions of a 't Hooft-like~\cite{'tHooft:1973jz} parameter $\lambda\equiv 4\pi^2 T^2$, where $T=R^2_{AdS}/2\pi\alpha'$ is the effective string tension.

The $T^4$ and $S^1$ backgrounds preserve 16 real supersymmetries\footnote{This is the maximal number of supersymmetries allowed for a background with an $AdS_3$ factor.} which combine with the bosonic Lie-algebra symmetries  into two different Lie super-algebras\footnote{The appearance of the super-algebra $\alg{d}(2,1;\alpha)$ in this setting was first observed in~\cite{Gauntlett:1998kc}.}
\begin{align*}
  M=T^4\qquad \qquad &: \qquad \qquad \alg{psu}(1,1|2)^2 \\
  M=S^3\times S^1\qquad \qquad &: \qquad \qquad \alg{d}(2,1;\alpha)^2\,.
\end{align*}
In two dimensions the conformal symmetry algebra is infinite-dimensional. The above Lie super-algebras are the finite-dimensional part of the full super-Virasoro algebras known as small and large $(4,4)$ super-conformal algebras for $T^4$ and $S^1$, respectively. Throughout this paper we will be only concerned with the Lie super-algebra symmetries.

The $CFT_2$ dual for $T^4$ is expected to be (a blow-up of) the $(4,4)$-supersymmetric sigma model $Sym^N(T^4)$. This is suggested by the following observation. The $T^4$ geometry is the near-horizon limit of a D1-D5 system in flat space. The open strings living on the D1-D5 intersection at low energies are described by a (non-conformal) 1+1-dimensional SYM. This gauge theory is expected to flow in the infrared to the $Sym^N(T^4)$ sigma-model. However, as has been known for some time~\cite{Seiberg:1999xz}, the point in moduli space described by the $Sym^N(T^4)$ orbifold is in many ways atypical, and it is not clear to what extent it can be regarded as the analogue of the free gauge theory in higher dimensions. In~\cite{Pakman:2009mi} an attempt was made to identify an integrable spin-chain in the sigma-model orbifold directly. However, perhaps because of a large amount of mixing, it is much harder to do this than in ${\cal N}=4$ SYM in four dimensions, and so a Minahan-Zarembo type spin-chain description of the $T^4$ CFT is still missing. The situation is more murky still in the case of the $CFT_2$ dual of $S^1$ where at present no suitable dual is known. A detailed discussion of this can be found in~\cite{Gukov:2004ym}. 

\subsection{The \texorpdfstring{$AdS_3/CFT_2$}{AdS3/CFT2} integrable systems and missing massless modes}

The above-mentioned obstacles make it difficult to study the $CFT_2$ side of the dualities directly. The string theory side of both the $T^4$ and $S^1$ backgrounds is better understood. For example, the string action, in a suitable kappa gauge, is known explicitly for both backgrounds. It consists of a Metsaev-Tseytlin coset action together with four/one extra free bosons describing the $T^4$/$S^1$ part of the geometry, respectively. As a result the equations of motion and Bianchi identities that follow from the action are equivalent to the flatness condition of a certain Lax connection. In other words, the theory is classically integrable. In~\cite{Babichenko:2009dk} finite gap equations were constructed from the monodromy matrix of the Lax connection. In~\cite{Babichenko:2009dk,OhlssonSax:2011ms} an all-loop Bethe ansatz (BA) was conjectured, which in the thermodynamic limit reduced to the string theory finite-gap equations. 

One might expect that, up to the knowledge of the so-called dressing phase and function $h(\lambda)$ which enters the dispersion relation, this all-loop BA should describe the spectral problem associated with both the $T^4$ and $S^1$ dual pairs. However, as was already noted in~\cite{Babichenko:2009dk}, the finite-gap equations obtained from the string theory monodromy matrix do not incorporate all stringy excitations. 

To see which states are missing from the finite-gap equations it is easiest to look at the BMN limit~\cite{Berenstein:2002jq}, where the string spectrum is known exactly~\cite{Berenstein:2002jq,Russo:2002rq,Lu:2002kw,Hikida:2002in,
Gomis:2002qi,Gava:2002xb,Sommovigo:2003kd}. In the BMN limit, any state in the string spectrum can be built up by acting on the groundstate $\ket{0}$ with bosonic and fermionic creation operators
\begin{equation}
  \alpha_{-n;m_i}^i\,, \qquad \psi_{-n;m_i}^i\,,
\end{equation}
and imposing a suitable level matching condition.
The creation operators are labeled by $n\in{\bf N}$, the Fourier-mode of a string coordinate, $i$ the target-space directions transverse to the light-cone, and $m_i$ which is called the {\em mass} of a particular excitation.\footnote{The mass $m_i$ does not depend on $n$.}
 For example, a state of the form
\begin{equation}
\psi_{-n;m_i}^i \ket{0}\,,
\end{equation}
has energy
\begin{equation}
\sqrt{m_i^2+n^2}\,.
\end{equation}
The $T^4$ theory has 
four target space directions $i$ for which the bosonic and fermionic excitations have $m_i=1$ (in suitable units) and four directions $i$ for which the bosonic and fermionic excitations have $m_i=0$. The $S^1$ theory
has two target space directions $i$ for which the bosonic and fermionic excitations have masses $m_i=1,\alpha,1-\alpha$ and $0$. The oscillators with $m_i=0$  lead to a degeneracy of the groundstate, since, for example, 
\begin{equation}
\psi_{n=0;m_i=0}^i\ket{0}\,,
\end{equation}
has zero energy.

It has been shown in~\cite{Babichenko:2009dk} that the finite-gap equations are incomplete. For example, they do not capture BMN-type solutions with $m_i=0$. In other words, we are ``missing'' four/two {\em massless} bosonic modes (and their fermionic superpartners) in the $T^4/S^1$ theories, respectively. As a result, the all-loop BA does not contain all the information about the spectral problem of these $AdS_3$ theories. Nevertheless, given that the all-loop BA is consistent -- for example, the S-matrix on which it is based satisfies the Yang-Baxter equation (YBE) -- we expect that  including the massless modes will modify the BA, rather than change its form entirely. 

Notice that as $\alpha\rightarrow 0$ the number of massless modes changes. In other words, the number of modes not captured by the all-loop BA changes in this limit. This observation will play an important role in what follows.

\subsection{Spin-chains for \texorpdfstring{$AdS_3/CFT_2$}{AdS3/CFT2}}

As we have already mentioned, the $CFT_2$ side of the duality is considerably less-well understood than the corresponding gauge theories of higher-dimensional dual pairs. A direct construction of an integrable spin-chain from the $CFT_2$ side seems challenging at present\cite{Gadde:2009dj,Gadde:2010zi,Gadde:2010ku,Liendo:2011xb,Liendo:2011wc}. In~\cite{OhlssonSax:2011ms}, a different approach was adopted: starting from the all-loop BAs of the $T^4$ and $S^1$ theories~\cite{Babichenko:2009dk,OhlssonSax:2011ms} small $\lambda$ BAs were extracted.\footnote{To do this one needs to assume that the dressing phase and the ubiquitous function $h(\lambda)$, which enters the dispersion relation, have suitable small $\lambda$ behavior. This assumption can be roughly thought of as an assertion that an integrable, local spin-chain description of the integrable system exists at small values of $\lambda$.} A homogeneous (respectively, alternating) integrable spin-chain was then constructed, whose spectrum could be computed using the weak coupling BA of the $T^4$ (respectively, $S^1$) theory. Regardless of their relation to $CFT_2$, these spin-chains can be viewed as a discretization of the string theory at small $\lambda$.
 
It was shown~\cite{OhlssonSax:2011ms} that certain solutions of the weak coupling BA become singular in the $\alpha\rightarrow 0$ limit. These solutions correspond to the increase in the number of missing modes in the BA as $\alpha\rightarrow 0$. What is remarkable however, is that the R-matrix, which encodes the integrable structure of the spin-chain remains well behaved as $\alpha\rightarrow 0$. In other words, the integrable structure of the spin-chain remains well-behaved in this limit. Since the number of missing modes changes at these values of $\alpha$, by keeping track of the integrable structure, we should learn about the fate of the missing modes. 
Indeed it was already understood in~\cite{OhlssonSax:2011ms} that in the $\alpha\rightarrow 0$ limit the alternating $S^1$ spin-chain and its R-matrix contained the homogeneous $T^4$ spin-chain and its R-matrix as part of a larger state space. 

\subsection{Plan of paper}

It is the purpose of this paper to investigate the spin-chains that arise in the $\alpha\rightarrow 0$ limit and show how one can understand the missing modes as excitations of such spin-chains. It will turn out that these spin-chains are somewhat unconventional. Firstly, the representations that enter the spin-chain are no longer irreducible at $\alpha= 0$. Secondly, for the limiting values of $\alpha$, there is no value of the spectral parameter where the R-matrix is proportional to the permutation operator. As such, we cannot immediately apply the conventional methods used in the study of integrable spin-chains. Nevertheless, the R-matrix encodes all the relevant physical information and we will use it to understand the $\alpha= 0$ spin-chain and its interactions.

This paper is organized as follows. In section~\ref{inof} we review the $\alg{sl}(2)$ representations that enter into the construction of the spin-chains paying particular attention to the $\alpha\rightarrow 0$ limit, where some of the representations are no longer irreducible. In order to understand better the R-matrix that enters our spin-chains in section~\ref{Yanu} we construct the Yangian for the relevant $\alg{sl}(2)$ representations using Drinfeld's second realisation.  In sections~\ref{ddddd} and~\ref{surmat}, 
 we show that the R-matrices given in~\cite{OhlssonSax:2011ms} (and reviewed in section~\ref{altern}), restricted to $\alg{sl}(2)$ and $\alg{sl}(2|1)$ subsectors can be obtained
from universal R-matrix expressions using the Yangians we have constructed in Drinfeld's second realisation. We take this as important evidence of the validity of our R-matrices, especially for $\alpha=0$. In section~\ref{altern} we review the construction of the spin-chains and R-matrices considered in~\cite{OhlssonSax:2011ms} and discuss some of their properties.
 In section~\ref{sec7} we investigate the spin-chain at $\alpha=0$. We argue that for this value of $\alpha$ the alternating spin-chain involves reducible representations, and, using the R-matrix as our guide, we identify a suitable notion of local interactions for such spin-chains. In section~\ref{sec8} we construct an Algebraic Bethe Ansatz (ABA) for the $\alpha=0$ spin-chain. 

Up to this point the spin-chain discussion will be in the small $\lambda$ regime. We will show that, as a result of working in the small $\lambda$ approximation, the spin-chain has a very large degeneracy of groundstates at $\alpha=0$. In section~\ref{sec9} we show that this is in fact an artifact of the small-lambda approximation, and that when higher order interactions are included, much of the degeneracy is lifted. In particular, we show that the remaining ground-state degeneracy is precisely what one expects from the BMN limit analysis. In this way we believe we are able to incorporate the missing modes  into the spin-chain description of the $AdS_3/CFT_2$ integrable system.

\section{A review of certain \texorpdfstring{$\alg{sl}(2)$}{sl(2)} representations}\label{inof}

In this section, we fix our conventions for the $\alg{sl}(2)$ Lie algebra. 
The motivation for treating $\alg{sl}(2)$ first, and only later turning to the $\alg{sl}(2|1)$ case of relevance to the superstring, is due to the fact that one can learn a great deal from this simplified setting. Many properties that will turn out to be quite crucial for the supersymmetric case are best observed when dealing with $\alg{sl}(2)$ representation, especially the distinction between the various modules at special points in the moduli space of the representation parameters and the issues related to unitarity. This treatment will also serve as an illustration of the Yangian algebra techniques we will be using, before applying them to  $\alg{sl}(2|1)$.

\subsection{Defining relations}

The basic commutation relations are given by
\begin{eqnarray}
\label{relazionizero}
&&[h,e]=2e, \qquad \qquad \, \, \, \, \, \, \, \, [h,f]=-2f, \qquad \qquad \, \, \, \, \, \, \, \, \, [e,f]=h,
\end{eqnarray}
for $h$ Cartan element, $e$ raising and $f$ lowering operator. The quadratic Casimir of the algebra is
\begin{eqnarray}
C_2 = \frac{h^2}{2}  \, + \, e f \, + \, f e. 
\end{eqnarray}
Let us focus on infinite-dimensional modules parameterized by a complex variable $s$, and choose one among the various representations available. This choice is not too restrictive, however, since in general one can find an isomorphism between different representations, as long as $s$ is strictly non-zero. One has (see for instance \cite{Perelomov:1986tf,Zhang:1990fy}) the following representation in terms of differential operators:
\begin{align}
\label{cho1}
&{P^{{\bf s}}}^3 = z \, \partial_z \, - s, \qquad {P^{{\bf s}}}^- = z^2 \, \partial_z \, - 2 \, s \, z, \qquad {P^{{\bf s}}}^+ = - \partial_z.
\end{align} 
One can prove that the identification with (\ref{relazionizero}) is done as follows:
\begin{align}
\label{assigno}
&h = 2 \, {P^{{\bf s}}}^3, \qquad e = {P^{{\bf s}}}^-, \qquad f = {P^{{\bf s}}}^+.
\end{align} 
The Casimir in this representation is equal to 
\begin{eqnarray}
\label{Casimiroinfinito}
C_2 = 2 \, s (s+1). 
\end{eqnarray}
Another representation with the same assignment (\ref{assigno}) (with ${P^{{\bf s}}}$ everywhere replaced by $ {S^{{\bf s}}}$) and same Casimir (\ref{Casimiroinfinito}) is given by
\begin{align}
\label{cho11}
& {S^{{\bf s}}}^3 = z \, \partial_z \, - s, \qquad  {S^{{\bf s}}}^- = -z, \qquad  {S^{{\bf s}}}^+ = -2\, s \, \partial_z \, + z \, \partial_z^2.
\end{align}
There exists a transformation between the two representations we just described, namely
\begin{eqnarray}
\label{mappato1}
\Psi^{-1} \,  {S^{{\bf s}}}^\pm \, \Psi \, = - {P^{{\bf s}}}^\pm, \qquad \Psi^{-1} \,  {S^{{\bf s}}}^3 \, \Psi \, = {P^{{\bf s}}}^3.
\end{eqnarray}
The formal expression for $\Psi$ is
\begin{eqnarray}
\Psi = \frac{1}{\Gamma \left(z \partial_z \, - 2 s\right)}.
\end{eqnarray}
This expression becomes well-defined on eigenstates of the `number operator'
\begin{eqnarray}
\label{numero}
\widehat{N} \equiv z \, \partial_z,
\end{eqnarray} 
given by 
\begin{eqnarray}
\label{osci1}
|n\rangle \equiv z^n, \qquad \qquad n\geq0.
\end{eqnarray} 
Another representation is given by (cf.~\cite{Holstein:1940zp})
\begin{align}
\label{cho12}
& {R^{{\bf s}}}^3 |n\rangle = (n-s)|n\rangle, \qquad  {R^{{\bf s}}}^- |n\rangle= - \sqrt{(n+1)(n-2s)}|n+1\rangle, \qquad  {R^{{\bf s}}}^+ |n\rangle = \sqrt{n \,(n-2s-1)} |n-1\rangle.
\end{align}
The assignment is still given by (\ref{assigno}) (with ${P^{{\bf s}}}$ everywhere replaced by $ {R^{{\bf s}}}$) and the Casimir still given by (\ref{Casimiroinfinito}). We can perform a similarity transformation on the representation (\ref{cho12}), namely
\begin{eqnarray}
\label{mappato22}
\gamma^{-1} \,  {R^{{\bf s}}}^\pm \, \gamma \, = \pm \, {R^{{\bf s}}}^\pm_{HO}, \qquad \gamma^{-1} \,  {R^{{\bf s}}}^3 \, \gamma \, = {R^{{\bf s}}}^3_{HO},
\end{eqnarray}
with 
\begin{eqnarray}
\label{trafo1}
\gamma = \sqrt{\frac{1}{\Gamma(\widehat{N}+1)}}.
\end{eqnarray}
The transformation (\ref{trafo1}) is regular on any state $|n\rangle$, since the eigenvalues of $\widehat{N}+1$ are strictly greater than zero on any such state. The resulting representation is closer to Holstein and Primakoff's original representation \cite{Holstein:1940zp}, and is manifestly unitary for $s$ real and $s<0$. In fact, if we  
introduce a set of 
oscillators $a \leftrightarrow \partial_z$, $a^\dag \leftrightarrow z$, such that
\begin{align}
\label{osci}
&[a,a^\dag ]=1, \qquad |n\rangle = {a^\dag}^n \, |0\rangle, \qquad n = 0,1,2,\dotsc\,.
\end{align} 
we immediately have
\begin{align}
\label{cho122}
{R^{{\bf s}}}^-_{HO} = a^\dag \, (a^\dag a - 2 s)^{\frac{1}{2}}, \qquad  {R^{{\bf s}}}^+_{HO} = (a^\dag a - 2 s)^{\frac{1}{2}} \,  a.
\end{align}
If we canonically assign the hermiticity property
\begin{eqnarray}
(a)^\dag = a^\dag
\end{eqnarray}
we see that $({R^{{\bf s}}}^-_{HO})^\dag = {R^{{\bf s}}}^+_{HO}$, while ${R^{{\bf s}}}^3_{HO} = \frac{1}{2} \, [{R^{{\bf s}}}^+_{HO},{R^{{\bf s}}}^-_{HO}]$ is hermitian.
 
There exists a transformation between the `$ {R^{{\bf s}}}$' representation (\ref{cho12}) and the `${P^{{\bf s}}}$' one (\ref{cho1}), namely
\begin{eqnarray}
\label{mappato2}
\xi^{-1} \,  {R^{{\bf s}}}^\pm \, \xi \, = - {P^{{\bf s}}}^\pm, \qquad \xi^{-1} \,  {R^{{\bf s}}}^3 \, \xi \, = {P^{{\bf s}}}^3.
\end{eqnarray}
The formal expression for $\xi$ is
\begin{eqnarray}
\label{trafo2}
\xi = \sqrt{\frac{\Gamma(\widehat{N}+1)}{\Gamma(\widehat{N}-2s)}},
\end{eqnarray}
which again becomes well defined on eigenstates $|n\rangle$ of the number operator. The transformation (\ref{trafo2}) is regular on all states $|n\rangle$ as long as $s \neq 0$, which implies via (\ref{mappato2}) the unitarity of the representation (\ref{cho1}) for $s$ real and $s<0$. From these observations we can already anticipate from the next section that for $s=0$ the representation (\ref{cho12}) will remain unitary, while the one in (\ref{cho1}) will not (similarly for the (\ref{cho11}) representation). 

In the list provided by \cite{Maldacena:2000hw}, the above equivalent representations all correspond to the one dubbed \emph{principal discrete} representation\footnote{This representation is referred to as \emph{lowest weight} in \cite{Maldacena:2000hw} (case (2), page 17). However, we will use the terminology \emph{highest weight} throughout this paper to refer to this very representation or to any representation with a highest or lowest weight, and speak about \emph{highest weight states} without distinction.}. One needs to assign 
\begin{align}
\label{assignoMO}
&h = 2 \, J^3, \qquad e = J^+, \qquad f = - J^-, \qquad s = -j, \qquad |0\rangle = |j;j\rangle,
\end{align} 
where $J^3$, $J^\pm$, $j$ and $|j;m\rangle$ are used in \cite{Maldacena:2000hw}, while $h$, $e$, $f$, $s$ and $|n\rangle$ are used in this paper. It follows from (\ref{assignoMO}) and from the description in \cite{Maldacena:2000hw} that unitary representations are obtained for $s$ real and $s<0$, which is what we have found by direct observation in (\ref{cho12}). Since we are working with the universal cover of $AdS_3$ (hence, time is a non-compact variable), the parameter $s$ can be any negative real number.

\subsection{\texorpdfstring{$s \to 0$}{s to 0} limit}

In this sub-section we consider the $s \to 0$ limit of the above representations.\footnote{We are grateful to Joe Chuang for explanations of this and related points.}
The representations that arise in this limit will play a central role throughout much of what follows. In the $s \to 0$ limit the similarity transformations $\Psi$, $\xi$ and $\Psi \, \xi^{-1}$ become singular when acting on the state $|0\rangle$, and the limiting representations, denoted respectively P, S and R, are no longer equivalent\footnote{
Notice that the three limiting modules still have the same value of the quadratic Casimir $C_2=0$ at $s=0$, since the modules before the limit do share the value (\ref{Casimiroinfinito}).}. We discuss them in more detail below. In particular, we show that all three modules become reducible.

\subsubsection{$P$ module}\label{Pmodulo}
Such representation has been intensively studied in the literature due to its connections with high-energy QCD (see for instance \cite{Korchemsky:2010kj} and references therein). Sometimes it is called the \emph{dual Verma module}. In terms of oscillators $a$, $a^\dag$ and states $|n\rangle$
one has
\begin{align}
&P^3 = a^\dag \, a, \qquad P^- = {a^\dag}^2 \, a, \qquad P^+ = - a.
\label{Pmod}
\end{align} 
One can prove that the identification with (\ref{relazionizero}) is done as follows:
\begin{align}
&h = 2 \, P^3, \qquad e = P^-, \qquad f = P^+.
\end{align} 
The action on states is given by
\begin{align}
\label{azin1}
&P^3 |n\rangle = n \, |n\rangle, \qquad P^- |n\rangle = n \, |n+1\rangle, \qquad P^+ |n\rangle = - n \, |n-1\rangle.
\end{align}
This module corresponds to an indecomposable representation, because all generators annihilate the state $|0\rangle$, however $P^+$ connects $|1\rangle$ and $|0\rangle$. 

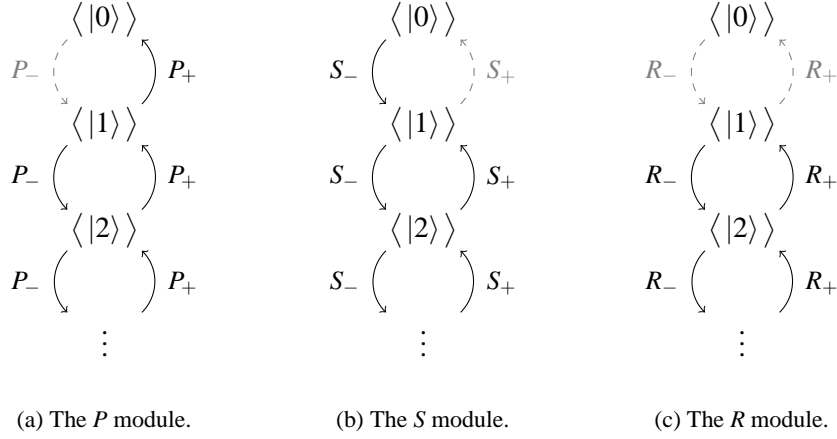
\begin{figure}
  \centering

  \subfloat[The $P$ module.\label{fig:P-module}]{
    \begin{tikzpicture}
      \useasboundingbox (-2cm,0.5cm) rectangle (2cm,-5cm);

      \node (phi0) at (0,-0\vertdist) [state] {$\big< \ket{0} \big>$};
      \node (phi1) at (0,-1\vertdist) [state] {$\big< \ket{1} \big>$};
      \node (phi2) at (0,-2\vertdist) [state] {$\big< \ket{2} \big>$};
      \node (phi3) at (0,-3\vertdist) [state] {\makebox[\nodewidth][c]{$\vdots$}};

      \draw [arrow down,gray,dashed] (phi0.south west) to (phi1.north west);
      \draw [arrow down] (phi1.south west) to (phi2.north west);
      \draw [arrow down] (phi2.south west) to (phi3.north west);

      \draw [arrow up] (phi1.north east) to (phi0.south east);
      \draw [arrow up] (phi2.north east) to (phi1.south east);
      \draw [arrow up] (phi3.north east) to (phi2.south east);

      \node at (-\hordist,-0.5\vertdist) [anchor=east,gray] {\small $P_-$};
      \node at (-\hordist,-1.5\vertdist) [anchor=east] {\small $P_-$};
      \node at (-\hordist,-2.5\vertdist) [anchor=east] {\small $P_-$};

      \node at (+\hordist,-0.5\vertdist) [anchor=west] {\small $P_+$};
      \node at (+\hordist,-1.5\vertdist) [anchor=west] {\small $P_+$};
      \node at (+\hordist,-2.5\vertdist) [anchor=west] {\small $P_+$};
    \end{tikzpicture}
  }
  \subfloat[The $S$ module.\label{fig:S-module}]{
    \begin{tikzpicture}
      \useasboundingbox (-2cm,0.5cm) rectangle (2cm,-5cm);

      \node (phi0) at (0,-0\vertdist) [state] {$\big< \ket{0} \big>$};
      \node (phi1) at (0,-1\vertdist) [state] {$\big< \ket{1} \big>$};
      \node (phi2) at (0,-2\vertdist) [state] {$\big< \ket{2} \big>$};
      \node (phi3) at (0,-3\vertdist) [state] {\makebox[\nodewidth][c]{$\vdots$}};

      \draw [arrow down] (phi0.south west) to (phi1.north west);
      \draw [arrow down] (phi1.south west) to (phi2.north west);
      \draw [arrow down] (phi2.south west) to (phi3.north west);

      \draw [arrow up,gray,dashed] (phi1.north east) to (phi0.south east);
      \draw [arrow up] (phi2.north east) to (phi1.south east);
      \draw [arrow up] (phi3.north east) to (phi2.south east);

      \node at (-\hordist,-0.5\vertdist) [anchor=east] {\small $S_-$};
      \node at (-\hordist,-1.5\vertdist) [anchor=east] {\small $S_-$};
      \node at (-\hordist,-2.5\vertdist) [anchor=east] {\small $S_-$};

      \node at (+\hordist,-0.5\vertdist) [anchor=west,gray] {\small $S_+$};
      \node at (+\hordist,-1.5\vertdist) [anchor=west] {\small $S_+$};
      \node at (+\hordist,-2.5\vertdist) [anchor=west] {\small $S_+$};
    \end{tikzpicture}
  }
  \subfloat[The $R$ module.\label{fig:R-module}]{
    \begin{tikzpicture}
      \useasboundingbox (-2cm,0.5cm) rectangle (2cm,-5cm);

      \node (phi0) at (0,-0\vertdist) [state] {$\big< \ket{0} \big>$};
      \node (phi1) at (0,-1\vertdist) [state] {$\big< \ket{1} \big>$};
      \node (phi2) at (0,-2\vertdist) [state] {$\big< \ket{2} \big>$};
      \node (phi3) at (0,-3\vertdist) [state] {\makebox[\nodewidth][c]{$\vdots$}};

      \draw [arrow down,gray,dashed] (phi0.south west) to (phi1.north west);
      \draw [arrow down] (phi1.south west) to (phi2.north west);
      \draw [arrow down] (phi2.south west) to (phi3.north west);

      \draw [arrow up,gray,dashed] (phi1.north east) to (phi0.south east);
      \draw [arrow up] (phi2.north east) to (phi1.south east);
      \draw [arrow up] (phi3.north east) to (phi2.south east);
      
      \node at (-\hordist,-0.5\vertdist) [anchor=east,gray] {\small $R_-$};
      \node at (-\hordist,-1.5\vertdist) [anchor=east] {\small $R_-$};
      \node at (-\hordist,-2.5\vertdist) [anchor=east] {\small $R_-$};

      \node at (+\hordist,-0.5\vertdist) [anchor=west,gray] {\small $R_+$};
      \node at (+\hordist,-1.5\vertdist) [anchor=west] {\small $R_+$};
      \node at (+\hordist,-2.5\vertdist) [anchor=west] {\small $R_+$};
    \end{tikzpicture}
  }
  
  \caption{Pictorial representations of the $P$, $S$ and $R$ modules. The grayed out
    and dashed lines indicates the action of generators that vanish on a
    specific state, rendering the corresponding module reducible.}
  \label{fig:PRS-modules}

\end{figure}

It is straightforward to show that the module (\ref{cho1}) tends to the $P$ module in the limit. The module (\ref{cho1}) corresponds in fact to an irreducible representation.
Its action on the states $|n\rangle$ is given by
\begin{align}
\label{gens1}
&{P^{{\bf s}}}^3 |n\rangle = (n-s) \, |n\rangle, \qquad {P^{{\bf s}}}^- |n\rangle = (n - 2 s) \, |n+1\rangle, \qquad {P^{{\bf s}}}^+ |n\rangle = - n \, |n-1\rangle.
\end{align}
When $s$ goes to zero, the state $|0\rangle$ is annihilated by all generators, and it generates an irreducible one-dimensional submodule (`singlet'). We have summarized the situation in 
Fig. (\ref{fig:PRS-modules}).

\subsubsection{$S$ module}\label{Smodulo}
The action on states in the representation (\ref{cho11}) is given by
\begin{align}
\label{gens2}
& {S^{{\bf s}}}^3 |n\rangle = (n-s) \, |n\rangle, \qquad  {S^{{\bf s}}}^- |n\rangle = \, - |n+1\rangle, \qquad  {S^{{\bf s}}}^+ |n\rangle = - n \, (2 s - n +1) \, |n-1\rangle.
\end{align}
The limit $s \to 0$ provides us with the following module, conventionally called the \emph{Verma module}. Utilizing the same oscillators and states (\ref{osci}), one has
\begin{align}
&S^3 = a^\dag \, a, \qquad S^- = a^\dag, \qquad S^+ = -a^\dag \, a^2.
\end{align} 
One obtains identification with the generators (\ref{relazionizero}) if one defines
\begin{align}
&h = 2 \, S^3, \qquad e = S^-, \qquad f = S^+.
\end{align} 
The action on states is given by
\begin{align}
\label{azin2}
&S^3 |n\rangle = n \, |n\rangle, \qquad S^- |n\rangle = |n+1\rangle, \qquad S^+ |n\rangle = - n \, (n-1) \, |n-1\rangle.
\end{align}
This module corresponds to an indecomposable representation, because no generator can have the state $|0\rangle$ as an outcome, however $S^-$ connects $|0\rangle$ and $|1\rangle$, as depicted in Fig. (\ref{fig:PRS-modules}). 
 
\subsubsection{$R$ module}\label{Rmodulo}
Taking the limit starting from the representation (\ref{cho12}), and performing a further similarity transformation \emph{regular on all states}, one obtain an $s=0$ module we call the $R$ module. More specifically,
\begin{eqnarray}
\zeta^{-1} \,  {R^{{\bf s}}}^\pm (s=0)\, \zeta \, = - R^\mp, \qquad \zeta^{-1} \,  {R^{{\bf s}}}^3 (s=0)\, \zeta \, = R^3,
\end{eqnarray}
with $\zeta$ formally given by
\begin{eqnarray}
\zeta = \sqrt{\Gamma(\widehat{N}+1)}.
\end{eqnarray}
One finds this way
\begin{align}
\label{uno}
&R^3 = a^\dag \, a, \qquad R^- = a^\dag \, \sqrt{(a^\dag \, a)}, \qquad R^+ = - \sqrt{(a^\dag \, a)} \, \, a.
\end{align} 
Identification with (\ref{relazionizero}) is done as follows:
\begin{align}
&h = 2 \, R^3, \qquad e = R^-, \qquad f = R^+.
\end{align} 
The action on states is given by
\begin{align}
\label{azin3}
&R^3 |n\rangle = n \, |n\rangle, \qquad R^- |n\rangle = \sqrt{n} \, |n+1\rangle, \qquad R^+ |n\rangle = - n \, \sqrt{n-1} \, |n-1\rangle.
\end{align}
This module corresponds to a decomposable (or `completely reducible') representation, because all generators annihilate the state $|0\rangle$, and no generator can have the state $|0\rangle$ as an outcome, as depicted in Fig. (\ref{fig:PRS-modules}). The two irreducible components are, respectively, a trivial one-dimensional representation and an infinite-dimensional one isomorphic to the $s=-1$ limit of the general (\ref{cho1}) representation.

\bigskip\bigskip\noindent
Let us mention that the $s\to 0$ limit for the $P$ and $S$ module results in a breakdown of unitarity. It is not possible to find an inner product which preserves good hermiticity properties of the generators in these two modules.  This is not the case for the $R$ module, as can clearly be seen from (\ref{uno}). As we will see later, the $R$ module will be of relevance to the $AdS/CFT$ correspondence. From the point of view of integrability it may however be interesting to investigate all three types of representations, and throughout the paper we will often present results for all three of them.

\subsection{The representations \texorpdfstring{$P\otimes P$}{P x P}, \texorpdfstring{$R\otimes R$}{R x R}, and 
\texorpdfstring{$S\otimes S$}{S x S}}
\label{sec23}

In this subsection we review the decomposition of the representations $P\otimes P$, $R\otimes R$ and $S\otimes S$ into indecomposable (and mostly irreducible) sub-modules. We will start by noting that the tensor product of an irreducible representation $\mathbf{s}$ ($\mathbf{s} < 0$) with itself, can be decomposed as
\begin{equation}
  \mathbf{s} \otimes \mathbf{s} = (\mathbf{2s}) \oplus (\mathbf{2s-1}) \oplus (\mathbf{2s-2}) \oplus (\mathbf{2s-3}) \oplus \dotsb \,.
\end{equation}
Each of the representations on the right-hand side denotes an irreducible $\alg{sl}(2)$ representation. In the following subsections we will write down the corresponding decomposition for the different $\mathbf{s} \to 0$ multiplets.

\subsubsection{$P\otimes P$}

The decomposition of $P\otimes P$ is given by
\begin{equation}
P\otimes P\cong P \oplus {\bf -1}\oplus {\bf -2}\oplus {\bf -3}\oplus\dotsb.
\label{PPirep}
\end{equation}
The highest weight states of the irreducible representations on the right-hand side of (\ref{PPirep}) are given by
\begin{equation}
\label{pphw}
\left|l\right>_{12}\equiv (\ad_1-\ad_2)^l\ket{0}_1\otimes\ket{0}_2\,.
\end{equation}
The $\left|l=0\right>_{12}$ state is part of the $P$ module, while the $\left|l>0\right>_{12}$ are highest weight states for $s=-l$ modules. The expression (\ref{pphw}) matches formula (\ref{aiwp}) in what follows (for $j=l$).

\subsubsection{$S\otimes S$}

The decomposition of $S\otimes S$ is given by
\begin{equation}
S\otimes S\cong S \oplus {\bf -1}\oplus {\bf -2}\oplus {\bf -3}\oplus\dotsb
\label{SSirep}
\end{equation}
On the right-hand side of (\ref{SSirep}), the two highest weight states of the module $S$ are
\begin{equation}
\left|l=0\right>_{12}\equiv
\ket{0}_1\otimes\ket{0}_2\,\qquad
\mbox{and}\qquad 
(S^-_1+S^-_2)\left|l=0\right>_{12}\,,
\end{equation}
while the highest weight state of the ${\bf -1}$ module is
\begin{equation}
\left|l=1\right>_{12}
=(\ad_1-\ad_2)\ket{0}_1\otimes\ket{0}_2.
\,
\end{equation}
The highest weight states of the irreps ${\bf -2},{\bf -3},\dotsc$ are given by
\begin{equation}
\label{shw}
\left|n\right>_{12}\equiv \sum_{l=1}^{n-1}
\frac{n!(n-2)! \, (-\ad_1)^l(\ad_2)^{n-l}}{l!(l-1)!(n-l)!(n-l-1)!} 
\ket{0}_1\otimes\ket{0}_2\,,
\end{equation}
for $n=2,3,\dotsc$. The expression (\ref{shw}) matches formula (\ref{hwS}) in what follows (for $j=n$).

\subsubsection{$R\otimes R$}

Recall that $R\cong {\bf 0}\oplus {\bf -1}$, so the irreducible decomposition of $R\otimes R$ is given by
\begin{equation}
R\otimes R\cong {\bf 0}\oplus {\bf -1}_S \oplus {\bf -1}_A \oplus {\bf -2}\oplus {\bf -3}\oplus\dotsb
\label{RRirep}
\end{equation}
The highest weight states of the first three irreps of the right-hand side are
\begin{equation}
  \begin{gathered}
    \ket{l=0}_{12} \equiv \ket{0}_1 \otimes \ket{0}_2 \,, \\
    \ket{l=1_S}_{12} \equiv \tfrac{1}{2} \left(\ket{1}_1 \otimes \ket{0}_2 + \ket{0}_1 \otimes \ket{1}_2 \right) \,, \qquad
    \ket{l=1_A}_{12} \equiv \tfrac{1}{2} \left(\ket{1}_1 \otimes \ket{0}_2 - \ket{0}_1 \otimes \ket{1}_2 \right) \,,
  \end{gathered}
\end{equation}
where we used the subscript ${}_S$ and ${}_A$ to distinguish between the symmetric and anti-symmetric ${\bf -1}$ representations. The highest weight states of the ${\bf -2},{\bf -3},\dotsc$ irreducible representations in the $R\otimes R$ decomposition are
\begin{equation}
\label{rhw}
\left|n\right>_{12}\equiv \sum_{l=1}^{n-1}
\frac{n!\sqrt{(n-2)!} \, (-\ad_1)^l(\ad_2)^{n-l}}{l!\sqrt{(l-1)!}(n-l)!\sqrt{(n-l-1)!}} 
\ket{0}_1\otimes\ket{0}_2\,,
\end{equation}
for $n=2,3,\dotsc$. The expression (\ref{rhw}) matches formula (\ref{hwR0}) in what follows (for $j=n$).

\section{Yangians}\label{Yanu}

We will now review the theory of Yangians relevant to our goals. In this section, we provide the defining relations. For details, we refer the reader to \cite{Chari:1994pz,Molev:2003,Torrielli:2011gg}.

\subsection{Drinfeld's first realization}\label{ssec:drinf1}

Let us first focus on bosonic Lie algebras. The Yangian $\Ya{\alg{g}}$  is a deformation of the universal enveloping algebra
of the loop algebra $\alg{g}[u]$ associated to a Lie algebra $\alg{g}$.\footnote{We remind the reader that $\alg{g}[u]$ is the algebra of
$\alg{g}$-valued polynomials in the complex variable $u$.} Let
$\alg{g}$ be a finite dimensional simple Lie algebra generated by
$\gen{J}^A$ with commutation relations
$\comm{\gen{J}^A}{\gen{J}^B} = f^{AB}_C\gen{J}^C$, equipped
with a non-degenerate invariant consistent supersymmetric bilinear form $\kappa^{AB}$ (such as the Killing form $\kappa^{AB} = f^{AC}_D \, f^{BD}_C$). The
Yangian is defined by the following commutation
relations between the level-zero generators $\gen{J}^A$ and the level-one generators $\genY{J}^A$: \begin{eqnarray}
\label{def1}\comm{\gen{J}^A}{\gen{J}^B} = f^{AB}_C\gen{J}^C,\nonumber\\
\label{rels}\comm{\gen{J}^A}{\genY{J}^B} = f^{AB}_C\genY{J}^C. \end{eqnarray} 
The original Lie algebra $\alg{g}$ is a subalgebra of $\Ya{\alg{g}}$. Higher level generators are defined recursively by subsequent
commutation of these basic generators, subject to the following Serre
relations (for $\alg{g} \neq \alg{sl}(2)$): 
\begin{eqnarray}
\label{Serr}
\comm{\genY{J}^{A}}{\comm{\genY{J}^B}{\gen{J}^{C}}} +
\comm{\genY{J}^{B}}{\comm{\genY{J}^C}{\gen{J}^{A}}} +
\comm{\genY{J}^{C}}{\comm{\genY{J}^A}{\gen{J}^{B}}} \eq
\frac{1}{4} f^{AG}_{D}f^{BH}_Ef^{CK}_{F}f_{GHK}
\gen{J}^{\{D}\gen{J}^E\gen{J}^{F\}}.
\end{eqnarray}
Curly brackets enclosing indices indicate complete symmetrization. Indices are raised or lowered with $\kappa^{AB}$ or its inverse, respectively. For the algebra $\alg{sl}(2)$, the above Serre relations are trivial, and one needs to impose a more complicated set of relations (cf.\@ section~2.1.1 of \cite{MacKay:2004tc}).    

The Yangian is equipped with a Hopf algebra structure. The coproduct is uniquely
determined for all generators by specifying it on the level-zero
and -one generators as follows: 
\begin{eqnarray}
\label{coptr}\copro (\gen{J}^A) \eq\gen{J}^A\otimes \mathds{1}+\mathds{1}\otimes\gen{J}^A, \\
\label{cop}\copro( \, \genY{J}^A )\eq\genY{J}^A\otimes \mathds{1}+\mathds{1}\otimes\genY{J}^A+\frac{1}{2}
\struc^{A}_{BC}\gen{J}^B\otimes\gen{J}^C.
\end{eqnarray}
Antipode and counit are easily obtained
from the Hopf algebra definitions. 

\subsection{Drinfeld's second realization}\label{ssec:drinf2}

Drinfeld's second realization explicitly solves the
recursion left implicit in the first realization. It defines
$\Ya{\alg{g}}$ in terms of (simple root) generators $\kappa_{i,m},
\xi^\pm_{i,m}$, $i=1,\dots, \text{rank} \alg{g}$, $m=0,1,2,\dots$,
and relations
\begin{eqnarray}
\label{def:drinf2}
&[\kappa_{i,m},\kappa_{j,n}]=0,\quad [\kappa_{i,0},\xi^\pm_{j,m}]=\pm a_{ij} \,\xi^+_{j,m},\nonumber\\
& \comm{\xi^+_{j,m}}{\xi^-_{j,n}}=\delta_{i,j}\, \kappa_{j,n+m},\nonumber\\
&[\kappa_{i,m+1},\xi^\pm_{j,n}]-[\kappa_{i,m},\xi^\pm_{j,n+1}] = \pm \frac{1}{2} a_{ij} \{\kappa_{i,m},\xi^\pm_{j,n}\},\nonumber\\
&\comm{\xi^\pm_{i,m+1}}{\xi^\pm_{j,n}}-\comm{\xi^\pm_{i,m}}{\xi^\pm_{j,n+1}} = \pm\frac{1}{2} a_{ij} \acomm{\xi^\pm_{i,m}}{\xi^\pm_{j,n}},\nonumber\\
&i\neq j,\, \, \, \, n_{ij}=1+|a_{ij}|,\, \, \, \, \, Sym_{\{k\}} [\xi^\pm_{i,k_1},[\xi^\pm_{i,k_2},\dots [\xi^\pm_{i,k_{n_{ij}}}, \xi^\pm_{j,l}]\dots]]=0.
\end{eqnarray}
In these formulas, $a_{ij}$ is the Cartan matrix, which we will assume to be symmetric.

Drinfeld's first and second realization are isomorphic to each other. Let $\gen{H}_i,
\gen{E}_i^\pm$ be a Chevalley-Serre basis for $\alg{g}$, and
denote by $\genY{H}_i, \genY{E}_i^\pm$ the corresponding
level-one generators in the first realization of the Yangian.
Drinfeld \cite{Drinfeld:1987sy} gave the isomorphism
\begin{eqnarray}
\label{def:isom}
&\kappa_{i,0}=\gen{H}_i,\quad \xi^+_{i,0}=\gen{E}^+_i,\quad \xi^-_{i,0}=\gen{E}^-_i,\nonumber\\
&\kappa_{i,1}=\genY{{H}}_i-v_i,\quad \xi^+_{i,1}=\genY{{E}}^+_i-w_i,\quad \xi^-_{i,1}=\genY{{E}}^-_i-z_i,
\end{eqnarray}
where
\begin{equation}
  \begin{aligned}\label{def:specialel}
    v_i &= \frac{1}{4} \sum_{\beta\in\Delta^+}\left(\alpha_i,\beta\right)(\gen{E}_\beta^-\gen{E}_\beta^+ +  \gen{E}_\beta^+\gen{E}_\beta^-) - \frac{1}{2}\gen{H}_i^2, \\
    w_i &= \frac{1}{4}\sum_{\beta\in\Delta^+}  \left(\gen{E}_\beta^-\adjb{{E}_i^+}{\gen{E}_\beta^+} + \adjb{{E}_i^+}{\gen{E}_\beta^+}\gen{E}_\beta^- \right) -  \frac{1}{4}\acomm{\gen{E}_i^+}{\gen{H}_i}, \\
    z_i &= \frac{1}{4}\sum_{\beta\in\Delta^+} \left(\adjb{\gen{E}_\beta^-}{{E}_i^-}\gen{E}_\beta^+ + \gen{E}_\beta^+ \adjb{\gen{E}_\beta^-}{{E}_i^-} \right) - \frac{1}{4}\acomm{\gen{E}_i^-}{\gen{H}_i} .
  \end{aligned}
\end{equation}
$\Delta^+$ denotes the set of positive root vectors, $\gen{E}^\pm_\beta$
are generators of the Cartan-Weyl basis constructed from $\gen{H}_i, \, \gen{E}_i^\pm$, and the
adjoint action is defined as $\adjb{x}{y} = [x,y]$.

To obtain a quasi-triangular Hopf algebra one needs the double of the Yangian, which is obtained by adding a second set of generators with `negative' level $\kappa_{i,m},
\xi^\pm_{i,m}$, $i=1,\dots, \text{rank} \alg{g}$, $m=-1,-2,\dots$, satisfying the same relations (\ref{def:drinf2}). In addition, one has a suitable pairing between positive and negative level generators, which is used to construct the universal R-matrix \cite{Khoroshkin:1994uk}. Following Drinfeld, the object constructed in this way provides solutions to the Yang-Baxter equation (YBE) when projected into representations of (\ref{def:drinf2}).  

The generalization to the supersymmetric case is straightforward. Commutators become graded commutators, and the relations with the `wrong statistics' on the right-hand side of (\ref{def:drinf2}) and (\ref{def:specialel}) also take up the wrong statistics in the graded case accordingly. We will spell out these relations in the specific example of $\alg{sl}(2|1)$ later on (see formula (\ref{def:drinf2-sl21})). 

\subsection{The Yangian of \texorpdfstring{$\alg{sl}(2)$}{sl(2)} in infinite-dimensional representations}\label{moduli}

Let us specialize Drinfeld's second realization of the Yangian to the $\alg{sl}(2)$ case. The map
between the first and the second realization
becomes in this case
\begin{align}
\label{def:isom2}
&h_{0}={h},\quad e_{0}={e},\quad f_{0}={f},\nonumber\\
&h_{1}=\hat{{h}}-v,\quad e_{1}=\hat{{e}}-w,\quad f_{1}=\hat{{f}} -z,
\end{align}
where
\begin{eqnarray}
v = \frac{1}{2} (\{ f,e\} - h^2 ), \qquad w = - \frac{1}{4} \{ e,h\}, \qquad z = - \frac{1}{4} \{ f,h\}.
\end{eqnarray}
The first realization is given by
\begin{eqnarray}
&&[h,e]=2e, \qquad \qquad \, \, \, \, \, \, \, \, [h,f]=-2f, \qquad \qquad \, \, \, \, \, \, \, \, \, [e,f]=h, \nonumber\\
&&[\hat{h},e]=[h,\hat{e}]=2\hat{e}, \qquad [\hat{h},f]=[h,\hat{f}]=-2\hat{f}, \qquad [\hat{e},f]=[e,\hat{f}]=\hat{h},
\end{eqnarray}
Let us consider a so-called \emph{evaluation} representation where $\hat{h} = u \, h$, $\hat{e} = u \, e$ and $\hat{f} = u \, f$. 
 By
applying Drinfeld's map to this 
representation, one first finds the level $0$ and $1$ generators
of the second realization. The generalization at all level $n$ is afterwards easily found as \cite{Khoroshkin:1994uk,Arutyunov:2009ce}
\begin{equation}
\label{exte}
e_n = e \Big( u +\frac{h+1}{2} \Big)^n,\qquad
f_n = f \Big( u +\frac{h-1}{2} \Big)^n,\qquad
h_n = e f_n - f e_n.
\end{equation}
It is easy to check that these generators satisfy the
correct defining relations stemming from (\ref{def:drinf2}):
\begin{equation}
  \label{relazionizeroY}
  \begin{gathered}
    [h_{m},h_{n}]=0, \qquad
    [e_{m},f_{n}]=\, h_{n+m}, \qquad
    [h_{0},e_{m}]= 2\,e_{m}, \qquad 
    [h_{0},f_{m}]=-  2\,f_{m}, \\
    \begin{aligned}
      [h_{m+1},e_{n}]-[h_{m},e_{n+1}] &=   \{h_{m},e_{n}\}, &
      [h_{m+1},f_{n}]-[h_{m},f_{n+1}] &= -  \{h_{m},f_{n}\}, \\
      [e_{m+1},e_{n}]-[e_{m},e_{n+1}] &=   \{e_{m},e_{n}\}, &
      [f_{m+1},f_{n}]-[f_{m},f_{n+1}] &= -   \{f_{m},f_{n}\}.
    \end{aligned}
  \end{gathered}
\end{equation}

We can extend the infinite-dimensional modules we have been discussing in section \ref{inof} to representations of $\Ya{\alg{sl}(2)}$ according to (\ref{exte}). This promptly produces, for $q=0,1,2,\dotsc$,
\begin{itemize}
\item {\bf Irreducible `${P^{{\bf s}}}$' module:}
\begin{equation}
\label{gensY}
e_q |n\rangle = \Big( u+n-s+\frac{1}{2}\Big)^q\, (n - 2 s) \, |n+1\rangle,
\quad
f_q |n\rangle = - \Big( u+n-s-\frac{1}{2}\Big)^q\, n \, |n-1\rangle.
\end{equation}

\item {\bf $P$ module:} 
\begin{equation}
\label{d2P}
\mathrlap{e_q |n\rangle = \Big( u+n+\frac{1}{2}\Big)^q\, n \, |n+1\rangle,}
\phantom{e_q |n\rangle = \Big( u+n-s+\frac{1}{2}\Big)^q\, (n - 2 s) \, |n+1\rangle,}
\quad 
\mathrlap{f_q |n\rangle = - \Big( u+n-\frac{1}{2}\Big)^q\, n \, |n-1\rangle.}
\phantom{f_q |n\rangle = - \Big( u+n-s-\frac{1}{2}\Big)^q\, n \, |n-1\rangle.}
\end{equation}

\item {\bf $S$ module:}
\begin{equation}
\mathrlap{e_q |n\rangle = \Big( u+n+\frac{1}{2}\Big)^q\, |n+1\rangle,}
\phantom{e_q |n\rangle = \Big( u+n-s+\frac{1}{2}\Big)^q\, (n - 2 s) \, |n+1\rangle,}
\quad
\mathrlap{f_q |n\rangle = - \Big( u+n-\frac{1}{2}\Big)^q\, n \, (n-1) \, |n-1\rangle.}
\phantom{f_q |n\rangle = - \Big( u+n-s-\frac{1}{2}\Big)^q\, n \, |n-1\rangle.}
\end{equation}

\item {\bf $R$ module:}
\begin{equation}
\mathrlap{e_q |n\rangle = \Big( u+n+\frac{1}{2}\Big)^q\, \sqrt{n} \, |n+1\rangle,}
\phantom{e_q |n\rangle = \Big( u+n-s+\frac{1}{2}\Big)^q\, (n - 2 s) \, |n+1\rangle,}
\quad
\mathrlap{f_q |n\rangle = - \Big( u+n-\frac{1}{2}\Big)^q\, n \, \sqrt{n-1} \, |n-1\rangle.}
\phantom{f_q |n\rangle = - \Big( u+n-s-\frac{1}{2}\Big)^q\, n \, |n-1\rangle.}
\end{equation}
\end{itemize}
Notice that the other `$ {S^{{\bf s}}}$' and `$ {R^{{\bf s}}}$' modules of section \ref{inof} are easily obtained from the `${P^{{\bf s}}}$' module (\ref{gensY}) by applying the very same transformations (\ref{mappato1}) and (\ref{mappato2}), respectively, to generators at \emph{arbitrary} Yangian level. In fact, (\ref{exte}) shows that the difference between level zero and higher level generators in this representation is always given by a factor which is diagonal on the $|n\rangle$ basis. This factor therefore commutes with the similarity transformation, which is also always diagonal on the $|n\rangle$ basis.

\section{Universal R-matrix}
\label{ddddd}

In the next two section we will show that the R-matrix used in \cite{OhlssonSax:2011ms} (see formulas (\ref{eq:R-matrix-ob-cc-aa}) and~(\ref{eq:R-matrix-ob-ca-ac}) in section \ref{altern}) can be re-derived from a Yangian construction of the type discussed in section \ref{Yanu}, by means of the universal R-matrix. Such a derivation will put the expressions for the R-matrix in terms of projectors on a much firmer footing. Our strategy will be as follows. In this section, we will review the notion of the universal R-matrix and the basic formulas for the $\alg{sl}(2)$ case. We will then specialize its action to the various modules discussed in section \ref{inof}, deriving exact formulas for the coefficients of such actions on arbitrary highest weight states. In the next section, we will generalize to the $\alg{sl}(2|1)$  super-algebra, where we will perform analogous calculations for the so-called \emph{chiral module}. 

The main motivation we have in performing the calculation for all three P, S, and R $\alg{sl}(2)$  modules is that we will \emph{a posteriori} observe that the R-matrices for these three modules coincide with one another when acting on the respective highest weight states. This will provide us with a justification to later on focus on one particular type of (unitary) $\alg{sl}(2|1)$ modules, without worrying too much about missing potentially interesting phenomena related to the other modules.    
  
\subsection{Universal formula}
The universal R-matrix solves the equation
\begin{eqnarray}
\label{definz}
\Delta^{op} (\alg{J}) \, R = R \, \Delta (\alg{J}) 
\end{eqnarray}
for any generator $\alg{J}$ of the Yangian. The coproduct is induced on the generators of Drinfeld's second realization by~\eqref{coptr},~\eqref{cop} via the map (\ref{def:isom}). The apex in $\Delta^{op}$ denotes the `opposite' coproduct $\Delta^{op} = \sigma \Delta$, with $\sigma$ the permutation operator $\sigma (a \otimes b) = (-)^{ab} \, b\otimes a$. 

The universal R-matrix for the double of the Yangian of
$\alg{sl}(2)$, solving (\ref{definz}) in any representation, reads \cite{Khoroshkin:1994uk}
\begin{eqnarray}
\label{univ} &&{ R}={ R}_E { R}_H { R}_F,
\end{eqnarray}
where
\begin{equation}
  \begin{aligned}
    \label{versal}
    R_E &= \prod_{n\ge 0}^{\rightarrow}\exp(- e_n\otimes f_{-n-1}), \\
    R_F &= \prod_{n\ge 0}^{\leftarrow}\exp(- f_n\otimes e_{-n-1}),  \\
    R_H & =\prod_{q\ge 0} \exp \left\{ \Res_{u=v} \left[
        \frac{\mathrm{d}}{\mathrm{d}\,t} (\log H^+(t)) \otimes \log H^-(v+2q+1)
      \right] \right\}.
\end{aligned}
\end{equation}
We have defined
\begin{eqnarray}
\label{eqn;Res}
&&\Res_{t=v}\left[A(t)\otimes B(v)\right]=\sum_k a_k\otimes b_{-k-1}
\end{eqnarray}
where $A(t)=\sum_k a_k t^{-k-1}$ and $B(u)=\sum_k b_k u^{-k-1}$, and the
so-called Drinfeld's currents (for the Cartan subalgebra) are given by
\begin{equation}
  \label{curr}
  H^{\pm}(t)=1\pm \sum_{\substack{n \ge 0 \\ n<0}} h_n \, t^{-n-1} \,.
\end{equation}
The arrows on the products in (\ref{versal}) indicate the so-called \emph{normal} ordering\footnote{Not to be confused with the normal ordering familiar from quantum mechanics.} prescription of \cite{Khoroshkin:1994uk}.  

Let us now evaluate the universal R-matrix on the various modules we have been describing in section \ref{moduli}. The calculational details are reported in appendix \ref{appA}, while here below we merely state the results of the action of $R$ on the respective highest weight states of the modules (the action on descendants being obtained by use of the $\alg{sl}(2)$ invariance).

\subsection{Action on highest weight states}

By taking into account formulas (\ref{eqn;RF21}), (\ref{defA}) and (\ref{cartaf}), we see that the successive action of the 3 factors in the universal R-matrix (\ref{univ}) on states gives
\begin{eqnarray}
\label{massiva}
&&R |m_1,m_2\rangle \, = \sum_{n=0}^{m_2+m} \sum_{m=0}^{m_1} A_m (m_1,m_2) \, \, R_H (m_1-m,m_2+m) \, \,  B_n (m_1-m,m_2+m) \, \,\times \nonumber \\
&&\qquad \qquad \qquad \qquad \qquad \qquad \qquad \qquad \qquad \times |m_1-m+n,m_2+m-n\rangle,
\end{eqnarray}
where we defined $|m,n\rangle \equiv |m\rangle \otimes |n\rangle$.
The quantities $A_m$, $R_H$ and $B_n$ are calculated in appendix \ref{appA} for the various modules of interest, with their definitions being provided by 
\begin{eqnarray}
\label{cartaf1}
&&R_H |n_1,n_2\rangle \equiv R_H (n_1, n_2) \, |n_1,n_2\rangle
\end{eqnarray}
and 
\begin{eqnarray}
&&R_E |m_1,m_2\rangle \equiv \sum_{m=0}^{m_2} B_m (m_1,m_2)\, \,  |m_1+m,m_2-m\rangle,\nonumber\\
&&R_F |m_1,m_2\rangle \equiv \sum_{m=0}^{m_1} A_m (m_1,m_2)\, \,  |m_1-m,m_2+m\rangle.
\end{eqnarray}
In the following, we will be focusing on the action of the R-matrix on highest weight states, which are particular linear combinations of states $|n_1,n_2\rangle$. One can therefore use the formula (\ref{massiva}) and suitably combine the results for different values of $n_1$ and $n_2$ to obtain the action on highest weight states.
\subsubsection{$P$ module}

We compute now the action of the R-matrix on highest weight states in the tensor product of two $P$ module representations. These states are annihilated by the (negative root) generator $f$,
\begin{eqnarray}
\label{notfixed}
\Delta(f) |hw\rangle_j = (f \otimes 1 + 1 \otimes f)  |hw\rangle_j = 0.
\end{eqnarray}
We obtain for the $P$ module (choosing a suitable normalization, as it is not fixed by  
(\ref{notfixed}))
\begin{eqnarray}
\label{aiwp}
|hw\rangle_j = \sum_{q=0}^j (-)^q \, \binom{j}{q} \, |q\rangle \otimes |j-q\rangle. 
\end{eqnarray}
One can then show that the action of the R-matrix is diagonal on these states\footnote{\label{notahw} Because of (\ref{definz}), and the fact that $\Delta^{op} (f) = \Delta (f)$ and $\Delta^{op} (h) = \Delta (h)$, one has
$$
\Delta (f) \, R |hw\rangle_j = R \Delta(f) |hw\rangle_j \, = 0,
$$
hence $R |hw\rangle_j$ is also a highest weight state. Moreover, because of (\ref{azin1}), 
$$
\Delta (h) \, R |hw\rangle_j = R \, \Delta(h) \, |hw\rangle_j = \, 2j \, R \, |hw\rangle_j,
$$
hence $R |hw\rangle_j$ is a highest weight state with total Cartan eigenvalue $2j$. It must then be proportional to $|hw\rangle_j$.}. By plugging formula (\ref{massiva}) into (\ref{aiwp}), one obtains, after a massive simplification\footnote{Computations are performed with the help of \textsf{Mathematica}.}, the final outcome 
\begin{eqnarray}
\label{mssv}
R |hw\rangle_j = {}_2F_1 (1 - j,-j,1 - j + u_1 - u_2,1) \, \, |hw\rangle_j = \prod_{k=1}^{j-1} \frac{u_1-u_2+k}{u_1-u_2-k} \, \, |hw\rangle_j \, \equiv \, R^{(P)}_j |hw\rangle_j.
\end{eqnarray}

\subsubsection{$S$ module}

We can again compute the action of the R-matrix on highest weight states in the tensor product of two $S$ module representations. In the $S$ module the highest weight states satisfying equation~(\ref{notfixed}) are
\begin{eqnarray}
\label{hwS}
|hw\rangle_j = \sum_{q=1}^{j-1} (-)^q \, \binom{j}{q} \, \binom{j-2}{q-1} \, |q\rangle \otimes |j-q\rangle
\end{eqnarray}
for $j>1$. Besides these states, the states $|0\rangle \otimes |0\rangle$, $|0\rangle \otimes |1\rangle$ and $|1\rangle \otimes |0\rangle$ are all annihilated by $\Delta(f)$ individually, hence they are all highest weights. The R-matrix acts as identity on these states.

The action of the R-matrix is diagonal on the states (\ref{hwS}), with the coefficient given in terms of another hypergeometric function,
\begin{eqnarray}
&&R |hw\rangle_j \, \equiv \, R^{(S)}_j \, |hw\rangle_j \, \\
&&=(u_1 - u_2) \, (1 + u_1 - u_2) \, \Gamma(1 - j + u_1 - u_2) \, \, {}_2\tilde{F}_1 (1 - j,2 - j,3 - j + u_1 - u_2,1) \, \, |hw\rangle_j,\nonumber
\end{eqnarray}
where ${}_2\tilde{F}_1 (a,b,c,x) = {}_2F_1 (a,b,c,x) / \Gamma(c)$. It so turns out that
\begin{eqnarray}
\label{eqPS}
R^{(S)}_j \, = R^{(P)}_j,
\end{eqnarray}
a relation which is also valid for the states $|0\rangle \otimes |0\rangle$, $|0\rangle \otimes |1\rangle$ and $|1\rangle \otimes |0\rangle$.

\subsubsection{$R$ module}

In the $R$ module the highest weight states satisfying equation~(\ref{notfixed}) are
\begin{eqnarray}
\label{hwR0}
|hw\rangle_j = \sum_{q=1}^{j-1} (-)^q \, \binom{j}{q} \, \sqrt{\binom{j-2}{q-1}} \, \, \, |q\rangle \otimes |j-q\rangle. 
\end{eqnarray}
for $j>1$. As in the $S$ module, besides these states, the states $|0\rangle \otimes |0\rangle$, $|0\rangle \otimes |1\rangle$ and $|1\rangle \otimes |0\rangle$ are all annihilated by $\Delta(f)$ are also individually highest weights. The R-matrix acts as identity on these states.

The action of the R-matrix is diagonal on the states (\ref{hwR0}), with the coefficient given in terms of a complicated combination of hypergeometric functions. We will not report the result explicitly here since it is not very illuminating, furthermore it again turns out that
\begin{eqnarray}
\label{eqPR}
R^{(R)}_j \, = R^{(P)}_j,
\end{eqnarray}
on the states (\ref{hwR0}) and on $|0\rangle \otimes |0\rangle$, $|0\rangle \otimes |1\rangle$ and $|1\rangle \otimes |0\rangle$. 

\subsection{Generic \texorpdfstring{$s$}{s} vs generic \texorpdfstring{$s$}{s}}\label{AppB2}
As a final example, we can consider the case where both factors in the tensor product are modules (\ref{gensY}) with generic $s_1 \neq 0$ and $s_2 \neq 0$. It is relatively straightforward, with the machinery built at the previous stages, to derive the following results. The highest weight states are given by (\ref{aiwp}), since the action of $f$ does not depend on $s$ in either factors of the tensor product and coincides with the action for the $P$ module. The action of the R-matrix is given by
\begin{eqnarray}
\label{generica}
&&R |hw\rangle_j \, \equiv \, R^{[s_1; s_2]}_j |hw\rangle_j \, \,= \\
&&\frac{2^{1 - 2 \delta u} \, \Gamma(j- s_1 -s_2- \delta u) \, \Gamma(1-j+ s_1 +s_2+ \delta u)}{\Gamma\big[ \frac{1}{2}(-s_1 - s_2 + \delta u)\big] \, \Gamma\big[ \frac{1}{2}1+s_1 - s_2 + \delta u)\big] \, \Gamma\big[\frac{1}{2}(1-s_1 + s_2 + \delta u)\big] \, \Gamma\big[\frac{1}{2}(2+s_1 + s_2 + \delta u)\big]} \, \, |hw\rangle_j,\nonumber
\nonumber\\
&& \qquad =\frac{\Gamma\big[\frac{1}{2}(1- s_1 - s_2 + \delta u) \big] \Gamma\big[\frac{1}{2}(1+ s_1 + s_2 + \delta u) \big]}{\Gamma\big[\frac{1}{2}(1+ s_1 - s_2 + \delta u) \big] \Gamma\big[\frac{1}{2}(1- s_1 + s_2 + \delta u) \big]}\prod_{k=0}^{j-1}\frac{\delta u - s_1 -s_2+ k}{\delta u + s_1 +s_2- k} \, \, |hw\rangle_j\nonumber
\end{eqnarray}
where
\begin{eqnarray}
\delta u = u_1 - u_2.
\end{eqnarray}
Notice that the systematic dependence of the R-matrices on the difference $\delta u$ of the evaluation parameters is a consequence of the shift automorphism of the Yangian \cite{Chari:1994pz,Molev:2003,Torrielli:2011gg}.

\bigskip\bigskip\noindent
To conclude our treatment of $\alg{sl}(2)$, in appendix \ref{appU} we perform some unitarity checks on the R-matrix actions we have derived in this section.

\section{The \texorpdfstring{$\mathfrak{sl}(2|1)$}{sl(2|1)} case}
\label{surmat}

Having derived the action of the $\alg{sl}(2)$ universal R-matrix on highest weight states, we still cannot directly compare to the formulas in \cite{OhlssonSax:2011ms}. These formulas hold for the $\mathfrak{sl}(2|1)$ case, where, as we will see shortly, fermionic degrees of freedom contribute non-trivially. Firstly, the highest weight states will contain fermionic excitations. Secondly, the coproducts used to determine highest weight states include fermionic algebra-generators. We therefore need to adapt the calculation to the supersymmetric case, which we will do in the rest of this section. We will first introduce the algebra and the \emph{chiral module}, then report the formula for the universal R-matrix and derive its action on highest weight states. We will explicitly compute the universal R-matrix action only on some low-lying highest weight states. The results reproduce the formulas of \cite{OhlssonSax:2011ms} for all highest weight states we can check analytically. This gives us confidence in the consistency of the R-matrix reported in section \ref{altern}. It would of course be desirable to obtain explicit results for a generic highest weight state starting from our Yangian and universal R-matrix formulas, although that appears at the moment as a quite challenging computational task. 

\subsection{The supersymmetric algebra}\label{disting}

The definition of the Lie superalgebra $\alg{sl}(2|1)$ in the Chevalley-Serre presentation, for a so-called \emph{distinguished} Dynkin diagram (one with the lowest number of fermionic nodes, see Fig. (\ref{fig:dynkin-diagram-sl21})), is obtained in terms of generators $\gen{h}_i, \gen{e}_i, f_i$, $i = 1,2$, standard commutation relations
\begin{eqnarray}\label{eq:sl2-comm-rel}
\comm{\gen{h}_i}{\gen{h}_j} = 0, \\
\comm{\gen{h}_i}{\gen{e}_j} = {a_{ij}}\gen{e}_j,  \\
\comm{\gen{h}_i}{\gen{f}_j} = -{a_{ij}}\gen{f}_j,  \\
\comm{\gen{e}_i}{\gen{f}_j } = \delta_{ij}{\gen{h}_i},
\end{eqnarray}
and the Serre relations
\begin{eqnarray}
\adj{{\gen{e}}_1}{2}{{\gen{e}}_2} = \adj{{\gen{e}}_2}{2}{{\gen{e}}_1} = 0, \\
\adj{{\gen{f}}_1}{2}{{\gen{f}}_2} = \adj{{\gen{f}}_2}{2}{{\gen{f}}_1} = 0, 
\end{eqnarray}
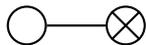
\begin{figure}
  \centering
  
    \begin{tikzpicture}
      [
      thick
      ]
      \useasboundingbox (-1.1cm,-0.7cm) rectangle (1.1cm,0.7cm);

      \node (v1) at (+0.65cm,  0.0cm) [dynkin node] {};
      \node (v2) at (-0.65cm,  0.0cm) [dynkin node] {};

      \draw (v1.south west) -- (v1.north east);
      \draw (v1.north west) -- (v1.south east);

      \draw (v2) -- (v1);
    \end{tikzpicture}

  \caption{The choice of (distinguished) Dynkin diagram of $\alg{sl}(2|1)$ used in the text.}
  \label{fig:dynkin-diagram-sl21}
\end{figure}

The Cartan matrix is given by 
\begin{equation}
\label{Cartana}
 a_{ij} = 
\begin{pmatrix}
2&-1\\
-1&0
\end{pmatrix}.
\end{equation}
The roots $\gen{e}_1, f_1$ are bosonic, the roots $\gen{e}_2, f_2$ are fermionic. Furthermore, the curly brackets $\acomm{}{}$ always denote the anticommutator, while the brackets $[,]$ denote the graded commutator, and $\adjbb{}$ denotes the super adjoint action \ $\adjb{x}{y} = x y - (-1)^{|x||y|}y x$ \ (with $|x|$ being the fermionic degree of $x$). 

The correspondence with the generators used in \cite{OhlssonSax:2011ms} is as follows:
\begin{align}
\gen{e}_1 &= \gen{J}^+, \qquad &\gen{f}_1 &= \gen{J}^- \qquad &\gen{h}_1 & = [\gen{e}_1,\gen{f}_1],\\
\gen{e}_2 &= \gen{S}^-, \qquad &\gen{f}_2 &= \gen{Q}^+, \qquad &\gen{h}_2 & = \{\gen{e}_2,\gen{f}_2\}. 
\end{align}
The generators corresponding to non-simple roots are given by 
\begin{align}
e_3 &= [\gen{e}_1,\gen{e}_2] = \gen{Q}^-, \qquad &f_3 &= [\gen{f}_1,\gen{f}_2] = \gen{S}^+,
\end{align}
where again the generators $Q^\pm$ and $S^\pm$ are the ones used in \cite{OhlssonSax:2011ms}.

We denote highest weight representations of $\alg{sl}(2|1)$ by $(s;b)$, where the two labels $s$ and $b$ give the eigenvalues of the highest weight state under the two charges $\frac{1}{2} \gen{h}_1$ and $\frac{1}{2} \gen{h}_1 + \gen{h}_2$, respectively. There are two kinds of short representations. A \emph{chiral} representation has a highest weight state that is annihilated by $\gen{f}_2$ and satisfies $b = s$. The charges of an \emph{anti-chiral} representation are given by $b = -s$. It has a highest weight state annihilated by the generator $\gen{e}_3$.

\subsection{Chiral representation}

The \emph{chiral} representation is given by\footnote{In order to have an easier comparison with the literature, we have switched to $-s$ for the supersymmetric case \textit{vs}.\@ the $\alg{sl}(2)$ case. Unitary representations are now obtained for positive real values of $s$, as one can see from \ref{represl21}.}
\begin{equation}
  \label{represl21}
  \begin{gathered}
  \begin{aligned}
    \gen{e}_1 \ket{\phi_n} &= - \sqrt{(n+2s)(n+1)} \ket{\phi_{n+1}} \,, &
    \gen{f}_1 \ket{\phi_n} &= + \sqrt{(n+2s-1)n} \ket{\phi_{n-1}} \,, \\
    \gen{e}_1 \ket{\psi_n} &= - \sqrt{(n+2s+1)(n+1)} \ket{\psi_{n+1}} \,, &
    \gen{f}_1 \ket{\psi_n} &= + \sqrt{(n+2s)n} \ket{\psi_{n-1}} \,, 
  \end{aligned}
  \\
    \gen{e}_2 \ket{\phi_n} = - \sqrt{n} \ket{\psi_{n-1}} \,, \qquad 
    \gen{e}_2 \ket{\psi_n} = 0 \,, \qquad
    \gen{f}_2 \ket{\phi_n} = 0 \,, \qquad
    \gen{f}_2 \ket{\psi_n} = + \sqrt{n+1} \ket{\phi_{n+1}} \,.
  \end{gathered}
\end{equation}

We impose the following conditions on the state $|hw\rangle_j$ in the tensor product of two chiral representations:
\begin{eqnarray}
\label{sl21hw}
\Delta(\gen{f}_1) \, |hw\rangle_j = 0, \qquad \Delta(f_3) \, |hw\rangle_j = 0, \qquad \Delta(\gen{e}_2) \, |hw\rangle_j = 0,  
\end{eqnarray}
with $\Delta(\gen{f}_1) = \gen{f}_1 \otimes \mathds{1} + \mathds{1} \otimes \gen{f}_1$, and similarly for the other two conditions in (\ref{sl21hw}). Notice that the condition with respect to $f_1$ is generated by anticommuting the remaining ones. This shows that, strictly speaking, the conditions (\ref{sl21hw}) are highest weight conditions for an all-fermionic Dynkin diagram, where the role of negative root generators is taken by $f_3$ and $e_2$, respectively. We will however continue working in the distinguished basis (cf.\@ section~\ref{disting}), since the R-matrix will act diagonally on $|hw\rangle_j$ by means of similar arguments as in footnote \ref{notahw}. One can prove that the conditions in~\eqref{sl21hw} are satisfied if we choose
\begin{eqnarray}
\label{chirale}
|hw\rangle_{j+1} = \, \sum_{q=0}^j \, \Big( \, a_{bf}(j,q) \, |\phi_q\rangle \otimes |\psi_{j-q}\rangle \, + \, a_{fb}(j,q) \, |\psi_q\rangle \otimes |\phi_{j-q}\rangle \, \Big), \qquad j=0,1,\dotsc
\end{eqnarray}
with
\begin{eqnarray}
\label{chirale2}
a_{bf}(j,q) = \beta_{bf}(j) \, \prod_{k=0}^{q-1} \, (-1) \frac{\sqrt{(j-k)(j-k+2 s_2)}}{\sqrt{(k+1)(k+2 s_1)}}, \nonumber\\
\nonumber\\
a_{fb}(j,q) = \beta_{fb}(j) \, \prod_{k=0}^{q-1} \, (-1) \frac{\sqrt{(j-k)(j-k-1+2 s_2)}}{\sqrt{(k+1)(k+1+2 s_1)}} 
\end{eqnarray}
and
\begin{eqnarray}
\label{sceltachirale}
\beta_{bf}(j) = - \beta_{fb}(j) \, \sqrt{\frac{2 s_1}{j + 2 s_2}}.
\end{eqnarray}
Besides these states, also the state 
\begin{eqnarray}
\label{chirale0}
|hw\rangle_0 \equiv |\phi_0\rangle \otimes |\phi_0\rangle
\end{eqnarray} 
is highest weight, as it satisfies (\ref{sl21hw}).

\subsection{The Yangian and the chiral-chiral R-matrix}\label{chch}

The Yangian in Drinfeld's second realization is given in terms of generators $\kappa_{i,n}$, $\xi^\pm_{i,n}$, $i=1,2$ and $n$ a non-negative integer, with $\kappa_{i,0}=\gen{h}_i$, $\xi^+_{i,0}=\gen{e}_i$ and $\xi^-_{i,0}=\gen{f}_i$, satisfying the following defining relations:
\begin{align}
\label{def:drinf2-sl21}
&[\kappa_{i,m},\kappa_{j,n}]=0,\quad [\kappa_{i,0},\xi^\pm_{j,m}]=\pm a_{ij} \,\xi^+_{j,m},\nonumber\\
& \comm{\xi^+_{i,m}}{\xi^-_{j,n}}=\delta_{i,j}\, \kappa_{j,n+m},\nonumber\\
&[\kappa_{i,m+1},\xi^\pm_{j,n}]-[\kappa_{i,m},\xi^\pm_{j,n+1}] = \pm \frac{1}{2} a_{ij} \{\kappa_{i,m},\xi^\pm_{j,n}\},\nonumber\\
&\comm{\xi^\pm_{i,m+1}}{\xi^\pm_{j,n}}-\comm{\xi^\pm_{i,m}}{\xi^\pm_{j,n+1}} = \pm\frac{1}{2} a_{ij} \acomm{\xi^\pm_{i,m}}{\xi^\pm_{j,n}}, \nonumber\\
&i\neq j,\, \, \, \, n_{ij}=1+|a_{ij}|,\, \, \, \, \, Sym_{\{k\}}
[\xi^\pm_{i,k_1},[\xi^\pm_{i,k_2},\dots [\xi^\pm_{i,k_{n_{ij}}},
\xi^\pm_{j,l}]\dots]]=0
\end{align}
with the Cartan matrix $a_{ij}$ given by (\ref{Cartana}). The all-level
representation corresponding to (\ref{represl21}) which solves all
the relations (\ref{def:drinf2-sl21}) is given by
\begin{eqnarray}
\label{represl21-yangian}
&&\xi_{1,p}^+ \, |\phi_n\rangle = -\sqrt{(n + 1) (2 s + n)} \, \, (\frac{1}{4} + n + s + u)^p \, \, |\phi_{n+1}\rangle,\nonumber\\
&&\xi_{1,p}^+ \, |\psi_n\rangle = -\sqrt{(n + 1) (2 s + 1 + n)} \, \, (\frac{5}{4} + n + s + u)^p \,\, |\psi_{n+1}\rangle,\nonumber \\
&&\xi_{1,p}^- \, |\phi_n\rangle = \sqrt{n \, (2 s -1 + n)} \, \, (-\frac{3}{4} + n + s + u)^p \,\, |\phi_{n-1}\rangle,\nonumber\\
&&\xi_{1,p}^- \, |\psi_n\rangle = \sqrt{n \, (2 s + n)} \, \, (\frac{1}{4} + n + s + u)^p \,\, |\psi_{n-1}\rangle, \\
&&\xi_{2,p}^+ \, |\phi_n\rangle = -\sqrt{n} \, \, (-\frac{1}{4} + s + u)^p \,\, |\psi_{n-1}\rangle, \qquad \xi_{2,p}^+ \, |\psi_n\rangle = 0,\nonumber\\
&&\xi_{2,p}^- \, |\phi_n\rangle = 0, \qquad \xi_{2,p}^- \, |\psi_n\rangle = \sqrt{n + 1} \, \, (-\frac{1}{4} + s + u)^p \,\, |\phi_{n+1}\rangle,\nonumber
\end{eqnarray}
where $|\phi_n\rangle$, ($|\psi_n\rangle$), for $n>0$, are an infinite tower of 
bosonic (fermionic) states. The Cartan generators $\kappa_{i,n}$ can be obtained from (\ref{def:drinf2-sl21}), for instance as $\kappa_{i,n} = \comm{\xi^+_{i,0}}{\xi^-_{i,n}}$.

The R-matrix related to this Yangian representation must satisfy
\begin{eqnarray}
\label{inv} \Delta^{op} ({\mathfrak{J}}) \, R = R \, \Delta
({\mathfrak{J}})
\end{eqnarray}
for any generator $\mathfrak{J}$ of the Yangian. The universal formula for $R$ is given by \cite{Khoroshkin:1994uk}
\begin{eqnarray}
\label{univ1} 
&&{ R}={ R}_2 \, \, R_{1+2} \, \, R_1 \, { R}_H \, { R}_{\bar{1}} \, \, { R}_{\bar{1}+\bar{2}} \, \, { R}_{\bar{2}},
\end{eqnarray}
where
\begin{eqnarray}
\label{versal1}
&&{ R}_1=\prod_{n\ge 0}^{\rightarrow}\exp(- \xi_{1,n}^+ \otimes \xi_{1,-n-1}^-),  \qquad { R}_2=\prod_{n\ge 0}^{\rightarrow}\exp(\xi_{2,n}^+ \otimes \xi_{2,-n-1}^-),  \nonumber\\
&&{ R}_{1+2}=\prod_{n\ge 0}^{\rightarrow}\exp(- [\xi_{1,0}^+,\xi_{2,n}^+] \otimes [\xi_{1,0}^-,\xi_{2,-n-1}^-]), \nonumber\\
&&{ R}_{\bar{1}}=\prod_{n\ge 0}^{\leftarrow}\exp(- \xi_{1,n}^- \otimes \xi_{1,-n-1}^+),  \qquad { R}_{\bar{2}}=\prod_{n\ge 0}^{\leftarrow}\exp(- \xi_{2,n}^- \otimes \xi_{2,-n-1}^+),  \nonumber\\
&&{ R}_{\bar{1}+\bar{2}}=\prod_{n\ge 0}^{\leftarrow}\exp([\xi_{1,0}^-,\xi_{2,n}^-] \otimes [\xi_{1,0}^+,\xi_{2,-n-1}^+]), \nonumber\\
&&{ R}_H=\exp \left\{ \Res_{t=v}\left[\sum_{i,j}
\frac{\mathrm{d}}{\mathrm{d}\,t}(\log H_i^+(t))\otimes { 
} D^{-1}_{ij} \log H_j^-(v)\right]\right\},
\end{eqnarray}
where $D_{ij} = - (T^{\frac{1}{2}} - T^{-\frac{1}{2}}) \, a_{ij} (T^{\frac{1}{2}})$, $a_{ij} (q) = \frac{q^{a_{ij}} - \, q^{-a_{ij}}}{q - q^{-1}}$ with $a_{ij}$ the Cartan matrix entries (\ref{Cartana}), and the operator $T$ is defined such that $T f(v) = f(v+1)$. The definition of $\Res$ and the Drinfeld currents $H^\pm_i(t)$ are given in~(\ref{eqn;Res}) and~(\ref{curr}).

The arrows on the products in (\ref{versal1}) indicate the ordering one has to follow in the multiplication, and are a consequence of the normal ordering prescription for the root factors in the universal R-matrix (see~\cite{Khoroshkin:1994uk}). The ordering of the factors $R_1, R_2, R_{1+2}$, and respective barred versions, is also prescribed. The ordering rule states that if two positive roots $\alpha_1$ and $\alpha_2$, corresponding to root generators $\xi_{\alpha_1}$ and $\xi_{\alpha_2}$, have already been ordered, \textit{ i.e.}, they satisfy $\alpha_1 < \alpha_2$ (where $<$ stands for the chosen ordering), then their sum must lie in between them, namely
\begin{eqnarray}
\alpha_1 < \alpha_1 + \alpha_2 < \alpha_2. 
\end{eqnarray}
Remember that the generator corresponding to the sum of the roots is the graded commutator 
\begin{eqnarray}
\xi_{\alpha_1 + \alpha_2} = [\xi_{\alpha_1},\xi_{\alpha_2}].
\end{eqnarray}
Fixing $\alpha_1$ and $\alpha_2$ to be the two positive simple roots of $\alg{sl}(2|1)$, and recalling that $\xi_{\pm \alpha_i} = \xi_i^\pm$, we obtain the ordering (\ref{univ1}).

In appendix \ref{appB} we give details of the calculation of the universal R-matrix action on the chiral module. Here below we focus on highest weight states, following the same rationale as in the $\alg{sl}(2)$ case. In particular, a formula analogous to equation~(\ref{massiva}) applies, although now with 7 terms stemming from the subsequent application of all the factors in equation~(\ref{univ1}).

\subsubsection{Action on highest weight states}\label{sl21hwe}
As it turns out to be quite cumbersome to deal with the generic expression (\ref{chirale}), we specialize to low-lying highest weight states first. Let us start with
the state
\begin{eqnarray}
|hw\rangle_0 = |\phi_0\rangle \otimes |\phi_0\rangle.
\end{eqnarray}
On this state, all root factors act as identity, and so does the Cartan factor $R_H$, hence 
\begin{eqnarray}
R |hw\rangle_0 = |hw\rangle_0.
\end{eqnarray}
Next, we consider
\begin{eqnarray}
|hw\rangle_1 = - \sqrt{s_1} \, |\phi_0\rangle \otimes |\psi_0\rangle \, + \sqrt{s_2} \, |\psi_0\rangle \otimes |\phi_0\rangle. 
\end{eqnarray}
where we have fixed $\beta_{fb}(0)=\sqrt{s_2}$ for convenience. On such a state, only a few factors give a contribution which is not just acting as the identity. The final result is
\begin{eqnarray}
R |hw\rangle_1 = - \frac{s_1+s_2+u_1-u_2}{s_1+s_2-u_1+u_2} \, \, |hw\rangle_1. 
\end{eqnarray} 
The calculation for the next highest weight (suitably normalized setting additionally $\beta_{fb}(1)=1$)
\begin{eqnarray}
|hw\rangle_2=-\sqrt{\frac{2s_1}{1 + 2 s_2}} \, |\phi_0\rangle \otimes |\psi_1\rangle 
+ |\phi_1\rangle \otimes |\psi_0\rangle + |\psi_0\rangle \otimes \phi_1\rangle -\sqrt{\frac{2s_2}{1 + 2 s_1}} \, |\psi_1\rangle \otimes |\phi_0\rangle
\end{eqnarray}
involves non-trivial contributions from all the root factors, and it is performed by mechanizing it into a {\textsf{Mathematica}} computer program. This program systematically deals with the subsequent action of all the seven factors of the universal R-matrix. After a massive simplification, one obtains
\begin{eqnarray}
R |hw\rangle_2 = \frac{(s_1 + s_2 + u_1 - u_2) (1 + s_1 + s_2 + u_1 - u_2)}{(s_1 + s_2 - u_1 + u_2) (1 + s_1 + s_2 - u_1 + u_2)} \, \, |hw\rangle_2. 
\end{eqnarray}
We have pushed the program to its limit\footnote{The result (\ref{eq:R-hw4}) is still obtained in analytic form, while subsequent highest weight states would require resorting to a numerical treatment. We have not checked formula (\ref{hw2}) for $j>4$.} by computing the action of $R$ on the next two (appropriately normalized) highest weight states :
\begin{eqnarray}
&&|hw\rangle_3=-\sqrt{\frac{s_1(1+2s_1)}{1 + s_2}} \, |\phi_0\rangle \otimes |\psi_2\rangle 
+ \sqrt{(1+2s_1)}|\psi_0\rangle \otimes |\phi_2\rangle + \sqrt{2(1+2s_1)}|\phi_1\rangle \otimes |\psi_1\rangle  \nonumber\\
&&\qquad - \sqrt{2(1+2s_2)}|\psi_1\rangle \otimes |\phi_1\rangle- \sqrt{(1+2s_2)}|\phi_2\rangle \otimes |\psi_0\rangle+\sqrt{\frac{s_2(1+2s_2)}{1 + s_1}} \, |\psi_2\rangle \otimes |\phi_0\rangle, 
\end{eqnarray} 
\begin{eqnarray}
R |hw\rangle_3 \, = - \, \frac{(s_1 + s_2 + u_1 - u_2) (1 + s_1 + s_2 + u_1 - u_2)(2+s_1 + s_2 + u_1 - u_2)}{(s_1 + s_2 - u_1 + u_2) (1 + s_1 + s_2 - u_1 + u_2)(2+s_1 + s_2 - u_1 + u_2)} \, \, |hw\rangle_3 
\end{eqnarray}
and
\begin{eqnarray}
&&|hw\rangle_4=-\sqrt{\frac{2 s_1 (1 + s_1) (1 + 2 s_1)}{(
  3 + 2 s_2}} \, \phi_0 \otimes \psi_3 + \sqrt{3 (1 + s_1) (1 + 2 s_1)} \phi_1 \otimes \psi_2 \, -\\
&&\sqrt{6 (1 + s_1) (1 + s_2)} \phi_2 \otimes \psi_1 + 
 \sqrt{(1 + s_2) (1 + 2 s_2)} \phi_3 \otimes \psi_0 - \sqrt{\frac{2 s_2 (1 + s_2) (1 + 2 s_2)}{(3 + 2 s_1)}} \psi_3 \otimes \phi_0 \, + \nonumber\\
 &&\sqrt{(1 + s_1) (1 + 2 s_1)} \psi_0 \otimes \phi_3 - \sqrt{6 (1 + s_1) (1 + s_2)} \psi_1 \otimes \phi_2 + \sqrt{3 (1 + s_2) (1 + 2 s_2)} \psi_2 \otimes \phi_1,\nonumber
\end{eqnarray}
\begin{eqnarray}
  \label{eq:R-hw4}
&&R |hw\rangle_4 \, = \\
&&\qquad \frac{(s_1 + s_2 + u_1 - u_2) (1 + s_1 + s_2 + u_1 - u_2)(2+s_1 + s_2 + u_1 - u_2)(3+s_1 + s_2 + u_1 - u_2)}{(s_1 + s_2 - u_1 + u_2) (1 + s_1 + s_2 - u_1 + u_2)(2+s_1 + s_2 - u_1 + u_2)(3+s_1 + s_2 - u_1 + u_2)} \, \, |hw\rangle_4.\nonumber
\end{eqnarray}

The above results are in agreement with the general formula \cite{OhlssonSax:2011ms,Derkachov:2000ne} 
\begin{eqnarray}
\label{hw2}
R |hw\rangle_j = \Bigg[ \prod_{k=0}^{j-1} \, \, \frac{(u_1 - u_2+s_1 + s_2 + k) }{(u_1 - u_2-s_1 - s_2 - k)}\Bigg] \, \, |hw\rangle_j \, ,
\end{eqnarray}
where an empty product is conventionally set to $1$.

Let us remark that the chiral $\alg{sl}(2|1)$ module, when restricted to the bosonic states $|\phi_n\rangle$ and taken at $s=0$, bears a resemblance to the $\alg{sl}(2)$ $R$ module introduced in section~\ref{Rmodulo}, hence it is physically the most relevant case (see discussion about unitarity at the very end of section~\ref{inof}). More precisely, upon identification $|\phi_n\rangle \sim |n\rangle$ , the similarity transformation $\chi(\widehat{N}) \equiv \sqrt{\Gamma(\widehat{N}+1)}$, $\widehat{N}$ being the number operator  $\widehat{N} \, | n\rangle = n \, |n\rangle$ (\textit{cf.} equation~(\ref{numero})), provides an isomorphism (up to a sign) between the restriction of the chiral module to bosonic states and the $\alg{sl}(2)$ $R$ module:

\begin{eqnarray}
\chi^{-1} (\widehat{N}) \, \xi_{1,0}^+ \, \chi(\widehat{N}) = - e, \qquad \chi^{-1} (\widehat{N}) \, \xi_{1,0}^- \, \chi(\widehat{N}) = - f
\end{eqnarray}
on any state $|n\rangle$. Such a similarity transformation is well-defined on all states since they are spanned by integers $n\geq0$, and the gamma function is therefore never singular.

\bigskip\bigskip\noindent
In principle one could repeat the R-matrix calculations we performed in this section for alternative $\alg{sl}(2|1)$ modules which, when seen as $\alg{sl}(2)$ modules, resemble more closely the $P$ and $S$ modules of sections \ref{Pmodulo} and \ref{Smodulo}, respectively. However, the experience with $\alg{sl}(2)$ suggests that the action of the universal R-matrix on highest weight states is the same for the three-types of modules (see for instance formulas (\ref{eqPS}) and (\ref{eqPR})). We have not checked this assumption for $\alg{sl}(2|1)$, also in view of the fact that, in what follows, we will be mostly interested in the R-type module (\ref{represl21-yangian}) we just described.

\subsection{Antichiral representation}

The \emph{antichiral} representation is given by 
\begin{equation}
  \label{represl21ac}
  \begin{gathered}
    \begin{aligned}
      \gen{e}_1 \ket{\phi_n} &= - \sqrt{(n+2s)(n+1)} \ket{\phi_{n+1}} \,, &
      \gen{f}_1 \ket{\phi_n} &= + \sqrt{(n+2s-1)n} \ket{\phi_{n-1}} \,, \\
      \gen{e}_1 \ket{\psi_n} &= - \sqrt{(n+2s+1)(n+1)} \ket{\psi_{n+1}} \,, &
      \gen{f}_1 \ket{\psi_n} &= + \sqrt{(n+2s)n} \ket{\psi_{n-1}} \,, 
    \end{aligned}
    \\
    \gen{e}_2 \ket{\phi_n} = 0 \,, \qquad
    \gen{e}_2 \ket{\psi_n} = + \sqrt{n+2s} \ket{\phi_{n}} \,, \qquad
    \gen{f}_2 \ket{\phi_n} = - \sqrt{n+2s} \ket{\psi_{n}} \,, \qquad
    \gen{f}_2 \ket{\psi_n} = 0 \,,
  \end{gathered}
\end{equation}
The all-level representation corresponding to (\ref{represl21ac}) which solves all the relations (\ref{def:drinf2-sl21}) is given by generators $\kappa_{i,n}$, $\xi^\pm_{i,n}$, $i=1,2$, with $n$ a non-negative integer, such that $\kappa_{i,0}=\gen{h}_i$, $\xi^+_{i,0}=\gen{e}_i$ and $\xi^-_{i,0}=\gen{f}_i$. One has
\begin{eqnarray}
\label{represl21-yangian2}
&&\xi_{1,p}^+ \, |\phi_n\rangle = -\sqrt{(n + 1) (2 s + n)} \, \, (\frac{1}{4} + n + s + u)^p \, \, |\phi_{n+1}\rangle,\nonumber\\
&&\xi_{1,p}^+ \, |\psi_n\rangle = -\sqrt{(n + 1) (2 s + 1 + n)} \, \, (\frac{1}{4} + n + s + u)^p \,\, |\psi_{n+1}\rangle,\nonumber \\
&&\xi_{1,p}^- \, |\phi_n\rangle = \sqrt{n \, (2 s -1 + n)} \, \, (-\frac{3}{4} + n + s + u)^p \,\, |\phi_{n-1}\rangle,\nonumber\\
&&\xi_{1,p}^- \, |\psi_n\rangle = \sqrt{n \, (2 s + n)} \, \, (-\frac{3}{4} + n + s + u)^p \,\, |\psi_{n-1}\rangle, \\
&&\xi_{2,p}^+ \, |\phi_n\rangle = 0, \qquad \xi_{2,p}^+ \, |\psi_n\rangle = \sqrt{n + 2 s} \, \, (-\frac{1}{4} - s + u)^p \, |\phi_n\rangle,\nonumber\\
&&\xi_{2,p}^- \, |\phi_n\rangle = - \sqrt{n + 2 s} \, \, (- \frac{1}{4} - s + u)^p \, |\psi_{n}\rangle, \qquad \xi_{2,p}^- \, |\psi_n\rangle = 0,\nonumber
\end{eqnarray}
for the same choice of Cartan matrix (\ref{Cartana}).
If we consider highest weight states corresponding to the conditions (\ref{sl21hw}), this time projected into an \emph{antichiral $\otimes$ antichiral} representation (namely, taking the coproducts with two representations of type (\ref{represl21ac}) in both factors of the tensor product), then one can verify that the corresponding highest weight states are still given by (\ref{chirale}), (\ref{chirale2}), (\ref{sceltachirale}) and (\ref{chirale0}). The reason is that the antichiral representation (\ref{represl21ac}) and the chiral one (\ref{represl21}) are related by the exchange of the generators $e_2 \leftrightarrow f_3$, $e_3 \leftrightarrow f_2$ (with $e_3 = [e_1,e_2]$ and $f_3 = [f_1,f_2]$ in both cases), which preserves the highest weight conditions (\ref{sl21hw}). By projecting the universal R-matrix (\ref{univ1}) into an \emph{antichiral $\otimes$ antichiral} representation, and performing a calculation analogous to the one in appendix \ref{appB}, we have observed that the action on the highest weight states coincides with (\ref{hw2}) for all the states we have checked.
 
\subsubsection{The mixed chiral-antichiral case} 

In this section we consider the tensor product of a chiral and an antichiral module, and the action of the universal R-matrix on highest weight states in this mixed tensor product representation. This means that we still impose  (\ref{sl21hw}), namely
\begin{eqnarray}
\label{sl21hw1chanch}
\Delta(f_1) \, |\omega\rangle_j = 0, \qquad \Delta(f_3) \, |\omega\rangle_j = 0, \qquad \Delta(e_2) \, |\omega\rangle_j = 0,
\end{eqnarray}
with $\Delta(f_1) = f_1 \otimes \mathds{1} + \mathds{1} \otimes f_1$, and similarly for the other two conditions. However, in the first factor of the tensor product we will use (\ref{represl21}), while in the second factor we will use (\ref{represl21ac}). We will always consider this mixed projection whenever we speak about tensor products in the rest of this section.

Conditions (\ref{sl21hw1chanch}) are satisfied by choosing (for $j=0,1,\dotsc$)
\begin{equation}
  \label{antichirale}
  \ket{\omega}_{j} = \sum_{q=0}^{j} \, a_{bb}(j,q) \, \ket{\phi_q} \otimes \ket{\phi_{j-q}}  + \sum_{q=0}^{j-1}  a_{ff}(j-1,q) \ket{\psi_q} \otimes \ket{\psi_{j-1-q}},
\end{equation}
with
\begin{equation*}
  \begin{gathered}
    a_{bb}(j,q) = \beta_{bb}(j) \, \prod_{k=0}^{q-1} \, (-1) \frac{\sqrt{(j-k)(j-k-1+2 s_2)}}{\sqrt{(k+1)(k+2 s_1)}} \,, \\
    a_{ff}(j,q) = \beta_{ff}(j) \, \prod_{k=0}^{q-1} \, (-1) \frac{\sqrt{(j-k)(j-k+2 s_2)}}{\sqrt{(k+1)(k+1+2 s_1)}} \,,
  \end{gathered}
\end{equation*}
and
\begin{eqnarray}
\beta_{ff}(j) = \beta_{bb}(j+1) \, \sqrt{\frac{j+1}{2 s_1}}.
\end{eqnarray}

In order to compute the R-matrix action on these highest weight states, we adopt the following line of reasoning. Fist, we notice that we can perform a map to another set of states\footnote{We consider $s_1$ and $s_2$ non-zero for the remainder of this section. If one of them is zero, then the map (\ref{newhw}) can be degenerate. More precisely, the only degenerate case turns out to be $\Delta(f_2)_{|s_2=0} |\omega\rangle_0 = \Delta(f_2)_{|s_2=0} |\phi_0\rangle \otimes |\phi_0\rangle= 0$. In this specific case, the action of the universal R-matrix (\ref{univ1}) on $|\omega\rangle_0$ can be computed directly without the need of resorting to (\ref{newhw}). In fact, all the generators in the second factor of each tensor product in (\ref{univ1}) annihilate $|\phi_0\rangle$ at $s_2=0$, hence $R|\omega\rangle_0 =|\omega\rangle_0$.}, namely 
\begin{eqnarray}
\label{newhw}
|\Omega\rangle_j = \Delta (f_2) \, |\omega\rangle_{j}.
\end{eqnarray}  
The states (\ref{newhw}) satisfy
\begin{eqnarray}
\label{sl21hw1chanch2}
\Delta(f_1) \, |\Omega\rangle_j = 0, \qquad \Delta(f_2) \, |\Omega\rangle_j = 0, \qquad \Delta(f_3) \, |\Omega\rangle_j = 0.
\end{eqnarray}
The first of conditions (\ref{sl21hw1chanch2}) is guaranteed by the nilpotency of the fermionic generator $f_2$, the second and third by the fact that $\Delta$ is still a Lie algebra homomorphism, hence $\Delta(f_1)$ and $\Delta(f_3)$ (anti)commute with $\Delta(f_2)$ and annihilate $|\omega\rangle_j$. The R-matrix will act diagonally on $|\Omega\rangle_j$ with the same eigenvalues as for $|\omega\rangle_j$ because of $\alg{sl}(2|1)$ invariance $\Delta(f_2) \, R = R \, \Delta(f_2)$. By explicitly acting on (\ref{antichirale}) as in (\ref{newhw}), we obtain (for $j=0,1,\dotsc$)
\begin{equation}
\label{antichiralexple}
\ket{\Omega}_{j} = \sum_{q=0}^{j} \left( a_{ff}(j-1,q-1) \sqrt{q} - a_{bb}(j,q) \sqrt{j-q}\right) \ket{\phi_q} \otimes \ket{\psi_{j-q}},
\end{equation}
with the coefficients being given by (\ref{sceltachirale}) and (\ref{chirale2}).

By inspecting (\ref{represl21}), (\ref{represl21ac}) and (\ref{antichirale}), one realizes that $|\Omega\rangle_j$ only contain states of the type $|\phi_m\rangle \otimes |\psi_n\rangle$. The action of $\Delta(f_3)$ and $\Delta(f_2)$ in the mixed representation on states $|\phi_m\rangle \otimes |\psi_n\rangle$ is actually identically zero, so the last two conditions of (\ref{sl21hw1chanch2}) are trivially satisfied. The only non-trivial constraint comes from the condition $\Delta(f_1) \, |\Omega\rangle_j$, which preserves states $|\phi_m\rangle \otimes |\psi_n\rangle$ but changes the values of $n$ and $m$. Since, from the point of view of $f_1$, the state $|\psi\rangle_n$ are as good as $|\phi\rangle_n$ as a basis for an $\alg{sl}(2)$ submodule, the condition  
\begin{eqnarray}
\Delta(f_1) \, |\Omega\rangle_j = 0
\end{eqnarray}
coincides with an $\alg{sl}(2) \otimes \alg{sl}(2)$ highest weight condition, for two $\alg{sl}(2)$ modules given by the action of $e_1$ and $f_1$ on $|\phi_m\rangle$ in the left (chiral) factor and $|\psi_m\rangle$ in the right (antichiral) factor of the tensor product, respectively.

The Yangian coproducts at level 1 are given by
\begin{equation}
  \label{coprid}
  \begin{aligned}
    \Delta(\kappa_{1,1}) &= \kappa_{1,1} \otimes \mathds{1} + \mathds{1} \otimes \kappa_{1,1} + h_1 \otimes h_1 - 2 \, f_1 \otimes e_1 + f_2 \otimes f_2 + f_3 \otimes e_3,\\
    \Delta(\kappa_{2,1}) &= \kappa_{2,1} \otimes \mathds{1} + \mathds{1} \otimes \kappa_{2,1} + h_2 \otimes h_2 + f_1 \otimes f_1 - f_3 \otimes e_3,\\
    \Delta(\xi^+_{1,1}) &= \xi^+_{1,1} \otimes \mathds{1} + \mathds{1} \otimes \xi^+_{1,1} + h_1 \otimes e_1 - f_2 \otimes e_3,\\
    \Delta(\xi^-_{1,1}) &= \xi^-_{1,1} \otimes \mathds{1} + \mathds{1} \otimes \xi^-_{1,1} +f_1 \otimes h_1 + f_3 \otimes e_2,\\
    \Delta(\xi^+_{2,1}) &= \xi^+_{2,1} \otimes \mathds{1} + \mathds{1} \otimes \xi^+_{2,1} + h_2 \otimes e_2 + f_1 \otimes e_3,\\
    \Delta(\xi^-_{2,1}) &= \xi^-_{2,1} \otimes \mathds{1} + \mathds{1} \otimes \xi^-_{2,1} + f_2 \otimes h_2 - f_3 \otimes e_1.
  \end{aligned}
\end{equation}
If we focus on $\Delta(\xi^\pm_{1,1})$, we see that the fermionic part of the coproduct tail acts as zero on states of the type $|\phi_m\rangle \otimes |\psi_n\rangle$, and the same holds for the opposite coproducts $\Delta^{op}(\xi^\pm_{1,1})$, by using again the explicit form of the representations (\ref{represl21}) and (\ref{represl21ac}). This means that the Yangian level 1 coproducts $\Delta(\xi^\pm_{1,1})$ effectively act on $|\phi_m\rangle \otimes |\psi_n\rangle$, hence on $|\Omega\rangle_j$, in the same way as those of the Yangian of $\alg{sl}(2)$, projected in the two representations $|\phi_m\rangle$ in the left (chiral) factor and $|\psi_m\rangle$ in the right (antichiral) factor of the tensor product, respectively. Such Yangian coproducts necessarily preserve states $|\phi_m\rangle \otimes |\psi_n\rangle$. 

This means that the Yangian R-matrix, when acting on the states $|\Omega\rangle_j$, has to effectively satisfy a set of $\alg{sl}(2)$ conditions
\begin{eqnarray}
\label{su2condition}
\Delta^{op} (\xi^\pm_{1,m}) \, R = R  \, \Delta (\xi^\pm_{1,m}), \qquad m=0,1
\end{eqnarray}
in the mixed chiral-antichiral $|\phi_m\rangle \otimes |\psi_n\rangle$ representation. The conditions (\ref{su2condition}) almost uniquely fix the R-matrix to coincide with the action of the $\alg{sl}(2)$ Yangian universal R-matrix in the two respective $\alg{sl}(2)$ modules, up to an overall scalar factor. Following this argument, we can simply compute such an $\alg{sl}(2)$ R-matrix in the same fashion as in section \ref{ddddd}. In fact, we can simply borrow the result of section~\ref{AppB2}. For $s_1$ and $s_2$ different from zero, the mixed chiral-antichiral $|\phi_m\rangle \otimes |\psi_n\rangle$ representation is isomorphic (up to a similarity transformation) to the tensor product of two $P^{\bf s}$ modules. The chiral part corresponds to 
\begin{eqnarray}
u_1 \rightarrow u_1 - \frac{1}{4}
\end{eqnarray}
in equation~(\ref{gensY}), followed by a similarity transformation (regular everywhere as long as $s_1\neq0$)
\begin{eqnarray}
-\rho_1 \, e_q \, \rho_1^{-1}, \qquad -\rho_1 \, f_q \, \rho_1^{-1}, \qquad \rho_1 |\phi_n\rangle = \sqrt{\frac{\Gamma(n+1)}{\Gamma(n + 2 s_1)}} |\phi_n\rangle, 
\end{eqnarray}
and finally a transformation
\begin{eqnarray}
s_1 \rightarrow - s_1.
\end{eqnarray}
The antichiral representation corresponds to 
\begin{eqnarray}
u_2 \rightarrow u_2 - \frac{3}{4}
\end{eqnarray}
in (\ref{gensY}), followed by a similarity transformation (regular everywhere for unitary representations)
\begin{eqnarray}
-\rho_2 \, e_q \, \rho_2^{-1}, \qquad -\rho_2 \, f_q \, \rho_2^{-1}, \qquad \rho_2 |\phi_n\rangle = \sqrt{\frac{\Gamma(n+1)}{\Gamma(n + 2 s_2+1)}} |\phi_n\rangle, 
\end{eqnarray} 
and finally a transformation
\begin{eqnarray}
s_2 \rightarrow - s_2 - \frac{1}{2}.
\end{eqnarray}

This means that the eigenvalues will be given by formula (\ref{generica}) with the appropriate substitutions\footnote{As a check, we have verified on a few low-lying states that the full universal R-matrix (\ref{univ1}) in the chiral-antichiral representation indeed coincides with formula (\ref{generica}) when acting on $\left(\rho_1^{-1} \otimes \rho_2^{-1} \, {|\Omega\rangle_j}\right)_{|s_1 \rightarrow -s_1, s_2 \rightarrow -s_2-\frac{1}{2}}$ (which, modulo an overall normalization, are the correspondent of (\ref{newhw}) after the maps described in the text are performed).}. In the final expression we obtain, we further shift $u_1 \rightarrow u_1 - \frac{1}{2}$ to make them suitable for comparison with the literature. This amounts to (apart from an overall scalar factor)
\begin{eqnarray}
\label{eq:R-mat-c-ac}
R \, |\Omega\rangle_j \, = \prod_{k=0}^{j-1}\frac{u_1 - u_2 + s_1 + s_2 +\frac{1}{2} + k}{u_1 - u_2 - s_1 - s_2 -\frac{1}{2}- k} \, \, |\Omega\rangle_j.
\end{eqnarray}

\section{The alternating \texorpdfstring{$\alg{d}(2,1;\alpha)$}{d(2,1;a)} spin-chain}\label{altern}

In this section we will briefly review the construction of a $\alg{d}(2,1;\alpha)$ symmetric spin-chain presented in~\cite{OhlssonSax:2011ms}. This spin-chain is proposed to describe the left-moving part of the spectrum of operators of the $\CFT_2$ dual to string theory in $\AdS_3 \times \Sphere^3 \times \Sphere^3 \times \Sphere^1$. As mentioned in the introduction, supersymmetry requires that the $\AdS$ radius $R_{\AdS}$, and the radii $R_+$ and $R_-$ of the two three-spheres satisfy
\begin{equation}
  \frac{1}{R_+^2} + \frac{1}{R_-^2} = \frac{1}{R_{\AdS}^2}.
\end{equation}
Hence, there is a one-parameter family of backgrounds, which can be parametrized by a parameter $\alpha$ in the range $0 < \alpha < 1$ defined by
\begin{equation}
  \alpha = \frac{R_{\AdS}^2}{R_+^2} = 1 - \frac{R_{\AdS}^2}{R_-^2}.
\end{equation}
The super-isometry of this string background is then given two copies of the exceptional $\alg{d}(2,1;\alpha)$ superalgebra, corresponding to the left- and right-moving sectors on $\AdS_3$. 

At weak coupling the left- and right-moving spin-chains decouple.\footnote{%
  At weak coupling the left- and right-movers interact only through the level matching condition, which says that for a physical state the total momentum is zero.%
} %
Here we will consider the $\alg{d}(2,1;\alpha)$ spin-chain describing the left-movers. This spin-chain is alternating, with odd and even sites transforming in two different short representations of the symmetry algebra.

\subsection{The \texorpdfstring{$\alg{d}(2,1;\alpha)$}{d(2,1;a)} algebra and its representations}
\label{sec:d21a-alg-rep}

In the relevant real form of $\alg{d}(2,1;\alpha)$, the bosonic subalgebra is given by $\alg{sl}(2)  \times \alg{su}(2)_+  \times \alg{su}(2)_-$, where we have added the subscripts $\pm$ to the distinguish the two $\alg{su}(2)$ algebras. We denote the corresponding triplets of generators by $\gen{S}_0$, $\gen{S}_\pm$ for the $\alg{sl}(2)$ algebra, $\gen{L}_5$, $\gen{L}_\pm$ for $\alg{su}(2)_+$ and $\gen{R}_8$, $\gen{R}_\pm$ for $\alg{su}(2)_-$. In addition to these bosonic charges there are eight supercharges $\gen{Q}_{b\beta\dot{\beta}}$, with each index taking values $\pm$, transforming in the $\mathbf{2}\otimes\mathbf{2}\otimes\mathbf{2}$ representation of the bosonic algebra. The full commutation relations for the $\alg{d}(2,1;\alpha)$ algebra are given in appendix~\ref{sec:d21a-algebra}.

We denote a highest weight representations of $\alg{d}(2,1;\alpha)$ by the weights of the bosonic sub-algebra. The even and odd sites of the alternating spin-chain transform in the representations $(-\tfrac{\alpha}{2};\tfrac{1}{2};0)$ and $(-\tfrac{1-\alpha}{2};0;\tfrac{1}{2})$, respectively. 

The $(-\tfrac{\alpha}{2};\tfrac{1}{2};0)$ representation consists of the bosonic fields\footnote{%
  In analogy with the spin-chain picture in $\mathcal{N}=4$ supersymmetric Yang-Mills theory we refer to the states of the representations at each site of the spin-chain as \emph{fields}, even though we do not have any direct interpretation of these states as fields in the dual two-dimensional CFT.%
} %
$\phi^{(n)}_{\pm}$ and the fermions $\psi^{(n)}_{\pm}$. The subscripts indicate that the bosons transform as a doublet under $\alg{su}(2)_+$ and the fermions as a doublet under $\alg{su}(2)_-$. The superscript $n$ gives the $\alg{sl}(2)$ level of the fields, with the corresponding generators acting as
\begin{equation}
    \begin{aligned}
    S_0 \ket{\phi_{\pm}^{(n)}} &= - \left( \tfrac{\alpha}{2} + n \right) \ket{\phi_{\pm}^{(n)}} \,, &
    S_0 \ket{\psi_{\pm}^{(n)}} &= - \left( \tfrac{\alpha}{2} + \tfrac{1}{2} + n \right) \ket{\psi_{\pm}^{(n)}} \,, \\
    S_- \ket{\phi_{\pm}^{(n)}} &= -\sqrt{(n + \alpha)(n + 1)} \ket{\phi_{\pm}^{(n+1)}} \,, &
    S_- \ket{\psi_{\pm}^{(n)}} &= -\sqrt{(n+1)(n + 1+ \alpha)} \ket{\psi_{\pm}^{(n+1)}} \,, \\
    S_+ \ket{\phi_{\pm}^{(n)}} &= +\sqrt{(n - 1 + \alpha) n} \ket{\phi_{\pm}^{(n-1)}} \,, &
    S_+ \ket{\psi_{\pm}^{(n)}} &= +\sqrt{n (n + \alpha)} \ket{\psi_{\pm}^{(n-1)}} \,,
  \end{aligned}
\end{equation}
In the notation of section~\ref{inof}, this corresponds to two $R^s$ representations, where $s=-\tfrac{\alpha}{2}$ for the bosons, and $s=-\tfrac{\alpha}{2}-\tfrac{1}{2}$ for the fermions.

The representation $(-\tfrac{1-\alpha}{2};0;\tfrac{1}{2})$, in which the even sites transform, is very similar to the above representation. Again there are two sets of fields, the bosons $\bar{\phi}^{(n)}_\pm$, which make up a doublet of $\alg{su}(2)_-$, and the fermions $\bar{\psi}^{(n)}_\pm$ transforming under $\alg{su}(2)_+$. Under $\alg{sl}(2)$ these fields transform in representations of spin $s=-\tfrac{1-\alpha}{2}$ and $s=-\tfrac{1-\alpha}{2}-\tfrac{1}{2}$, respectively.

The two representations at the odd and even sites of the alternating spin-chain are short representations of $\alg{d}(2,1;\alpha)$. The highest weight states $\phi^{(0)}_+$ and $\bar{\phi}^{(0)}_+$ are annihilated by the supercharges\footnote{%
  In our conventions a highest weight state is annihilated by the generators $\gen{S}_+$, $\gen{L}_+$, $\gen{R}_+$ and $\gen{Q}_{+\pm\pm}$.%
} %
$\gen{Q}_{-+\pm}$ and $\gen{Q}_{-\pm+}$, respectively. A state of the alternating spin-chain of length $L$ transforms in the $L$-fold tensor product
\begin{equation}
  \big( (-\tfrac{\alpha}{2};\tfrac{1}{2};0) \otimes (-\tfrac{1-\alpha}{2};0;\tfrac{1}{2}) \big)^{\otimes L} = (-\tfrac{L}{2};\tfrac{L}{2};\tfrac{L}{2}) \oplus \dotsb \,.
\end{equation}
On the right-hand side of the above equation we have given the leading term in the decomposition of the tensor product into irreducible $\alg{d}(2,1;\alpha)$ representations. The highest weight state of this $(-\tfrac{L}{2};\tfrac{L}{2};\tfrac{L}{2})$ representation is given by $(\phi^{(0)}_+ \bar{\phi}^{(0)}_+)^L$. This state is annihilated by the supercharge $\gen{Q}_{-++}$. Hence the $(-\tfrac{L}{2};\tfrac{L}{2};\tfrac{L}{2})$ representation is a short $1/4$-BPS state, which we take as the spin-chain groundstate.

\subsection{Closed subsectors}
\label{sec:closed-subsectors}

To understand the alternating spin-chain it is often helpful to restrict to a closed subsector. We need to make sure that the subsector is closed under interactions. To do this, we want to construct a semi-positive definite charge $\gen{J}$ that commutes with the Cartan generators of the $\alg{d}(2,1;\alpha)$ algebra, and in particular with the left-moving Hamiltonian $-\gen{S}_0 + \alpha\gen{L}_5 + (1-\alpha)\gen{R}_8$. The subsector consists of all fields annihilated by $\gen{J}$. We construct such a charge as a linear combination of generators $\gen{J} = c_1 \gen{S}_0 + c_2 \gen{L}_5 + c_3 \gen{R}_8 + c_4 \gen{B}$. In addition to the Cartan elements of the bosonic $\alg{sl}(2) \times \alg{su}(2) \times \alg{su}(2)$ algebra, we have also the ``baryonic'' charge $\gen{B}$ which takes values $+1/2$ on odd sites and $-1/2$ on even sites.\footnote{%
  Note that a physical spin-chain states have equal number of odd and even sites and hence has $\gen{B}=0$.%
} %
The closed subsectors are summarized in table~\ref{tab:subsectors}. This analysis of subsectors is similar to the discussion in~\cite{Beisert:2003jj} about subsectors in $\mathcal{N}=4$ super-Yang-Mills theory.
\begin{table}
  \centering
  \begin{tabular}{ccc}
    \toprule
    Charge ($-\gen{J}$) & Fields & Sector \\
    \midrule
    $\gen{S}_0 + \alpha\gen{L}_5 + (1-\alpha)\gen{R}_8$ & $\phi_+^{0}$, $\bar{\phi}_+^{(0)}$ & $1/4$-BPS \\
    $\gen{S}_0 + \gen{R}_8 + \alpha\gen{B}$ & $\phi_\pm^{(0)}$, $\psi_+^{(0)}$, $\bar{\phi}_+^{(0)}$ & $\alg{su}(2|1)_+$ \\
    $\gen{S}_0 + \gen{L}_5 - (1-\alpha)\gen{B}$ & $\phi_+^{(0)}$, $\bar{\phi}_\pm^{(0)}$, $\bar{\psi}_+^{(0)}$, & $\alg{su}(2|1)_-$ \\
    \midrule
    $\gen{L}_5 + \gen{R}_8 - \gen{L}$ & $\phi_+^{(n)}$, $\psi_+^{(n)}$, $\bar{\phi}_+^{(n)}$, $\bar{\psi}_+^{(n)}$ & $\alg{sl}(2|1)$ \\
    $\gen{L}_5 - \gen{L}$ & $\phi_+^{(n)}$, $\bar{\psi}_+^{(n)}$ & $\alg{sl}(2)_+$ \\
    $\gen{R}_8 - \gen{L}$ & $\psi_+^{(n)}$, $\bar{\phi}_+^{(n)}$ & $\alg{sl}(2)_-$ \\
    $2\gen{S}_0 - (1-2\alpha) \gen{B} + \gen{L}$ & $\phi_\pm^{(0)}$, $\bar{\phi}_\pm^{(0)}$ & $\alg{su}(2) \times \alg{su}(2)$ \\
    $\gen{S}_0 + \gen{L}_5 + 2\gen{R}_8 + \alpha \gen{B} - \gen{L}$ & $\phi_+^{(0)}$, $\psi_+^{(0)}$, $\bar{\phi}_+^{(0)}$ & $\alg{su}(1|1)_+$ \\
    $\gen{S}_0 + 2\gen{L}_5 + \gen{R}_8 - (1-\alpha) \gen{B} - \gen{L}$ & $\phi_+^{(0)}$, $\bar{\phi}_+^{(0)}$, $\bar{\psi}_+^{(0)}$ & $\alg{su}(1|1)_-$ \\
    \bottomrule
  \end{tabular}
  \caption{Closed subsectors of the alternating $\alg{d}(2,1;\alpha)$ spin-chain. The first column gives the semi-definite charge $\gen{J}$ that identifies the sector. The first three sectors in the table are closed to all loops. For the other sectors the charge $\gen{J}$ is conserved only to leading order at strong coupling.}
  \label{tab:subsectors}
\end{table}

The simplest sector is the 1/4-BPS sector consisting of only the groundstate, and is obtained by choosing the charge $\gen{J}$ to be the spin-chain Hamiltonian $-\gen{S}_0-\alpha\gen{L}_5-(1-\alpha)\gen{R}_8$. The states in this sector are constructed out of the fields $\phi_+^{(0)}$ and $\bar{\phi}_+^{(0)}$. 

The next subsector is the $\alg{su}(2|1)_+$ sector where, in addition to the groundstate, we have one bosonic excitation $\phi_-^{(0)}$ and one fermionic excitation $\psi_+^{(0)}$. Note that both these excitations live on the odd sites of the spin-chain. The only other sector that is closed at all loops is the conjugate $\alg{su}(2|1)_-$ sector, where one bosonic and one fermionic excitation sit at the even sites.

To leading order at weak coupling there are no length changing interactions in the spin-chain Hamiltonian. We can then include the extra charge $\gen{L}$ measuring the length of the chain. This allows us to construct additional closed subsectors. The two $\alg{su}(2|1)$ sectors can now be split into two bosonic $\alg{su}(2)$ sectors as well as two $\alg{su}(1|1)$ sectors with a single fermionic excitation. Moreover, the two $\alg{su}(2)$ sectors can be combined to an $\alg{su}(2) \times \alg{su}(2)$ sector.

Additionally, there are two types non-compact subsectors. The $\alg{sl}(2|1)$ sector consists of the fields $\phi_+^{(n)}$, $\psi_+^{(n)}$, $\bar{\phi}_+^{(n)}$ and $\bar{\psi}_+^{(n)}$. The final type of sector that is closed as long as the length of the chain is preserved is an $\alg{sl}(2)$ sector which consist of the bosons $\phi_+^{(n)}$ on the odd sites and the fermions $\bar{\psi}_+^{(n)}$ on the even sites. Note that the usual groundstate is not part of this last sector. This situation is very similar to the $\alg{sl}(2)$ sector of the ABJM spin-chain~\cite{Zwiebel:2009vb}.

\subsection{The \texorpdfstring{$\alg{sl}(2|1)$}{sl(2|1)} subsector}
\label{sec:sl21-subsector}

We obtain the $\alg{sl}(2|1)$ subalgebra from $\alg{d}(2,1;\alpha)$ by defining the generators
\begin{equation}
  \gen{e}_1 = \gen{S}_- \,, \qquad
  \gen{f}_1 = \gen{S}_+ \,, \qquad
  \gen{e}_2 = \gen{Q}_{++-} \,, \qquad
  \gen{f}_2 = \gen{Q}_{-++} \,.
\end{equation}
Using the commutation relations in appendix~\ref{sec:d21a-algebra}, it is straightforward to show that the $\alg{sl}(2|1)$ algebra in~\eqref{eq:sl2-comm-rel} is satisfied. From the $\alg{d}(2,1;\alpha)$ representations for the fields we find that the above generators act on the fields $\phi_n \equiv \phi^{(n)}_+$ and $\psi_n \equiv \psi^{(n)}_+$ as
\begin{equation}
  \label{eq:sl21-sector-rep1}
  \begin{gathered}
    \begin{aligned}
      \gen{e}_1 \ket{\phi_n} &= - \sqrt{(n+\alpha)(n+1)} \ket{\phi_{n+1}} \,, &
      \gen{f}_1 \ket{\phi_n} &= + \sqrt{(n+\alpha-1)n} \ket{\phi_{n-1}} \,, \\
      \gen{e}_1 \ket{\psi_n} &= - \sqrt{(n+\alpha+1)(n+1)} \ket{\psi_{n+1}} \,, &
      \gen{f}_1 \ket{\psi_n} &= + \sqrt{(n+\alpha)n} \ket{\psi_{n-1}} \,, 
    \end{aligned}
    \\
    \gen{e}_2 \ket{\phi_n} = 0 \,, \qquad
    \gen{e}_2 \ket{\psi_n} = + \sqrt{n+\alpha} \ket{\phi_{n}} \,, \qquad
    \gen{f}_2 \ket{\phi_n} = - \sqrt{n+\alpha} \ket{\psi_{n}} \,, \qquad
    \gen{f}_2 \ket{\psi_n} = 0 \,,
  \end{gathered}
\end{equation}
while the action on the fields $\bar{\phi}_n \equiv \bar{\phi}^{(n)}_+$ and $\bar{\psi}_n \equiv \bar{\psi}^{(n)}_+$ is given by
\begin{equation}
  \label{eq:sl21-sector-rep2}
  \begin{gathered}
    \begin{aligned}
      \gen{e}_1 \ket{\bar{\phi}_n} &= - \sqrt{(n+1-\alpha)(n+1)} \ket{\bar{\phi}_{n+1}} \,, &
      \gen{f}_1 \ket{\bar{\phi}_n} &= + \sqrt{(n+1-\alpha-1)n} \ket{\phi_{n-1}} \,, \\
      \gen{e}_1 \ket{\bar{\psi}_n} &= - \sqrt{(n+1-\alpha+1)(n+1)} \ket{\bar{\psi}_{n+1}} \,, &
      \gen{f}_1 \ket{\bar{\psi}_n} &= + \sqrt{(n+1-\alpha)n} \ket{\bar{\psi}_{n-1}} \,, 
    \end{aligned}
    \\
    \gen{e}_2 \ket{\bar{\phi}_n} = - \sqrt{n} \ket{\bar{\psi}_{n-1}} \,, \qquad 
    \gen{e}_2 \ket{\bar{\psi}_n} = 0 \,, \qquad
    \gen{f}_2 \ket{\bar{\phi}_n} = 0 \,, \qquad
    \gen{f}_2 \ket{\bar{\psi}_n} = + \sqrt{n+1} \ket{\bar{\phi}_{n+1}} \,.
  \end{gathered}
\end{equation}
Comparing the above relations to equations~\eqref{represl21} and~\eqref{represl21ac} we see that~\eqref{eq:sl21-sector-rep1} corresponds to an \emph{anti-chiral} representation spin $s_+=\tfrac{\alpha}{2}$, and~\eqref{eq:sl21-sector-rep2} to a \emph{chiral} representation of spin $s_-=\tfrac{1-\alpha}{2}$.

Using the results in section~\ref{surmat} we can write down R-matrices acting on these representations. There are four R-matrices acting on the various combinations of $\alg{sl}(2|1)$ representations at the odd and even sites. Each such R-matrix can be written as a sum over projectors onto irreducible representations in the tensor products of two sites. For two anti-chiral representations $(s_+;-s_+)$ or two chiral representations $(s_-;+s_-)$ this decomposition is given by
\begin{align}
  (s_+;-s_+) \otimes (s_+;-s_+) &= (2s_+;-2s_+) \oplus \bigoplus_{n=1}^{\infty} (2s_+ - \tfrac{1}{2} + n; -2s + \tfrac{1}{2}) \,, \label{eq:sl2-aa-decomp} \\
  (s_-;+s_-) \otimes (s_-;+s_-) &= (2s_-;+2s_-) \oplus \bigoplus_{n=1}^{\infty} (2s_- - \tfrac{1}{2} + n; +2s - \tfrac{1}{2}) \,. \label{eq:sl2-cc-decomp}
\end{align}
The first representation appearing on the right-hand side of~\eqref{eq:sl2-aa-decomp} and~\eqref{eq:sl2-cc-decomp} is short, while all other representations in these decompositions are long. In the case of one anti-chiral and one chiral representation decomposes into a sum of long representations
\begin{equation} \label{eq:sl2-ac-decomp}
  (s_+;-s_+) \otimes (s_-;+s_-) = \bigoplus_{n=0}^{\infty} (s_- + s_+ + n; s_- - s_+) \,.
\end{equation}

Indicating the anti-chiral and chiral representations by $\pm$, we find the R-matrix for two identical representation from~\eqref{hw2} by setting $s_1 = s_2 = s_\pm$
\begin{equation}\label{eq:R-matrix-ob-cc-aa}
  R_{\pm\pm}(u) = \sum_{n=0}^{\infty} \, \prod_{k=0}^{n-1} \frac{u + 2s_{\pm} + k}{u - 2s_{\pm} - k} \, \Pi_n^{\pm\pm} \,,
\end{equation}
where $\Pi_n^{\pm,\pm}$ denotes the projectors onto the representations appearing in the decompositions in~\eqref{eq:sl2-aa-decomp} and~\eqref{eq:sl2-cc-decomp}.
Similarly we set $s_1 + s_2 = s_+ + s_-$ in equation~\eqref{eq:R-mat-c-ac} to get the R-matrix for the mixed chiral--anti-chiral case,
\begin{equation}\label{eq:R-matrix-ob-ca-ac}
  R_{\pm\mp}(u) = \sum_{n=0}^{\infty} \prod_{k=0}^{n-1} \frac{u + s_+ + s_- + \tfrac{1}{2} + k}{u - s_+ - s_- - \tfrac{1}{2} - k} \, \Pi_n^{\pm\mp} \,,
\end{equation}
where $\Pi_n^{\pm\mp}$ projects onto the representations appearing in~\eqref{eq:sl2-ac-decomp}. Inserting the relevant values for $s_\pm$ and rescaling of the spectral parameter $u \to u/2$, and simplifying~\eqref{eq:R-matrix-ob-ca-ac} using $s_+ + s_- = 1/2$, we find that the above expressions perfectly agrees with the corresponding result in~\cite{OhlssonSax:2011ms}.

\subsection{The \texorpdfstring{$\alpha \to 0$}{alpha to 0} limit of the alternating spin-chain}
\label{sec:alpha-0-limit}

In the $\alpha \to 0$ limit one of the three-spheres decompactifies in the string background. Together with the $\Sphere^1$, this sphere forms a (partially decompactified) $\Torus^4$. At the same time the $\alg{d}(2,1;\alpha)^2$ isometry reduces to $\alg{psu}(1,1|2)^2$. As discussed in~\cite{OhlssonSax:2011ms}, the limit of the $\alg{d}(2,1;\alpha)$ algebra is taken in such a way that the generators of the $\alg{su}(2)_+$ subalgebra become commuting. We will now study what happens to the representations at the spin-chain sites in the above limit, by restricting to the $\alg{sl}(2|1)$ subsector.

Let us consider the anti-chiral $\alg{sl}(2|1)$ representation~\eqref{eq:sl21-sector-rep1} in the $\alpha \to 0$ limit. On the bosonic fields $\phi_n$ the generators now act as
\begin{equation*}
  \gen{e}_1 \ket{\phi_n} = - \sqrt{(n+1)n} \ket{\phi_{n+1}} \,, \quad
  \gen{f}_1 \ket{\phi_n} = + \sqrt{n(n-1)} \ket{\phi_{n+1}} \,, \quad
  \gen{e}_2 \ket{\phi_n} = 0 \,, \quad
  \gen{f}_2 \ket{\phi_n} = - \sqrt{n} \ket{\psi_n} \,.
\end{equation*}
In particular, all generators annihilate the state $\ket{\phi_0}$. In order to identify the rest of the module we introduce a new field $\varphi_n \equiv -\phi_{n+1}$. The limit of the relations in~\eqref{eq:sl21-sector-rep1} then take the form
\begin{equation}
  \label{eq:sl21-sector-rep1-alpha-0}
  \begin{gathered}
  \begin{aligned}
    \gen{e}_1 \ket{\psi_n} &= - (n+1) \ket{\psi_{n+1}} \,, &
    \gen{f}_1 \ket{\psi_n} &= + n \ket{\psi_{n-1}} \,, \\
    \gen{e}_1 \ket{\varphi_n} &= - \sqrt{(n+2)(n+1)} \ket{\varphi_{n+1}} \,, &
    \gen{f}_1 \ket{\varphi_n} &= + \sqrt{(n+1)n} \ket{\varphi_{n-1}} \,,
  \end{aligned}
  \\
    \gen{e}_2 \ket{\psi_n} = - \sqrt{n} \ket{\varphi_{n-1}} \,, \qquad
    \gen{e}_2 \ket{\varphi_n} = 0 \,, \qquad
    \gen{f}_2 \ket{\psi_n} = 0 \,, \qquad
    \gen{f}_2 \ket{\varphi_n} = + \sqrt{n} \ket{\psi_{n+1}} \,.
  \end{gathered}
\end{equation}
Comparing the above expression with~\eqref{represl21} we see that this looks like a \emph{chiral} $\alg{sl}(2|1)$  representation with spin $s=1/2$, but where the bosonic and fermionic fields have switched roles. Taking the same $\alpha \to 0$ limit of the chiral representation in~\eqref{eq:sl21-sector-rep2}, we find an ordinary chiral representation with spin $s=1/2$. Hence the representations living on the odd and even sites of the limit of the alternating $\alg{sl}(2|1)$ spin-chain are almost the same except that there is an additional singlet state at the odd sites, and that the statistics of the fields on the odd and even sites are switched around.

It is instructive to also consider the first few states in the tensor product of two odd sites. For general $\alpha$ this tensor product decomposes as
\begin{equation}\label{eq:ac-ac-tensor-decomp}
  (\tfrac{\alpha}{2};-\tfrac{\alpha}{2}) \otimes (\tfrac{\alpha}{2};-\tfrac{\alpha}{2}) =
  (\alpha;-\alpha) \oplus (\alpha+\tfrac{1}{2};-\alpha+\tfrac{1}{2}) \oplus (\alpha+\tfrac{3}{2};-\alpha+\tfrac{1}{2}) \oplus \dotsb \,.
\end{equation}
The first representation on the right hand side is anti-chiral, with the rest of the representation in the sum being long. The highest weight state of the anti-chiral representation is given by $\ket{\phi_0 \phi_0}$. The supercharge $\gen{e}_3 = \acomm{\gen{e}_1}{\gen{e}_2}$ acts on this state as
\begin{equation}
  \gen{e}_3 \ket{\phi_0 \phi_0} = -\sqrt{\alpha}(\ket{\psi_0 \phi_0} + \ket{\phi_0 \psi_0}) \,.
\end{equation}
For $\alpha = 0$, the state $\ket{\phi_0 \phi_0}$ becomes a singlet, and the state on the right above is the highest weight state in a new \emph{chiral} representation $(\tfrac{1}{2};\tfrac{1}{2})$. In other words, $(\alpha;-\alpha)$ decomposes in the limit to
\begin{equation}
  (\alpha;-\alpha) \to \rep{1} \oplus (\tfrac{1}{2};\tfrac{1}{2})_S \,,
\end{equation}
where $\rep{1}$ is the trivial representation and the subscript $S$ in the second term indicates that the representation lives in the symmetric part of the tensor product.

We can analyse the $(\alpha+\tfrac{1}{2};-\alpha+\tfrac{1}{2})$ in a similar way. For $\alpha \neq 0$ this is a long representation with highest weight state $\ket{hw}_1 = \ket{\psi_0 \phi_0} - \ket{\phi_0 \psi_0}$. Acting with $\gen{e}_3$ on this state gives
\begin{equation}
  \gen{e}_3 \ket{hw}_1 = 2\sqrt{\alpha} \ket{\psi_0 \psi_0} \,.
\end{equation}
At $\alpha = 0$ this representation splits into two short multiplets. The original highest weight state $\ket{hw}_1$ generates another copy of the chiral representation $(\tfrac{1}{2};\tfrac{1}{2})$, while $\ket{\psi_0 \psi_0}$ is the highest weight state of a new chiral representation with charges $(1;1)$. The decomposition of the second state on the right hand side of~\eqref{eq:ac-ac-tensor-decomp} is
\begin{equation}
  (\alpha+\tfrac{1}{2};-\alpha+\tfrac{1}{2}) \to (\tfrac{1}{2};\tfrac{1}{2})_A \oplus (1;1) \,,
\end{equation}
where the subscript $A$ indicates an anti-symmetric state. All the other representations in the decomposition~\eqref{eq:ac-ac-tensor-decomp} remain long in the $\alpha \to 0$ limit. Hence we can write the full decomposition in this limit as
\begin{equation}
  \big( \rep{1} \oplus (\tfrac{1}{2};\tfrac{1}{2}) \big) \otimes \big( \mathbf{1} \oplus (\tfrac{1}{2};\tfrac{1}{2}) \big)
  =
  \big( \rep{1} \oplus (\tfrac{1}{2};\tfrac{1}{2})_S \big) \oplus \big( (\tfrac{1}{2};\tfrac{1}{2})_A \oplus (1;1) \big) \oplus (\tfrac{3}{2};\tfrac{1}{2}) \oplus \dotsc \,,
\end{equation}
where we have grouped terms that originate from the same $\alpha \neq 0$ representation.

The R-matrix in~\eqref{eq:R-matrix-ob-cc-aa} is written as a sum over projectors acting on two anti-chiral spin-chain sites. Each such projector acts on one of the representations in~\eqref{eq:ac-ac-tensor-decomp}. In the $\alpha \to 0$ limit the projectors acting on the first two representation in the decomposition split into two projectors, corresponding to the split of these multiplets when the limit is taken. However, according to~\eqref{eq:R-matrix-ob-cc-aa} these representations still have the same eigenvalues under the R-matrix. Setting $\alpha = 0$ in these eigenvalues, we find that the R-matrix acts as the identity on the trivial representation $\rep{1}$ as well as on the two $(\tfrac{1}{2};\tfrac{1}{2})$ representations. On the other representations in the above decomposition, the eigenvalues of the R-matrix perfectly agrees with the result for the chiral R-matrix in~\eqref{hw2} provided we set $s_1 = s_2 = 1/2$.

Similarly, the tensor product of the representations on an odd and an even site of the spin-chain gives the decomposition
\begin{equation}\label{eq:ac-c-tensor-decomp}
  (\tfrac{\alpha}{2};-\tfrac{\alpha}{2}) \otimes (\tfrac{1-\alpha}{2};\tfrac{1-\alpha}{2}) =
  (\tfrac{1}{2};\tfrac{1}{2}-\alpha) \oplus (\tfrac{3}{2};\tfrac{1}{2}-\alpha) \oplus (\tfrac{5}{2};\tfrac{1}{2}-\alpha) \oplus \dotsc \,.
\end{equation}
For $\alpha > 0$ all these representations are long, but for $\alpha = 0$ the first representation splits into two chiral representations
\begin{equation}
  (\tfrac{1}{2};\alpha-\tfrac{1}{2}) \to (\tfrac{1}{2};\tfrac{1}{2}) + (1;1) \,.
\end{equation}
The R-matrix again acts trivially on the $(\tfrac{1}{2};\tfrac{1}{2})$ multiplet, and has the same eigenvalues as in~\eqref{hw2} (with $s_1 = s_2 = 1/2$) when acting on the rest of the tensor product. In particular, $R(0)$ acts as the identity if the one of the states is a singlet, and as a permutation otherwise.

This structure also appears in the limit of the full $\alg{d}(2,1;\alpha)$ spin-chain. At the even sites of the chain we have a irreducible $\alg{su}(1,1|2)$ representation denoted by $(-\tfrac{1}{2};\tfrac{1}{2})$. On the odd sites we have a reducible representation. In the full spin-chain there are two singlets, originating from the $\alg{su}(2)_+$ doublet $\phi^{(0)}_\pm$. The rest of the fields on the odd sites transform in the same $(-\tfrac{1}{2};\tfrac{1}{2})$ representation as the fields on the even sites, but with the opposite statistics.

In the following section we will study an alternating spin-chain with a very similar structure as the spin-chain discussed here. In order to simplify the calculations we will consider an $\alg{sl}(2)$ spin-chain with even sites in the spin $s=-1$ representations and odd sites in a reducible representation containing a singlet and a $s=-1$ multiplet. Apart from some minus signs originating from the fermion statistics this $\alg{sl}(2)$ spin-chain can be obtained from the $\alg{sl}(2|1)$ spin-chain by restricting to the fields in the $(-\tfrac{1}{2};\tfrac{1}{2})$ representation with $\alg{sl}(2)$ spin $s=-1,-2,\dotsc$. This is not a closed sector of the full $\alg{su}(1,1|2)$ spin-chain, but the analysis still captures the structure of the reducible spin-chain discussed above. It is straightforward to generalize the following construction to the full superalgebra case.

\section{\texorpdfstring{$R$}{R} module spin-chains}
\label{sec7}

In the preceding sections we have shown that the R-matrix presented in~\cite{OhlssonSax:2011ms} can be obtained from a Yangian universal R-matrix construction. In particular, this confirms that in the $\alpha\rightarrow 0$ limit the integrable structure of the alternating $\alg{d}(2,1;\alpha)$ spin-chain is preserved. In the present section we will investigate this spin-chain in the $\alpha\equiv 2s\rightarrow 0$ limit. In order to focus on the key features that the spin-chain exhibits in this limit, we will restrict ourselves in this section mostly to an $\alg{sl}(2)$ subsector. The generalizations to the complete spin-chain are mostly straightforward and we will comment on them at the end of the section.

As we have seen in section~\ref{inof} there are three possible $s\rightarrow 0$ representations that may arise, but that only one of these, the reducible $R$ module, is a unitary representation. Nevertheless,  representation theory on its own does not tell us whether the $s\rightarrow 0$ spin-chain will involve $P$,  $R$ or $S$ modules. In fact, we have found sensible R-matrices for spin-chains made out of $P$,  $R$ or $S$ modules. As a result, we may investigate the integrability properties of the three types of spin-chains. It turns out that spin-chains involving the $P$ or $S$ modules have unconventional integrability properties, which we consider undesirable for our purposes. We present a more detailed description of this in section~\ref{sec71} and delegate some of the computational details to the appendices. 

The observations made in section~\ref{sec71} lead us to conclude that in the $s\rightarrow 0$ limit the spin-chains we wish to investigate involve the $R$ module. Since there appears to be little literature on spin-chains with reducible representations we will consider some toy-models first before focusing on the alternating chain in question. In section~\ref{sec:homogenous-toy-model} we consider the simplest  toy-model: a homogenous spin-chain with a three-dimensional reducible $\alg{su}(2)$ representation ${\bf 1}\oplus {\bf 2}$ at each site. In section~\ref{sec:homogenous-R-spin-chain} we generalize this to a homogenous $\alg{sl}(2)$ spin-chain with the $R$ module at each site. In section~\ref{sec:alternating-toy-model} we generalize the $\alg{su}(2)$ toy model to an alternating chain with odd sites in the reducible ${\bf 1}\oplus {\bf 2}$ representation, and even sites transforming in ${\bf 2}$. Finally, in section~\ref{sec:alternating-R-spin-chain} we consider the alternating chain with an $R$ module at even lattice sites and an ${\bf s}=-1$ module at odd sites; as we have discussed in the previous section, this spin-chain is closely related to the spin-chain discussed in~\cite{OhlssonSax:2011ms}.

Before proceeding further it is appropriate to point out a technical reason for why we cannot simply apply the many results on integrable spin-chains for irreducible representations to the present cases. It appears that these results, to a large extent, depend on the technical assumption that (for a homogeneous spin-chain) the R-matrix evaluated at a privileged value of the spectral parameter (typically at $u=0$) is proportional to the permutation operator. On highest-weight states $\left|j\right>_{01}$ this implies that  we must have\footnote{Analogous conditions exist for alternating chains.}
\begin{equation}
R_{01}(u=0)\left|j\right>_{01}\propto (-1)^j\left|j\right>_{01}\,.
\label{permop}
\end{equation}
While this holds for the R-matrix we have been considering in this paper away from $ s\neq 0$, when $s=0$ it is easy to see that 
\begin{equation}
R_{01}(u=0)\left|j=0\right>_{01}=R_{01}(u=0)\left|j=1\right>_{01}\,.
\end{equation}
In other words, the R-matrix at $u=0$ is {\em not} proportional to the permutation operator. It seems that as a result, many of the conventional integrable spin-chain techniques do not immediately apply, and one has to obtain information about such spin-chains from first principles.

\subsection{Integrability for \texorpdfstring{$S$}{S} or \texorpdfstring{$P$}{P} module spin-chains}
\label{sec71}

In this subsection we review some of the integrability properties of spin-chains where (some of) the sites transform in the $S$ or $P$ representations. We will simplify the problem slightly by considering only homogeneous $S$ or $P$ module spin-chains. These will already have the peculiar features which we referred to above. It is then easy to convince oneself that alternating spin-chains involving these modules will also have such features.

Conventionally, the first step in investigating an integrable spin-chain is to define the monodromy matrix $T(u)$. For a homogeneous spin-chain this is given by
\begin{equation}
T(u)=R_{01}(u)R_{02}(u)\ldots R_{0N}(u)\,,
\end{equation}
where the subscripts $1,2,\ldots,N$ label the sites in the spin-chain, and the subscript $0$ corresponds to the auxiliary space which is taken  to be in the same representation as all the sites of the spin-chain. The R-matrix for both the homogeneous $P$ and $S$ module spin-chains can be expressed as a sum of projectors
\begin{equation}
R(u)=R^*(u)\Pi^*+
\sum_{l=1}^\infty R^l(u)\Pi^{(l)}
\end{equation}
where 
\begin{equation}
R^l(u)=
(-1)^{l+1} \frac{\Gamma(1-u)\Gamma(l+u)}{\Gamma(1+u)\Gamma(l-u)}\,,\qquad\mbox{and}\qquad R^*(u)=1\,.
\end{equation}
Above, $\Pi^*$ and $\Pi^{(l)}$ are projectors onto irreps in the decomposition of $P\otimes P$ or $S\otimes S$ given in equations~(\ref{PPirep}) and~(\ref{SSirep}); in particular $\Pi^*$ projects onto the $P$ or $S$ module on the right-hand side of those equations.
In the appendix we find explicit expressions for these projectors in terms of the oscillator representations introduced in section~\ref{inof}. 

Using this form of the R-matrices, we may compute the action of the monodromy matrix $T(u)$ on short spin-chain states. This is further facilitated by the fact that the monodromy matrix is manifestly length preserving, and it turns out also to preserve the overall number of excitations on the spin-chain state. This allows one to consider subsectors of the spin-chain state-space, with a fixed length and fixed number of excitations. The details of the calculation are presented in the appendix, but the important observation is that $T(u)$ has non-trivial Jordan blocks, and so is not diagonalisable. This feature also persists when one computes the transfer matrix
\begin{equation}
\tau(u)=\tr_0(R_{01}(u)R_{02}(u)\ldots R_{0N}(u))\,.
\end{equation}

While each of the matrices $R_{0i}(u)$ has no non-trivial Jordan blocks the product $R_{0i}(u)R_{0j}(u)$ does. For example, given the explicit form of the R-matrices one can show that  $R_{0i}(u)R_{0j}(u)$ acting on a three-site chain with three excitations has a single non-trivial $2\times 2$ Jordan block. Denoting by $\ket{m_0,m_1,m_2}$ a three site state
\begin{equation}
(a_0^\dagger)^{m_0}(a_1^\dagger)^{m_1}(a_2^\dagger)^{m_2} \ket{0,0,0}
\end{equation}
with $m_0+m_1+m_2$ excitations, one can show that in the $S$ module
\begin{equation}
  R_{01}(0)R_{02}(0) \ket{1,1,1} = \ket{1,1,1} \,, \qquad
  R_{01}(0)R_{02}(0) \ket{0,1,2} = \ket{0,1,2} + 2 \ket{1,1,1} \,.
\end{equation}
Similarly, in the $P$ module the two states
\begin{equation}
  \begin{aligned}
    \ket{\psi_1}_P &= \tfrac{1}{2}\big(\ket{0,3,0} - \ket{3,0,0} + (\ket{0,1,2} - \ket{0,2,1}) + (\ket{2,0,1} - \ket{1,0,2}) - 2 (\ket{1,2,0} - \ket{2,1,0})\big) \,, \\
    \ket{\psi_2}_P &= \ket{0,2,1} + \ket{1,0,2} - 2 \ket{1,1,1} \,,
  \end{aligned}
\end{equation}
satisfy
\begin{equation}
  R_{01}(0)R_{02}(0) \ket{\psi_1}_P = \ket{\psi_1}_P \,, \qquad
  R_{01}(0)R_{02}(0) \ket{\psi_2}_P = \ket{\psi_2}_P+ 2\ket{\psi_1}_P \,.
\end{equation}
Longer chains or chains with more excitations have more complicated Jordan blocks.  The presence of Jordan blocks in these settings for $P$ module spin-chains was already noted in~\cite{Korchemsky:1994um}. It turns out that the space of all spin-chain states with at least one excitation at each site is a closed subsector of the full integrable chain. One can show that in this subsector non-trivial Jordan blocks do not arise. We also note here that for $P$ and $S$ module spin-chains, Jordan blocks also appear for the matrix $\rho$ (defined in equation~\ref{rhomat} below) which is used in the Algebraic Bethe Ansatz construction. Some explicit examples of this are given in appendix~\ref{sec:jordan-blocks}.

Having non-trivial Jordan blocks in the full spin-chain would be unphysical - after all its energies are meant to correspond to the spectrum of physical string excitations at small $\lambda$. One may wonder whether such unphysical states somehow decouple in the full spin-chain.\footnote{We thank Matthias Gaberdiel for an interesting discussion of related issues in CFTs.} One can check that analyzing the complete alternating chain which includes both left- and right-movers does not alter the presence of Jordan blocks - this is in contrast to what happens in logarithmic CFTs, see for example the review~\cite{Gaberdiel:2001tr}. The only other way to remove such states would be by introducing an additional projection into the spin-chain at $\alpha=0$. The only natural projection that we could identify comes from the zero momentum condition one imposes in the $\alpha\neq 0$ spin-chain. Each state in the $\alpha=0$ spin-chain has a corresponding state in the $\alpha\neq 0$ spin-chain. We might then decide at $\alpha=0$ to keep only those states that come from momentum-conserving states in the $\alpha\neq 0$ theory. However, one may check that this does not eliminate all the non-trivial Jordan blocks. We do not know of any other physically motivated projections. 

To summarize, the results of this subsection lead us to the conclusion that, from the point of view of the $AdS_3/CFT_2$ correspondence, we should be interested in spin-chains based on the reducible $R$ module and its super-algebra generalizations. In the rest of this section we will discuss in detail integrable spin-chains with reducible representations at some of their sites.

\subsection{The homogeneous \texorpdfstring{$\mathbf{1}\oplus \mathbf{2}$}{1+2} spin-chain}
\label{sec:homogenous-toy-model}

Given the results of the previous section we would like to investigate the $\alpha\rightarrow 0$ limit of the alternating $\alg{d}(2,1;\alpha)$ spin-chain. In this limit some of the representations become reducible. As we showed in the discussion around equation~(\ref{permop}),  spin-chains involving reducible representations cannot be treated using conventional integrability techniques. As we saw in the first part of the paper, however, the integrable properties of such chains are well established through the Yangian construction of the R-matrix. We want to understand better their ``local'' properties. In the rest of this section we will consider spin-chains with reducible representations at (some) sites and use their R-matrices to deduce the ``local'' physics. We will start with simpler examples which share many of the essential features of the $\alpha\rightarrow 0$ limit of the alternating $\alg{d}(2,1;\alpha)^2$ spin-chain in which we are ultimately interested in. 

In this subsection we consider a homogenous $\alg{su}(2)$ spin-chain with ${\bf r}\equiv{\bf 1}\oplus {\bf 2}$, the three-dimensional reducible representation, at each site.
A convenient basis for this vector space is given by
\begin{equation}
\ket{0}\,,\qquad
\left|1\right>\,,\qquad
\left|2\right>\,,
\end{equation}
where the first state is the singlet. The $\alg{su}(2)$ generators can be represented as the following matrices
\begin{equation}
  J^i=
  \begin{pmatrix}
    0 & 0 \\
    0 & \tfrac{i}{2} \sigma^i
  \end{pmatrix} \,.
\end{equation}
The decomposition into irreps of the tensor product of two of these representations is given by
\begin{equation}
{\bf r}\otimes {\bf r}\cong {\bf 1}\oplus {\bf 2}\oplus {\bf 2}\oplus {\bf 3}\oplus {\bf 1}\,.
\label{irrepdecomptrunc}
\end{equation}
The highest weight states of the irreps on the right-hand side above are given by
\begin{equation}
\left|0,0\right>_{12}\,,\,\,\,
\left|1,0\right>_{12}\,,\,\,\,
\left|0,1\right>_{12}\,,\,\,\,
\left|1,1\right>_{12}\,,\,\,\,
\left|2,1\right>_{12}-\left|1,2\right>_{12}\,.
\end{equation}
For completeness we note that the modules ${\bf 2}$ and ${\bf 3}$ span the following subsets of
the nine-dimensional vector space
\begin{align*}
{\bf 2} &= \mbox{span}\left\{\left|1,0\right>_{12}\,,\, \left|2,0\right>_{12}\right\}\,, \\
{\bf 2} &= \mbox{span}\left\{\left|0,1\right>_{12}\,,\, \left|0,2\right>_{12}\right\}\,, \\
{\bf 3} &= \mbox{span}\left\{\left|1,1\right>_{12} \,,\,\left|2,1\right>_{12}+\left|1,2\right>_{12} \left|2,2\right>_{12} \right\}\,.
\end{align*}
The analogue of the R-matrix considered in the previous sections can be written as a $9\times 9$ matrix
\begin{equation}
R_{{\bf r}\otimes {\bf r}}(u)=
\left(\begin{array}{ccccccccc} 
1 & 0 & 0 & 0 & 0 & 0 & 0 & 0 & 0 \\
0 & 1 & 0 & 0 & 0 & 0 & 0 & 0 & 0 \\
0 & 0 & 1 & 0 & 0 & 0 & 0 & 0 & 0 \\
0 & 0 & 0 & 1 & 0 & 0 & 0 & 0 & 0 \\
0 & 0 & 0 & 0 & r_3(u) & 0 & 0 & 0 & 0 \\
0 & 0 & 0 & 0 & 0 & \dfrac{r_3(u)+r_1(u)}{2} & 0 & \dfrac{r_3(u)-r_1(u)}{2}  & 0 \\
0 & 0 & 0 & 0 & 0 & 0 & 1 & 0 & 0 \\
0 & 0 & 0 & 0 & 0 & \dfrac{r_3(u)-r_1(u)}{2}  & 0 & \dfrac{r_3(u)+r_1(u)}{2}  & 0 \\
0 & 0 & 0 & 0 & 0 & 0 & 0 & 0 & r_3(u) 
\end{array}\right)\,,
\end{equation}
where
\begin{equation}
r_1(u)= -(1+u) \,,\qquad
r_3(u)= +(1-u) \,.
\end{equation}
and the R-matrix above acts in the basis
\begin{equation}
\left(\left|0,0\right>_{12}\,,\,\left|0,1\right>_{12}\,,\,\left|0,2\right>_{12}\,,\,
\left|1,0\right>_{12}\,,\,\left|1,1\right>_{12}\,,\,\left|1,2\right>_{12}\,,\,
\left|2,0\right>_{12}\,,\,\left|2,1\right>_{12}\,,\,\left|2,2\right>_{12}
\right)^t
\end{equation}
with ${}^t$ denoting the transpose. In terms of projectors the R-matrix can be written as
\begin{equation}
  \label{RRRmatasproj}
  R_{\mathbf{r} \otimes \mathbf{r}}(u) = \Pi_{\mathbf{1}} + \Pi_{\mathbf{2}} + \Pi_{\mathbf{2}} + r_3(u)\Pi_{\mathbf{3}} + r_1(u) \Pi_{\mathbf{1}}\,,
\end{equation}
where $\Pi_{{\bf j}}$ projects onto the ${\bf j}$ representation on the right-hand side of equation~(\ref{irrepdecomptrunc}). In this form the similarity to the R-matrices considered in the first part of this paper is most easily noted.
One can quickly check that the YBE is satisfied\footnote{In fact one could pick a more general R-matrix consistent with the YBE of the form
\begin{equation}
R(u)=\Pi_{{\bf 1}}\oplus b(u)\Pi_{{\bf 2}}\oplus c(u)
\Pi_{{\bf 2}}\oplus r_2(u)\Pi_{{\bf 3}}\oplus r_1(u)\Pi_{{\bf 1}}\,,
\end{equation}
where $b(u)$ and $c(u)$ are arbitrary functions of the spectral parameter. However, the analogy with the R-matrix considered in the previous sections is most apt when $b=c=1$.}
\begin{equation}
R_{{\bf r}\otimes{\bf r},12}(u-v)R_{{\bf r}\otimes{\bf r},13}(u)
R_{{\bf r}\otimes{\bf r},23}(v)
=
R_{{\bf r}\otimes{\bf r},23}(v)
R_{{\bf r}\otimes{\bf r},13}(u)
R_{{\bf r}\otimes{\bf r},12}(u-v)\,.
\label{RYBE}
\end{equation}
We find it convenient to write the R-matrix as
\begin{equation}
R_{{\bf r}\otimes {\bf r}}(u) \equiv I^\perp + R_{{\bf 2}\otimes {\bf 2}}(u)\,,
\label{Rmatrrsum}
\end{equation}
where
\begin{align}
I^\perp &\equiv \Pi_{{\bf 1}}\oplus \Pi_{{\bf 2}}\oplus\Pi_{{\bf 2}} \equiv I^{\perp\,1}+I^{\perp\,2}\,, \\
R_{{\bf 2}\otimes {\bf 2}}(u) &= \frac{r_3(u)+r_1(u)}{2}(0\oplus 1_2)\otimes (0\oplus 1_2) + \dfrac{r_3(u)-r_1(u)}{2}(0\oplus \sigma^i)\otimes (0\oplus \sigma^i)\,.
\end{align}
Above, 
\begin{equation}
I^{\perp\,1}=\left(\begin{array}{ccc} 0 & 0 & 0 \\ 0 & 1 & 0 \\ 0 & 0 & 1 \end{array}\right)\otimes
\left(\begin{array}{ccc} 1 & 0 & 0 \\ 0 & 0 & 0 \\ 0 & 0 & 0 \end{array}\right)\,,\qquad
I^{\perp\,2}=\left(\begin{array}{ccc} 1 & 0 & 0 \\ 0 & 0 & 0 \\ 0 & 0 & 0 \end{array}\right)\otimes  
\left(\begin{array}{ccc} 1 & 0 & 0 \\ 0 & 1 & 0 \\ 0 & 0 & 1 \end{array}\right)\,,
\end{equation}
and, for example,
\begin{equation}
0\oplus \sigma^1=\left(\begin{array}{ccc} 0 & 0 & 0 \\ 0 & 0 & 1 \\ 0 & 1 & 0 \end{array}\right)\,,
\end{equation}
In particular, $R_{{\bf 2}\otimes {\bf 2}}(u)$ is the conventional 
($4\times 4$) R-matrix of the $XXX_{1/2}$ spin-chain ``enlarged'' to a $9\times 9$ matrix with a bunch of zeros.
The monodromy matrix can be defined in the usual way
\begin{equation}
T(u)=R_{01}(u)R_{02}(u)\ldots R_{0N}(u)\,,
\end{equation}
where the subscripts $1,2,\ldots,N$ label the sites in the spin-chain, and the subscript $0$ corresponds to the auxiliary space which is also an ${\bf r}$ module.

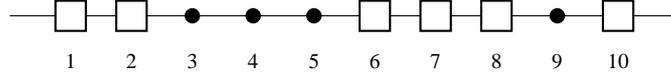
\begin{figure}

\centering
\begin{tikzpicture}
  \coordinate (start) at (-1\sitedist,0);
  \node (n1) at (0\sitedist,0) [odd site] {};
  \node (n2) at (1\sitedist,0) [odd site] {};
  \node (n3) at (2\sitedist,0) [null site] {};
  \node (n4) at (3\sitedist,0) [null site] {};
  \node (n5) at (4\sitedist,0) [null site] {};
  \node (n6) at (5\sitedist,0) [odd site] {};
  \node (n7) at (6\sitedist,0) [odd site] {};
  \node (n8) at (7\sitedist,0) [odd site] {};
  \node (n9) at (8\sitedist,0) [null site] {};
  \node (n10) at (9\sitedist,0) [odd site] {};
  \coordinate (end) at (10\sitedist,0) {};

  \draw [chain line] (start) -- (n1) -- (n2) -- (n3) -- (n4) -- (n5) -- (n6) -- (n7) -- (n8) -- (n9) -- (n10) -- (end);

  \node at ($(n1)-(0,\labeldist)$) {$\scriptstyle 1$};
  \node at ($(n2)-(0,\labeldist)$) {$\scriptstyle 2$};
  \node at ($(n3)-(0,\labeldist)$) {$\scriptstyle 3$};
  \node at ($(n4)-(0,\labeldist)$) {$\scriptstyle 4$};
  \node at ($(n5)-(0,\labeldist)$) {$\scriptstyle 5$};
  \node at ($(n6)-(0,\labeldist)$) {$\scriptstyle 6$};
  \node at ($(n7)-(0,\labeldist)$) {$\scriptstyle 7$};
  \node at ($(n8)-(0,\labeldist)$) {$\scriptstyle 8$};
  \node at ($(n9)-(0,\labeldist)$) {$\scriptstyle 9$};
  \node at ($(n10)-(0,\labeldist)$) {$\scriptstyle 10$};
\end{tikzpicture}

  \caption{An example of a state in a reducible homogeneous spin-chain. The squares indicate sites in a non-trivial irrep, such as ${\bf 2}$ in section~\ref{sec:homogenous-toy-model} or  the $R$ module in section~\ref{sec:homogenous-R-spin-chain}. The dots indicate sites in a trivial (singlet) representation. In the notation introduced in equations~(\ref{partition}) and~(\ref{ntildedef}) this state has ${\bf N}=\{1,2,3,4,5,6,7,8,9,10\}$, ${\bf n}=\{3,4,5,9\}$ and ${\bf {\tilde n}}=\{1,2,7,8,10\}$ as well as $N=10$, $n=4$ and ${\tilde n}=6$. The states are drawn according to the ordering given by the site number.
}
  \label{fig:redchain}

\end{figure}

It is convenient to denote the states $\left| 1\right>$ and $\left| 2\right>$ as $\left|\downarrow\right>$ or $\left|\uparrow\right>$ to emphasize that they are an $\alg{su}(2)$ doublet. Collectively we will refer to these doublet states as $\left|\updownarrow\right>$. 
A basis of states of length $N$ can be constructed from all two-cell partitions of the set ${\bf N}\equiv \left\{1,2,\dots,N\right\}$
\begin{equation}
{\bf r}^N=\mbox{span }
\bigcup_{
{\bf n}\subset {\bf N} } \{\left|{\bf n}\right>\}\,.
\label{rbasis1}
\end{equation}
By an abuse of terminology we allow ${\bf n}=\emptyset, {\bf N}$, as well as all proper subsets of ${\bf N}$. The state $\left|{\bf n}\right>$ is defined as follows. Given a two-cell partition defined by ${\bf n}\subset {\bf N}$, we label the elements of ${\bf n}$ and its complement
${\tilde {\bf n}}\equiv {\bf N}\setminus {\bf n}$ as
\begin{equation}
{\bf n}=\left\{n_1\,,\,\ldots\,,\,n_{n}\right\}\,,\qquad
{\tilde {\bf n}}=\left\{{\tilde n}_1\,,\,\ldots\,,\,{\tilde n}_{N-n}\right\}\,,
\label{partition}
\end{equation}
where $n\equiv |{\bf n}|$\footnote{Below, we will often refer to these two-cell partitions simply as partitions, since we will never need partitions with more then two cells.} and
\begin{equation}
{\tilde n}\equiv |{\tilde{\bf n}}|=N-n\,.
\label{ntildedef}
\end{equation}
The basis-states are then given by
\begin{equation}
\left|{\bf n}\right>\equiv {\cal P}\left(\bigotimes_{j=1}^n\ket{0}_{n_j}\bigotimes_{k=1}^{{\tilde n}}\left|\updownarrow\right>_{{\tilde n}_k}\right)\,,
\label{rbasis4}
\end{equation}
where ${\cal P}(\cdots)$ orders the states according to their position in the spin-chain. 
An example of a state in this spin-chain is given in figure~(\ref{fig:redchain})

Acting on the basis states with the monodromy matrix gives 
\begin{equation}
  \label{mon1}
  T(u)\left|{\bf n}\right> = \prod_{k\in {\bf n}} I^\perp_{0k} \prod_{l\in {\tilde {\bf n}}} 
  \left(I^{\perp}_{0l}+R_{{\bf 2}\otimes {\bf 2},0l}(u)\right)\left|{\bf n}\right>\,.
\end{equation}
Above, we have used the following identities
\begin{equation}
  \begin{aligned}
    R_{{\bf r}\otimes {\bf r},\,0i}(u)\ket{0}_i & = I^\perp_{0i}\ket{0}_i \,, \\
    \left[ I^\perp_{0i}\,,\, R_{{\bf 2}\otimes {\bf 2},\,0j}(u) \right] &= 0\,,\qquad\mbox{for all } i\mbox{ and } j\,.
\end{aligned}
\end{equation}
Next we note that
\begin{equation}
  \begin{aligned}
    I^\perp_{0i}\ket{0}_i &= \mbox{Id}_0 \ket{0}_i\,, \\
    I^{\perp}_{0i}\left|\updownarrow\right>_i &= I^{\perp\,2}_{0i}\left|\updownarrow\right>_i\,,
  \end{aligned}
\end{equation}
where $\mbox{Id}_0$ is a $3\times 3$ identity matrix acting on the auxiliary space labeled by ${}_0$. This allows us to re-write
equation~(\ref{mon1})  to get
\begin{align}
T(u)\left|{\bf n}\right> &= \prod_{k\in {\bf n}} \mbox{Id}_0 \prod_{l\in {\tilde{\bf n}}}
\left(I^{\perp\,2}_{0l}+R_{{\bf 2}\otimes {\bf 2},\,0l}(u)\right)\left|{\bf n}\right> \\
&=\prod_{l\in {\tilde{\bf n}}}
\left(I^{\perp\,2}_{0l}+R_{{\bf 2}\otimes {\bf 2},\,0l}(u)\right)\left|{\bf n}\right>\,.
\label{mon2}
\end{align}
For ${\bf n}={\bf N}$ we see immediately that the monodromy matrix is just the identity. For all other states we can use the identity
\begin{equation}
I^{\perp\,2}_{0i}R_{{\bf 2}\otimes {\bf 2},\,0j}(u)=0\,,\qquad\mbox{for all } i\mbox{ and } j\,
\end{equation}
and re-write equation~(\ref{mon2}) as
\begin{align}
T(u)\left|{\bf n}\right> &= (1-\delta_{n,N})
\left(\left(\prod_{l\in {\tilde{\bf n}}}
I^{\perp\,2}_{0l}\right) + \left(\prod_{l\in {\tilde{\bf n}}}
R_{{\bf 2}\otimes {\bf 2},\,0l}(u)\right)\right)
\left|{\bf n}\right>
+\delta_{n,N}\left|{\bf n}\right>
\\
&= (1-\delta_{n,N})\left((\mbox{Id}_0-1_2)_0 + \prod_{l\in {\tilde {\bf n}}}
R_{{\bf 2}\otimes {\bf 2},\,0l}(u)\right)\left|{\bf n}\right>
+\delta_{n,N}\left|{\bf n}\right>
\,. \label{mon3}
\end{align}  
For completeness we remind the reader that
\begin{equation}
\mbox{Id}_0-1_2=\left(\begin{array}{ccc} 1 & 0 & 0 \\ 0 & 0 & 0 \\ 0 & 0 & 0 \end{array}\right)\,.
\end{equation}
We note at this point that this monodromy matrix by construction satisfies the Fundamental Commutation Relation (FCR)
\begin{equation}
R_{a_1a_2}(u-v)T_{a_1}(u)T_{a_2}(v)=T_{a_2}(v)T_{a_1}(u)R_{a_1a_2}(u-v)\,,
\end{equation}
where $a_1$ and $a_2$ label two distinct auxiliary spaces. 

We have arrived at a very explicit expression for the monodromy matrix; in particular this expression makes it is easy to take the trace over the three-dimensional auxiliary space to get an explicit expression for the transfer matrix
\begin{equation}
\tau(u)\left|{\bf n}\right>=\tr_0(T(u))\left|{\bf n}\right>=
(1-\delta_{n,N})\tau_{{\bf 2}\otimes {\bf 2}}(u)\left|{\bf n}\right>
+(1+2\delta_{n,N})\left|{\bf n}\right>
\,,
\label{rtr}
\end{equation}
where 
\begin{equation}
\tau_{{\bf 2}\otimes {\bf 2}}(u)\left|{\bf n}\right>\equiv
\mbox{tr}'_0\left(
\prod_{l\in {\tilde {\bf n}}}
R_{{\bf 2}\otimes {\bf 2},\,0l}(u)\right)\left|{\bf n}\right>\,.
\end{equation}
Above, $\mbox{tr}'_0$ denotes the trace over the two-dimensional auxiliary space of the $XXX_{1/2}$ R-matrix.
The transfer matrix in turn is the generating 
object for the main observables of the spin-chain. What remains is to find the most sensible Hamiltonian and shift operator for the system. 

From equation~(\ref{rtr}), we see that $\tau$ does not change the length $|{\bf N}|$ of a state and that it ``preserves the partition ${\bf n}$'', \textit{i.e.}, acts diagonally on each sub-space spanned by $\left|{\bf n}\right>$. Moreover, let us define the linear map
\begin{equation}
S(\left|{\bf n}\right>)\equiv\bigotimes_{k=1}^{{\tilde n}}\left|\updownarrow\right>_{{\tilde n}_k}\,,
\end{equation}
which acts as a surjection from an ${\bf r}^{\otimes N}$ module into a ${\bf 2}^{\otimes {\tilde n}}$ module. It is easy to see that for two states $\left|{\bf n}\right>$ and $\left|{\bf n'}\right>$
of lengths $N$ and $N'$, respectively, which satisfy
\begin{equation}
S(\left|{\bf n}\right>)=S(\left|{\bf n'}\right>)\,,
\end{equation}
we also have
\begin{equation}
\tau_{{\bf N}}(u)\left|{\bf n}\right>=\tau_{{\bf N'}}(u)\left|{\bf n'}\right>\,.
\label{trimp}
\end{equation}
In other words, the charges contained in the transfer matrix are determined exclusively by the ${\tilde{\bf n}}$ parts. Any $\ket{0}$ parts are simply impurities which do not change the charges encoded in the tranfer matrix. The transfer matrix contains all the physical information about the spin-chain as encoded in a set of commuting conserved charges. For a physical interpretation of the reducible spin-chain we would like to  define the notion of a {\em local} operator and find a conserved charge from the transfer matrix which is local with respect to this notion of locality. 

In a conventional $XXX_{1/2}$ chain, for example, one can find the shift operator amongst the operators in the transfer matrix. Recall that the shift operator ${Sh}$ takes a state at site $k$ and maps it to a corresponding state at site $k+1$
\begin{equation}
Sh(\left|v_1\right>_1\left|v_2\right>_2\dots \left|v_n\right>_n)
=\left|v_n\right>_1\left|v_1\right>_2\dots \left|v_{n-1}\right>_n\,,
\label{sh}
\end{equation}
and satisfies $Sh^n=\mbox{Id}_n$.
A {\em local} operator ${\cal O}$ is then defined with respect to $Sh$ as
\begin{equation}
{\cal O}=\sum_{k=0}^{n-1}Sh^k{\cal O}_{12}\,.
\end{equation}
We would like to identify a corresponding notion of locality for the reducible spin-chain with transfer matrix~(\ref{rtr}). For states with no $\ket{0}$ impurities (in other words for states with ${\bf n}=\emptyset$) the notion of locality reduces to the one used in a conventional $XXX_{1/2}$ spin-chain described above. This follows since the tranfer matrix reduces to the $XXX_{1/2}$ transfer matrix for such states (see equation~(\ref{rtr})). Starting with any such state $\left|{\bf N}={\tilde {\bf n}}\right>$, we can construct a new state $\left|{\bf n'}\right>$ by adding some $\ket{0}$ impurities (and so increasing ${\bf N}$ and ${\bf n}$). Equation~(\ref{trimp}) tells us that, no matter where or how many of such impurities we add,  the conserved charges encoded in the transfer matrix will be the same for $\left|{\bf N}={\tilde {\bf n}}\right>$ and $\left|{\bf n'}\right>$. In other words, any operator constructed out of the transfer matrix will act in the same way on these two states. On the state $\left|{\bf N}={\tilde {\bf n}}\right>$ the shift operator $Sh$ 
(cf.\@ equation~(\ref{sh})) defines the notion of locality. As we just saw, for states $\left|{\bf n'}\right>$, there is no way to non-trivially modify $Sh$ through some operator extracted from the transfer matrix, since the transfer matrix does not ``notice'' the $\ket{0}$ impurities.

The argument in the above paragraph leads us to conclude that the only possible notion of locality in the reducible spin-chain comes via the generalized shift operator, which we continue denoting by $Sh$. This  operator moves an excitation at site ${\tilde n}_{k}$ to an excitation at site ${\tilde n}_{k+1}$ (cf.\@ equation~(\ref{partition}))
\begin{equation}
Sh\left(\left|{\bf n}\right>\right)\equiv Sh\left({\cal P}\left(\bigotimes_{j=1}^n\ket{0}_{n_j}\bigotimes_{k=1}^{{\tilde n}}\left|v_{k}\right>_{{\tilde n}_k}\right)\right)={\cal P}\left(\bigotimes_{j=1}^n\ket{0}_{n_j}\bigotimes_{k=1}^{{\tilde n}}\left|v_{k+1}\right>_{{\tilde n}_k}\right)\,,
\label{Shred}
\end{equation}
where $v_k=\uparrow,\downarrow$ at site $k$ and $v_{N-n+1}\equiv v_1$.  In figure~\ref{fig:redchainloc} we show pictorially what is the notion of locality introduced by the operator $Sh$ on the state from figure~\ref{fig:redchain}.

\begin{figure}

\centering
\begin{tikzpicture}
  \coordinate (start) at (-1\sitedist,0);
  \node (n1) at (0\sitedist,0) [odd site] {};
  \node (n2) at (1\sitedist,0) [odd site] {};
  \coordinate (m3) at (1.5\sitedist,0);
  \node (n3) at (1.5\sitedist,0.5\sitedist) [null site] {};
  \node (n4) at (1.5\sitedist,1.25\sitedist) [null site] {};
  \node (n5) at (1.5\sitedist,2.0\sitedist) [null site] {};
  \node (n6) at (2\sitedist,0) [odd site] {};
  \node (n7) at (3\sitedist,0) [odd site] {};
  \node (n8) at (4\sitedist,0) [odd site] {};
  \coordinate (m9) at (4.5\sitedist,0);
  \node (n9) at (4.5\sitedist,0.5\sitedist) [null site] {};
  \node (n10) at (5\sitedist,0) [odd site] {};
  \coordinate (end) at (6\sitedist,0) {};

  \draw [chain line] (start) -- (n1) -- (n2) -- (m3) -- (n6) -- (n7) -- (n8) -- (m9) -- (n10) -- (end);
  \draw [chain line] (m3) -- (n3) -- (n4) -- (n5);
  \draw [chain line] (m9) -- (n9);

  \node at ($(n1)-(0,\labeldist)$) {$\scriptstyle 1$};
  \node at ($(n2)-(0,\labeldist)$) {$\scriptstyle 2$};
  \node at ($(m3)-(0,1.5\labeldist)$) {$\scriptstyle 3$};
  \node at ($(m3)-(0,2.25\labeldist)$) {$\scriptstyle 4$};
  \node at ($(m3)-(0,3.0\labeldist)$) {$\scriptstyle 5$};
  \node at ($(n6)-(0,\labeldist)$) {$\scriptstyle 6$};
  \node at ($(n7)-(0,\labeldist)$) {$\scriptstyle 7$};
  \node at ($(n8)-(0,\labeldist)$) {$\scriptstyle 8$};
  \node at ($(m9)-(0,1.5\labeldist)$) {$\scriptstyle 9$};
  \node at ($(n10)-(0,\labeldist)$) {$\scriptstyle 10$};
\end{tikzpicture}

  \caption{The spin-chain state from figure~\ref{fig:redchain} drawn in a way that reflects the notion of locality as dictated by the integrable structure. The nearest-neighbors are lined up along the horizontal direction. For example,  the nearest neighbors of the site 2 are sites 1 and 6, and the nearest neighbors of the site 7 are sites 6 and 8. The Hamiltonian of the spin-chain is local with respect to this notion of locality. The monodromy matrix $\tau(u)$ acts non-trivially only on the sites on which the non-singlet representations sit, which, in this case are $1,2,6,7,8,10$.}
  \label{fig:redchainloc}

\end{figure}

We can check that the operator $Sh$ can be extracted from the transfer matrix. To this end we note that with some suitable rescaling of the spectral parameter 
\begin{equation} 
v\equiv(iu-2)/4
\end{equation} 
as well as an overall rescaling of the R-matrix by a scalar function of the spectral parameter, the monodromy
matrix is a polynomial of order $N$ in the new spectral parameter $v$. We will expand around $v=0$ 
\begin{equation} 
T(v)=\sum_{k=0}^N v^k T^{(k)}\,,\qquad \tau(v)=\sum_{k=0}^N v^k \tau^{(k)}\,. 
\end{equation} 
It is
then straightforward to show that for a general basis state $\left|{\bf n}\right>$ (with ${\tilde n}\neq 0$) 
\begin{equation}
T^{(0)}\left|{\bf n}\right>=(\mbox{Id}_0-1_2)\left|{\bf n}\right>+ \prod_{k=1}^{{\tilde n}}P_{0{\tilde n}_k}\left|{\bf n}\right>\,, 
\end{equation} 
where
\begin{equation} 
P_{0{\tilde n}_k}=\frac{1}{2}\left(1_2\otimes 1_2+\sigma^i\otimes \sigma^i\right)_{0{\tilde n}_k}\,, 
\end{equation} 
with the
subscript ${}_{0j_k}$ indicating the vector spaces on which the operator acts. From this we find
\begin{equation}
\tau^{(0)}\left|{\bf n}\right>= (1+\prod_{k=1}^{{\tilde n}-1}P_{{\tilde n}_k{\tilde n}_{k+1}})\left|{\bf n}\right>=
(1+Sh)\left|{\bf n}\right>\,,
\label{shfromtr}
\end{equation} 
where we have used the identity
\begin{equation} 
Sh\left|{\bf n}\right>=\prod_{l=1}^{{\tilde n}-1}P_{{\tilde n}_l{\tilde n}_{l+1}}\left|{\bf n}\right>\,.
\end{equation} 
The
next term in the expansion is given by
\begin{equation}
T^{(1)}\left|{\bf n}\right>= -\sum_{l=1}^{{\tilde n}}\prod_{k=1\, \\ \,k\neq l
}^{{\tilde n}}P_{0{\tilde n}_k}\left|{\bf n}\right>\,, 
\end{equation} 
and one can then show that 
\begin{equation}
\tau^{(1)}\left|{\bf n}\right>
=-(\tau^{(0)}-1)\sum_{l=1}^{{\tilde n}-1}P_{{\tilde n}_l{\tilde n}_{l+1}}\left|{\bf n}\right>
=-Sh\left(\sum_{l=1}^{{\tilde n}-1}P_{{\tilde n}_l{\tilde n}_{l+1}}\left|{\bf n}\right>\right)\,. 
\end{equation} 
Higher charges can be worked out analogously. In fact, $Sh$ is invertible
\begin{equation} 
(Sh)^{-1}\left|{\bf n}\right>=\prod_{l={\tilde n}-1}^1P_{{\tilde n}_l{\tilde n}_{l+1}}\left|{\bf n}\right> \,,
\end{equation} 
so we can define the following charge
\begin{equation} 
H\left|{\bf n}\right>\equiv(\tau^{(0)}-1)^{-1}\tau^{(1)}\left|{\bf n}\right>=
-\sum_{l=1}^{{\tilde n}-1}P_{{\tilde n}_l{\tilde n}_{l+1}}\left|{\bf n}\right> \equiv \sum_{l=1}^{{\tilde n}-1}H^{XXX_{1/2}}_{{\tilde n}_l{\tilde n}_{l+1}}\left|{\bf n}\right>\,.
\end{equation} 
As is guaranteed by the FCR, the operator $H$ above commutes with the charges contained in the transfer matrix. The final equality in the above equation shows that $H$ can be thought of as just the $XXX_{1/2}$ Hamiltonian, acting on the ${\bf{\tilde n}}$ part of the spin-chain.

In summary then we use $Sh$ to define the notion of local operators on the ${\bf r}$ spin-chain, and we take $H$ to be the Hamiltonian of this spin-chain. It is easy to check that $H$ is indeed local with respect to $Sh$.

\subsection{A homogeneous \texorpdfstring{$R$}{R} module spin-chain}
\label{sec:homogenous-R-spin-chain}

In the previous subsection we presented in some detail the homogeneous integrable $\alg{su}(2)$ spin-chain with each site in the reducible finite-dimensional ${\bf 1}\oplus {\bf 2}$ representation. In this section we will consider a homogeneous $\alg{sl}(2)$ spin-chain with the reducible infinite-dimensional module $R={\bf 0}\oplus{\bf -1}$ at each site.\footnote{The module ${\bf 0}$ is the trivial one-dimensional module.} The basis-states for a length $N$ spin-chain of this kind can be thought of in almost the same way as that of the ${\bf r}$ spin-chain of the previous subsection. In particular, basis states correspond to two-cell partitions of  the set ${\bf N}=\left\{1,2,\dots,N\right\}$, just as in equations~(\ref{rbasis1})-(\ref{rbasis4}), but now states belonging to ${\tilde{\bf n}}$ are not in the doublet $\left|\updownarrow\right>$, but rather in the ${\bf s}={\bf -1}$ infinite dimensional representation.

We remind the reader that the representation $R\otimes R$ can be decomposed into irreducible modules as in equation~(\ref{RRirep}). As a result, we can take the R-matrix to be
\begin{equation}
R_{R\otimes R}(u)=\Pi_{{\bf 0}} + \Pi_{{\bf -1}_1} + \Pi_{{\bf -1}_2} + \sum_{j=2}^{\infty} r_{j}(u)\Pi_{-j}\,,
\end{equation}
with $\Pi_{-j}$ the projector onto the $\mathbf{s} = -j$ representation on the right-hand side of equation~(\ref{RRirep}) and 
\begin{equation}
r_{j}(u)=\prod_{k=0}^{j-1}\frac{u+2k}{u-2k}\,.
\end{equation}
This expression is in agreement with the general expression derived in equation~(\ref{mssv}) upon setting $u = u_1 - u_2$ and $j=\abs{s}$.
Explicit expressions for the above projection operators are given in appendix~\ref{sec:sl2-projectors}.
Just as in the previous subsection we can write the R-matrix as
\begin{equation}
R_{R\otimes R}(u) \equiv I^\perp + R_{{\bf -1}\otimes {\bf -1}}(u)\,,
\end{equation}
where
\begin{equation}
I^\perp\equiv\Pi_{{\bf 0}} +  \Pi_{{\bf -1}_1} + \Pi_{{\bf -1}_2}\,,
\end{equation}
and $R_{{\bf -1}\otimes {\bf -1}}(u)$ is the conventional R-matrix for the $XXX_{-1}$ spin-chain ``augmented by some zeros'' so it acts on the full module $R$ and not just on the ${\bf -1}\otimes {\bf -1}$ sub-module; this is in analogy with equation~(\ref{Rmatrrsum}). 

Acting on the basis states 
with the monodromy matrix gives 
\begin{equation}
\label{mon1-R}
T(u)\left|{\bf n}\right> = \prod_{k\in {\bf n}} I^\perp_{0k} \prod_{l\in {\tilde {\bf n}}} 
\left(I^{\perp}_{0l}+R_{{\bf -1}\otimes {\bf -1},0l}(u)\right)\left|{\bf n}\right>\,.
\end{equation}
Above, we have used the following identities
\begin{equation}
  \begin{aligned}
    R_{R\otimes R,\,0i}(u)\ket{0}_i & = I^\perp_{0i}\ket{0}_i \,, \\
    \left[ I^\perp_{0i}\,,\, R_{{\bf -1}\otimes {\bf -1},\,0j}(u) \right]&= 0\,,\qquad\mbox{for all } i\mbox{ and } j\,.
\end{aligned}
\end{equation}
It is easy to check that these identities hold, using the explicit expressions for the projectors $\Pi_{{\bf s}}$ given in appendix~\ref{sec:sl2-projectors}. Using these expressions we note also that
\begin{equation}
  \begin{aligned}
    I^\perp_{0i}\ket{0}_i &= \mbox{Id}_0 \ket{0}_i\,, \\
    I^{\perp}_{0i}\left|v\right>_i &= (\Pi_{{\bf 0}}+\Pi_{{\bf -1}_2}) \left|v\right>_i\,,
  \end{aligned}
\end{equation}
where $\mbox{Id}_0$ is the identity matrix acting on the auxiliary space, and $\left|v\right>_i$ is any state in the ${\bf -1}$ sub-module of the $R$ module at site $i$. This allows us to re-write
equation~(\ref{mon1-R})  to get
\begin{align}
T(u)\left|{\bf n}\right> &= \prod_{k\in {\bf n}} \mbox{Id}_0 \prod_{l\in {\tilde{\bf n}}}
\left(\Pi_{{\bf 0}}+\Pi_{{\bf -1}_2}+R_{{\bf -1}\otimes {\bf -1},\,0l}(u)\right)\left|{\bf n}\right> \\
&= \prod_{l\in {\tilde{\bf n}}}
\left(\Pi_{{\bf 0}}+\Pi_{{\bf -1}_2}+R_{{\bf -1}\otimes {\bf -1},\,0l}(u)\right)\left|{\bf n}\right>\,.
\label{mon2-R}
\end{align}
For ${\bf n}={\bf N}$ we see immediately that 
\begin{equation}
T(u)\left|{\bf n}={\bf N}\right>=\left|{\bf n}={\bf N}\right>\,,
\end{equation}
For all other states ${\bf n}\neq{\bf N}$ we can use the identity
\begin{equation}
(\Pi_{{\bf 0}}+\Pi_{{\bf -1}_2})_{0i}R_{{\bf -1}\otimes {\bf -1},\,0j}(u)=0\,,\qquad\mbox{for all } i\mbox{ and } j
\end{equation}
to simplify the monodormy matrix further. Combining the two parts we obtain the following expression for the monodromy matrix
\begin{align}
T(u)\left|{\bf n}\right> &= (1-\delta_{n,N})
\left(\left(\prod_{l\in {\tilde{\bf n}}}
\Pi_{{\bf 0}}+\Pi_{{\bf -1}_2}\right) + \left(\prod_{l\in {\tilde{\bf n}}}
R_{{\bf -1}\otimes {\bf -1},\,0l}(u)\right)\right)
\left|{\bf n}\right>
+\delta_{n,N}\left|{\bf n}\right>
\\
&= (1-\delta_{n,N})\left(\ket{0}_{0}\left<0\right|_0 + \prod_{l\in {\tilde {\bf n}}}
R_{{\bf -1}\otimes {\bf -1},\,0l}(u)\right)\left|{\bf n}\right>
+\delta_{n,N}\left|{\bf n}\right>
\,.
\label{monR3}
\end{align}  
In the above it is useful to recall that
\begin{equation}
\Pi_{{\bf 0},0k}+\Pi_{{\bf -1}_2,0k}=\ket{0}_{0}\left<0\right|_0 \,,
\end{equation}
for any site $k$. This identity can be explicitly verified using the projector expressions given in appendix~\ref{sec:sl2-projectors}. We can now trace over the auxiliary space to get an explicit expression for the transfer matrix
\begin{equation}
\tau(u)\left|{\bf n}\right>=\tr_0(T(u))\left|{\bf n}\right>=
(1-\delta_{n,N})\tau_{{\bf -1}\otimes {\bf -1}}(u)\left|{\bf n}\right>
+(1+(Z-1)\delta_{n,N})\left|{\bf n}\right>
\,,
\label{Rtr}
\end{equation}
where 
\begin{equation}
\tau_{{\bf -1}\otimes {\bf -1}}(u)\left|{\bf n}\right>\equiv
\mbox{tr}'_0\left(
\prod_{l\in {\tilde {\bf n}}}
R_{{\bf -1}\otimes {\bf -1},\,0l}(u)\right)\left|{\bf n}\right>\,,
\end{equation}
and $Z$ is a $u$-independent (infinite) constant
\begin{equation}
Z=\tr_R(\mbox{Id})=\sum_{n=0}^\infty 1\,.
\end{equation}
Above, $\mbox{tr}'_0$ denotes the trace over the auxiliary space of a conventional $XXX_{-1}$ R-matrix.
As in the previous subsection, we see that the transfer matrix acts identically on two states which differ from one another by an addition/removal of $\left| 0\right>$ impurities (cf.\@ equation~(\ref{trimp})). This in turn implies that the only notion of a generalized shift operator is given by the same formal expression as the one in equation~(\ref{Shred}), but with $\left|v_k\right>_{{\tilde n}_k}$ now belonging to the ${\bf -1}$ submodule.
We can perform a similar analysis to the one in the previous subsection to show that the generalized shift operator $Sh$ can be extracted from the transfer matrix (cf.\@ equation~(\ref{shfromtr})). We can also extract from the transfer matrix a local conserved charge which we will call the Hamiltonian of the system
\begin{equation} 
H\left|{\bf n}\right> =\sum_{l=1}^{{\tilde n}-1}H^{XXX_{-1}}_{{\tilde n}_l{\tilde n}_{l+1}}\left|{\bf n}\right>
\,,
\end{equation}
where
\begin{equation}
H^{XXX_{-1}}_{{\tilde n}_l{\tilde n}_{l+1}}=2\sum_{j=-2}^{-\infty}h(1-j) (\Pi_{-j})_{{\tilde n}_l{\tilde n}_{l+1}}\,,
\end{equation} 
and $h(j)$ is the $j$-th harmonic number.
This Hamiltonian acts only on the ${\bf{\tilde n}}$ part of the spin-chain through the action of a conventional $XXX_{-1}$ spin-chain Hamiltonian, and is, by construction, local with respect to the generalized shift operator $Sh$.

\subsection{The alternating \texorpdfstring{$\mathbf{r}\otimes \mathbf{2}$}{r x 2} spin-chain}
\label{sec:alternating-toy-model}

\begin{figure}

\centering
\begin{tikzpicture}
  \coordinate (start) at (-1\sitedist,0);
  \node (n1) at (0\sitedist,0) [odd site] {};
  \node (n2) at (1\sitedist,0) [odd site] {};
  \node (n3) at (2\sitedist,0) [odd site] {};
  \node (n4) at (3\sitedist,0) [null site] {};
  \node (n5) at (4\sitedist,0) [odd site] {};
  \node (n6) at (5\sitedist,0) [null site] {};
  \node (n7) at (6\sitedist,0) [odd site] {};
  \node (n8) at (7\sitedist,0) [odd site] {};
  \node (n9) at (8\sitedist,0) [odd site] {};
  \node (n10) at (9\sitedist,0) [null site] {};
  \coordinate (end) at (10\sitedist,0) {};

  \draw [chain line] (start) -- (n1) -- (n2) -- (n3) -- (n4) -- (n5) -- (n6) -- (n7) -- (n8) -- (n9) -- (n10) -- (end);

  \node at ($(n1)-(0,\labeldist)$) {$\scriptstyle 1$};
  \node at ($(n2)-(0,\labeldist)$) {$\scriptstyle 2$};
  \node at ($(n3)-(0,\labeldist)$) {$\scriptstyle 3$};
  \node at ($(n4)-(0,\labeldist)$) {$\scriptstyle 4$};
  \node at ($(n5)-(0,\labeldist)$) {$\scriptstyle 5$};
  \node at ($(n6)-(0,\labeldist)$) {$\scriptstyle 6$};
  \node at ($(n7)-(0,\labeldist)$) {$\scriptstyle 7$};
  \node at ($(n8)-(0,\labeldist)$) {$\scriptstyle 8$};
  \node at ($(n9)-(0,\labeldist)$) {$\scriptstyle 9$};
  \node at ($(n10)-(0,\labeldist)$) {$\scriptstyle 10$};
\end{tikzpicture}
  
  \caption{An example of a state in a reducible alternating spin-chain of the type discussed in sections~\ref{sec:alternating-toy-model} and~\ref{sec:alternating-R-spin-chain}. For such spin-chains the odd sites are in the ${\bf r}$ or $ R$ representations respectively (denoted in the figure by a square). The even sites are either singlets (denoted by a dot) or in the  ${\bf r}$ (respectively  $R$) representation.
 In the notation introduced in equations~(\ref{partition}) and~(\ref{ntildedef}) this state has ${\bf N}=\{1,2,3,4,5\}$, ${\bf n}=\{2,3,5\}$ and ${\bf {\tilde n}}=\{1,4\}$ as well as $N=5$, $n=3$ and ${\tilde n}=1$. 
 In the notation introduced in equations~(\ref{eqMdef}) and~(\ref{eqmdef}) this state has ${\bf M}=\{1,2,3,4,5,6,7,8,9,10\}$, ${\bf n}=\{4,6,10\}$ and ${\bf {\tilde n}}=\{1,2,3,5,7,8,9\}$ as well as $M=10$, $m=3$ and ${\tilde m}=7$. 
The states are drawn according to the ordering given by the site number.
}
  \label{fig:redaltchain}

\end{figure}
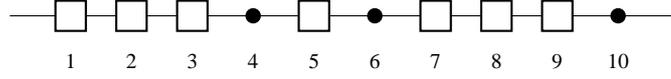

In this subsection we consider an alternating $\alg{su}(2)$ spin-chain, where the odd/even sites are in the ${\bf 2}$/${\bf r}$ representation, respectively. The general procedure we follow is outlined in~\cite{Minahan:2008hf}. We will denote the even (odd) sites by an un-bared (bared) index ${}_a$ (${}_{{\bar a}}$), respectively. A convenient basis of spin-chain states of length $2N$ can be constructed using two-cell partitions of ${\bf N}=\left\{1,2,\dots,N\right\}$ just as in section~\ref{sec:homogenous-toy-model}, but now replacing each $\left| 0\right>$ or $\left|\updownarrow\right>$ state with 
$\left|\updownarrow\right>\otimes\left| 0\right>$ or $\left|\updownarrow\right>\otimes\left|\updownarrow\right>$. In particular we will have
\begin{equation}
({\bf 2}\otimes {\bf r})^N=\mbox{span }
\bigcup_{
{\bf n}\subset {\bf N} } \{\left|{\bf n}\right>\}\,,
\label{altb1}
\end{equation}
where now 
\begin{equation}
\left|{\bf n}\right>\equiv {\cal P}\left(\bigotimes_{j=1}^n\left(\left|\updownarrow\right>_{n_j}\otimes\ket{0}_{n_j}\right)\bigotimes_{k=1}^{{\tilde n}}\left(
\left|\updownarrow\right>_{{\tilde n}_k}\otimes
\left|\updownarrow\right>_{{\tilde n}_k}\right)\right)\,.
\label{altb2}
\end{equation}
There is also a second way to parametrize the states $\left|{\bf n}\right>$ which will also be useful below. If we define 
\begin{equation}
{\bf M}\equiv{\bf 2N}=\left\{1,2,\dots,2N\right\}\,,
\label{eqMdef}
\end{equation}
and the two cell partition of ${\bf M}$ as
\begin{equation}
{\bf m}=\left\{2n_1,2n_2,\dots,2n_n\right\}\,,\qquad
{\bf {\tilde m}}\equiv {\bf M}\setminus{\bf m}\,,
\label{eqmdef}
\end{equation}
then the state $\left|{\bf n}\right>$ can also be thought of as
\begin{equation}
\left|{\bf n}\right>\equiv 
\left|{\bf m}\right>\equiv {\cal P}\left(\bigotimes_{j=1}^m\ket{0}_{m_j}\bigotimes_{k=1}^{{\tilde m}}\left|\updownarrow\right>_{{\tilde m}_k}\right)\,,
\label{Raltbasis}
\end{equation}
where ${\cal P}(\cdots)$ orders the states according to their position in the spin-chain as counted by ${\bf M}$, $m=|{\bf m}|$
and ${\tilde m}=|{\bf {\tilde m}}|$. Above, the $m_j$ denote the elements of ${\bf m}$, and the ${\tilde m}_k$ the elements of ${\bf {\tilde m}}$. This alternate basis just reflects the fact that a generic basis state $\left|{\bf n}\right>$ in the alternating chain consists of a particular ordering of $n$ singlet states $\ket{0}$ and $2N-n$ states in the ${\bf 2}$ representation. Hence, such a basis state can be thought of as a length $2N$ basis state of the homogeneous ${\bf r}$ spin chain discussed in section~\ref{sec:homogenous-toy-model}. An example of a state in this spin-chain is shown in figure~\ref{fig:redaltchain}.

Let us turn to the integrable structure of this alternating spin chain.
 In such a model there are three R-matrices: $R_{{\bf r}\otimes{\bf r}}$ (given in equation~(\ref{RRRmatasproj})), the conventional $XXX_{1/2}$ R-matrix $R_{{\bf 2}\otimes{\bf 2}}$ and a further R-matrix
\begin{equation}
R_{{\bf r}\otimes{\bf 2}}(u)=\Pi_2\oplus R_{{\bf 2}\otimes{\bf 2}}(u)\,.
\end{equation}
Above, $R_{{\bf 2}\otimes{\bf 2}}(u)$ is the conventional $XXX_{1/2}$ R-matrix augmented by some zeros so that it now acts on ${\bf r}\otimes {\bf 2}$. The R-matrices now satisfy a number of YBE equations: in addition to~(\ref{RYBE}), and the $XXX_{1/2}$ YBE for the $4\times 4$ $R_{{\bf 2}\otimes{\bf 2}}(u)$, they also satisfy
\begin{align}
R_{{\bar 1}{\bar 2}}(u-v)R_{{\bar 1}3}(u)
R_{{\bar 2}3}(v)
&=
R_{{\bar 2}3}(v)
R_{{\bar 1}3}(u)
R_{{\bar 1}{\bar 2}}(u-v)\,,
\label{altYBE1}\\
R_{12}(u-v)R_{1{\bar 3}}(u)
R_{2{\bar 3}}(v)
&=
R_{2{\bar 3}}(v)
R_{1{\bar 3}}(u)
R_{12}(u-v)\,,
\label{altYBE2}\\
R_{1{\bar 2}}(u-v)R_{13}(u)
R_{{\bar 2} 3}(v)
&=
R_{{\bar 2} 3}(v)
R_{13}(u)
R_{1{\bar 2}}(u-v)\,,
\label{altYBE3}\\
R_{{\bar 1}2}(u-v)R_{{\bar 1}{\bar 3}}(u)
R_{2{\bar 3}}(v)
&=
R_{2{\bar 3}}(v)
R_{{\bar 1}{\bar 3}}(u)
R_{{\bar 1}2}(u-v)\,.
\label{altYBE4}
\end{align}
We may define two monodromy matrices
\begin{equation}
  \begin{aligned}
T_a(u) &= C R_{a1}(u)R_{a{\bar 1}}(u)R_{a2}(u)R_{a{\bar 2}}(u)\dots R_{aN}(u)R_{a{\bar N}}(u) \\
T_{{\bar a}}(u) &= C R_{{\bar a}1}(u)R_{{\bar a}{\bar 1}}(u)R_{{\bar a}2}(u)R_{{\bar a}{\bar 2}}(u)\dots R_{{\bar a}N}(u)R_{{\bar a}{\bar N}}(u)\,,
\end{aligned}
\end{equation}
where ${}_a$ and ${}_{{\bar a}}$ denote auxiliary spaces in the ${\bf 2}$ and ${\bf r}$ representations; $\alpha$ and $C$ are constants. From the above YBE relations the following fundamental commutation relations (FCRs) follow
\begin{equation}
  \begin{aligned}
    R_{ab}(u-v)T_a(u)T_b(v) &= T_b(v)T_a(u)R_{ab}(u-v)
    \\
    R_{{\bar a}{\bar b}}(u-v)T_{{\bar a}}(u)T_{{\bar b}}(v) &= T_{{\bar b}}(v)T_{{\bar a}}(u)R_{{\bar a}{\bar b}}(u-v)\,.
  \end{aligned}
\end{equation}
As a result, if we define the transfer matrices
\begin{equation}
\tau(u)=\tr_a T_a(u)\,,\qquad
{\bar \tau}(u)=\tr_{{\bar a}} T_{{\bar a}}(u)\,,
\end{equation}
they commute amongst themselves for different values of the spectral parameter
\begin{equation}
\left[\tau(u)\,,\,\tau(v)\right]=0
\,,\qquad
\left[{\bar \tau}(u)\,,\,{\bar \tau}(v)\right]=0
\end{equation}
To show that  $\tau(u)$ and ${\bar\tau}(u)$ also commute, 
\begin{equation}
\left[\tau(u)\,,\,{\bar\tau}(v)\right]=0\,,
\end{equation}
we need a further FCR 
\begin{equation}
R_{a{\bar b}}(u-v)T_a(u)T_{{\bar b}}(v)
=T_{{\bar b}}(v)T_a(u)R_{a{\bar b}}(u-v)\,,
\end{equation}
to hold. In fact, such an FCR can hold as long as the Yang-Baxter equations~\eqref{altYBE3} and~\eqref{altYBE4} are satisfied.

\begin{figure}

\centering
\begin{tikzpicture}
  \coordinate (start) at (-1\sitedist,0);
  \node (n1) at (0\sitedist,0) [odd site] {};
  \node (n2) at (1\sitedist,0) [odd site] {};
  \node (n3) at (2\sitedist,0) [odd site] {};
  \coordinate (m4) at (2.5\sitedist,0);
  \node (n4) at (2.5\sitedist,0.5\sitedist) [null site] {};
  \node (n5) at (3\sitedist,0) [odd site] {};
  \coordinate (m6) at (3.5\sitedist,0);
  \node (n6) at (3.5\sitedist,0.5\sitedist) [null site] {};
  \node (n7) at (4\sitedist,0) [odd site] {};
  \node (n8) at (5\sitedist,0) [odd site] {};
  \node (n9) at (6\sitedist,0) [odd site] {};
  \coordinate (m10) at (6.5\sitedist,0);
  \node (n10) at (6.5\sitedist,0.5\sitedist) [null site] {};
  \coordinate (end) at (7\sitedist,0) {};

  \draw [chain line] (start) -- (n1) -- (n2) -- (n3) -- (m4) -- (n5) -- (m6) -- (n7) -- (n8) -- (n9) -- (m10) -- (end);
  \draw [chain line] (m4) -- (n4);
  \draw [chain line] (m6) -- (n6);
  \draw [chain line] (m10) -- (n10);

  \node at ($(n1)-(0,\labeldist)$) {$\scriptstyle 1$};
  \node at ($(n2)-(0,\labeldist)$) {$\scriptstyle 2$};
  \node at ($(n3)-(0,\labeldist)$) {$\scriptstyle 3$};
  \node at ($(m4)-(0,1.5\labeldist)$) {$\scriptstyle 4$};
  \node at ($(n5)-(0,\labeldist)$) {$\scriptstyle 5$};
  \node at ($(m6)-(0,1.5\labeldist)$) {$\scriptstyle 6$};
  \node at ($(n7)-(0,\labeldist)$) {$\scriptstyle 7$};
  \node at ($(n8)-(0,\labeldist)$) {$\scriptstyle 8$};
  \node at ($(n9)-(0,\labeldist)$) {$\scriptstyle 9$};
  \node at ($(m10)-(0,1.5\labeldist)$) {$\scriptstyle 10$};
\end{tikzpicture}

  \caption{The reducible alternating spin-chain state from figure~\ref{fig:redaltchain} drawn in a way that reflects locality as dictated by the integrable structure. The nearest-neighbors are lined up along the horizontal direction. For example,  the nearest neighbors of the site 2 are sites 1 and 3, and the nearest neighbors of the site 5 are sites 3 and 7. The Hamiltonian of the spin-chain is local with respect to this notion of locality. The monodromy matrix $\tau(u)$ acts non-trivially only on the sites on which the non-singlet representations sit, which, in this case are $1,2,3,5,7,8,9$. Notice that, unlike the homogeneous spin-chain, there can be at most one defect between two local sites.
}
  \label{fig:redaltchainloc}
\end{figure}
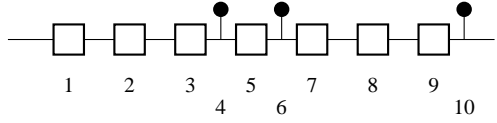

We can now expand the transfer matrices to extract the generalized shift operator $Sh$. In fact it is easiest to write down this operator explicitly in the $\left|{\bf m}\right>$ basis. We find 
\begin{equation}
  \begin{aligned}
    \tau^{(0)}\left|{\bf m}\right>&= \prod_{k=1}^{{\tilde m}-1}P_{{\tilde m}_k{\tilde m}_{k+1}}\left|{\bf m}\right>\equiv Sh\left|{\bf m}\right>\,,
    \\
    {\bar \tau}^{(0)}\left|{\bf m}\right>&= (1+\prod_{k=1}^{{\tilde m}-1}P_{{\tilde m}_k{\tilde m}_{k+1}})\left|{\bf m}\right>= (1+Sh)\left|{\bf m}\right>\,.
    \label{altshfromtr}
  \end{aligned}
\end{equation} 
The $Sh$ operator defines the notion of locality for the alternating chain (see corresponding discussion in section~\ref{sec:homogenous-toy-model}). In figure~\ref{fig:redaltchainloc} we draw the state depicted in figure~\ref{fig:redaltchain} in a way  that reflects locality induced by $Sh$.
In the $\left|{\bf m}\right>$ basis one can extract an operator from the transfer matrices which is local with respect to the above $Sh$ operator - this will be our Hamiltonian. Its explicit form is
\begin{equation} 
H\left|{\bf m}\right>\equiv
-\sum_{l=1}^{{\tilde m}-1}P_{{\tilde m}_l{\tilde m}_{l+1}}\left|{\bf m}\right> =
\sum_{l=1}^{{\tilde m}-1}H^{XXX_{1/2}}_{{\tilde m}_l{\tilde m}_{l+1}}\left|{\bf m}\right>
\,.
\end{equation} 

In the alternating spin-chain in~\cite{OhlssonSax:2011ms}, the a Hamiltonian similar to the above was constructed as
\begin{equation}
  H = \left.\frac{\mathrm{d}}{\mathrm{d}\,u} \log(\tau(u) \bar{\tau}(u))\right|_{u=0}
  = (\tau(0) \bar{\tau}(0))^{-1} (\tau'(0) \bar{\tau}(0) + \tau(0) \bar{\tau}'(0)) \,.
\end{equation}
We can repeat the same construction in the reducible spin-chain, provided we shift $\bar{\tau}(0)$ by $-1$ to obtain the invertible shift operator $Sh$. The discussion above shows that both $\tau(u)$ and $\bar{\tau}(u)$ provide transfer matrices that effectively act on the $\mathbf{s}=-1$ part of the spin-chain. Moreover, these transfer matrices commute. Hence the resulting Hamiltonian simply consists of two copies of $H^{XXX_{1/2}}$ from above.

\subsection{The alternating \texorpdfstring{$R$}{R} module spin-chain}
\label{sec:alternating-R-spin-chain}

In this subsection we consider an alternating integrable $\alg{sl}(2)$ spin-chain, where the odd sites are in a conventional ${\bf -1}$ representation and the even sites are an $R$ module. Having gained some experience with the construction of local quantities for reducible spin-chains in the last three subsections, we can be quite brief here. In particular the construction is analogous to the one in the previous subsection, upon replacing any ${\bf 2}$ representation of $\alg{su}(2)$ with a ${\bf -1}$ representation of $\alg{sl}(2)$. For example, a basis for this spin-chain can be constructed 
\begin{equation}
({\bf 2}\otimes {\bf R})^N=\mbox{span }
\bigcup_{
{\bf n}\subset {\bf N} } \{\left|{\bf n}\right>\}\,,
\end{equation}
where now 
\begin{equation}
\left|{\bf n}\right>\equiv {\cal P}\left(\bigotimes_{j=1}^n\left(\left|v\right>_{n_j}\otimes\ket{0}_{n_j}\right)\bigotimes_{k=1}^{{\tilde n}}\left(
\left|v\right>_{{\tilde n}_k}\otimes
\left|v\right>_{{\tilde n}_k}\right)\right)\,,
\end{equation}
where $\left|v\right>$ is a generic state in the ${\bf -1}$ representation. As above, there is a second way to parametrize the states $\left|{\bf n}\right>$. Defining ${\bf M}$, ${\bf m}$, ${\bf {\tilde m}}$, $m_j$ and ${\tilde m}_k$ as in the previous subsection, the states $\left|{\bf n}\right>$ can be re-expressed as
\begin{equation}
\left|{\bf n}\right>\equiv 
\left|{\bf m}\right>\equiv {\cal P}\left(\bigotimes_{j=1}^m\ket{0}_{m_j}\bigotimes_{k=1}^{{\tilde m}}\left|v\right>_{{\tilde m}_k}\right)\,.
\end{equation}
Using the expressions for projection operators in appendix~\ref{sec:sl2-projectors}, we can construct the transfer and monodromy matrices for this alternating chain in a manner similar to the previous subsection. In the end we find that the generalized shift operator is 
\begin{equation}
Sh\left|{\bf m}\right>\equiv
 \prod_{k=1}^{{\tilde m}-1}P_{{\tilde m}_k{\tilde m}_{k+1}}\left|{\bf m}\right>
\end{equation}
and the Hamiltonian, local with respect to $Sh$, is
\begin{equation} 
H\left|{\bf m}\right>\equiv
\sum_{l=1}^{{\tilde n}-1}H^{XXX_{-1}}_{{\tilde m}_l{\tilde m}_{l+1}}\left|{\bf m}\right>
\,,
\end{equation} 
where $H^{XXX_{-1}}_{{\tilde m}_l{\tilde m}_{l+1}}$ is the conventional nearest-neighbor Hamiltonian of an $XXX_{-1}$ spin-chain
\begin{equation}
H^{XXX_{-1}}_{{\tilde m}_l{\tilde m}_{l+1}}=2\sum_{j=-2}^{-\infty}h(1-j) (\Pi_{-j})_{{\tilde m}_l{\tilde m}_{l+1}}\,.
\end{equation}

\subsection{The \texorpdfstring{$\alg{d}(2,1;\alpha=0)$}{d(2,1;a)} spin-chain}
\label{sec76}

The generalization of the above to the alternating  $\alg{d}(2,1;\alpha=0)=\alg{psu}(1,1|2)$ chain is straightforward. The odd sites are in a $(-\frac{1}{2};\frac{1}{2})$ irreducible short representation, and the even sites are in a $0^{\oplus 2}\oplus(-\frac{1}{2};\frac{1}{2})$ representation. Compared with the alternating $\alg{sl}(2)$ spin-chain 
discussed in the previous subsection, the $\alg{d}(2,1;\alpha=0)$ chain has two main differences. Firstly, at the even sites, where the reducible representations sit, there are now {\em two} singlet states. Secondly, the highest weight states of the $(-\frac{1}{2};\frac{1}{2})$ modules at the even/odd sites are fermions/bosons, respectively. The fermionic statistics of the even site groundstates is a consequence of these states being fermionic descendants in the $(-\frac{\alpha}{2};\frac{1}{2};0)$ irreducible module at $\alpha\neq 0$.
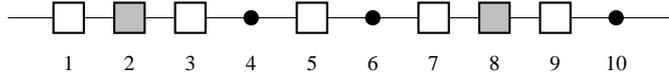
\begin{figure}

\centering
\begin{tikzpicture}
  \coordinate (start) at (-1\sitedist,0);
  \node (n1) at (0\sitedist,0) [odd site] {};
  \node (n2) at (1\sitedist,0) [even site] {};
  \node (n3) at (2\sitedist,0) [odd site] {};
  \node (n4) at (3\sitedist,0) [null site] {};
  \node (n5) at (4\sitedist,0) [odd site] {};
  \node (n6) at (5\sitedist,0) [null site] {};
  \node (n7) at (6\sitedist,0) [odd site] {};
  \node (n8) at (7\sitedist,0) [even site] {};
  \node (n9) at (8\sitedist,0) [odd site] {};
  \node (n10) at (9\sitedist,0) [null site] {};
  \coordinate (end) at (10\sitedist,0) {};

  \draw [chain line] (start) -- (n1) -- (n2) -- (n3) -- (n4) -- (n5) -- (n6) -- (n7) -- (n8) -- (n9) -- (n10) -- (end);

  \node at ($(n1)-(0,\labeldist)$) {$\scriptstyle 1$};
  \node at ($(n2)-(0,\labeldist)$) {$\scriptstyle 2$};
  \node at ($(n3)-(0,\labeldist)$) {$\scriptstyle 3$};
  \node at ($(n4)-(0,\labeldist)$) {$\scriptstyle 4$};
  \node at ($(n5)-(0,\labeldist)$) {$\scriptstyle 5$};
  \node at ($(n6)-(0,\labeldist)$) {$\scriptstyle 6$};
  \node at ($(n7)-(0,\labeldist)$) {$\scriptstyle 7$};
  \node at ($(n8)-(0,\labeldist)$) {$\scriptstyle 8$};
  \node at ($(n9)-(0,\labeldist)$) {$\scriptstyle 9$};
  \node at ($(n10)-(0,\labeldist)$) {$\scriptstyle 10$};
\end{tikzpicture}
  
  \caption{An example of a state in a reducible alternating spin-chain discussed in section~\ref{sec76}. The odd sites are in an irreducible representation of the super-group, with a bosonic highest weight state. These are denoted by a white square. The even sites can either be in a singlet representation (denoted by a dot), or in an irreducible representation of the same type as the odd sites, but with opposite boson/fermion grading (denoted by a grey box). In a $\alg{d}(2,1;\alpha=0)$ spin-chain, there are two singlets, one bosonic and one fermionic, we don't distinguish between these in the diagram. In the $\alg{sl}(2|1)$ sub-chain there is only one bosonic singlet state.}

  \label{fig:redaltsuchain}

\end{figure}

Modulo these two differences, the alternating  $\alg{d}(2,1;\alpha=0)$ spin-chain is quite similar in form to the alternating spin-chain of the previous subsection. It is still useful to define the basis elements $\left|{\bf m}\right>$
\begin{equation}
\left|{\bf m}\right>\equiv {\cal P}\left(\bigotimes_{j=1}^m\left|0_{m_j}\right>_{m_j}\bigotimes_{k=1}^{{\tilde m}}\left|v\right>_{{\tilde m}_k}\right)\,,
\end{equation}
where now $\left| 0_{m_j}\right>$ can be either of the two possible singlets.
$\left|v\right>$ denotes a generic state in a $(-\frac{1}{2};\frac{1}{2})$ module, 
taking into account the bosonic/fermionic nature of the groundstate. The generalized shift operator is 
\begin{equation}
Sh\left|{\bf m}\right>\equiv
 \prod_{k=1}^{{\tilde m}-1}{\hat P}_{{\tilde m}_k{\tilde m}_{k+1}}\left|{\bf m}\right>\,,
\end{equation}
where ${\hat P}$ is the graded permutation operator. The Hamiltonian, local with respect to $Sh$, is
\begin{equation} 
H\left|{\bf m}\right>\equiv
\sum_{l=1}^{{\tilde m}-1}{\hat H}^{(-1/2;1/2)}_{{\tilde m}_l{\tilde m}_{l+1}}\left|{\bf m}\right>
\,,
\label{halgred}
\end{equation} 
where $H^{(-1/2;1/2)}_{{\tilde m}_l{\tilde m}_{l+1}}$ is the conventional nearest-neighbor Hamiltonian of an $\alg{psu}(1,1|2)$ homogeneous spin-chain with $(-\frac{1}{2};\frac{1}{2})$ irreps at each site~\cite{Beisert:2003jj}. The ``hat'' on ${\hat H}$ indicates the extra grading required due to the switched statistics of the even-site irreps.  We can write this Hamiltonian as a sum of projectors by lifting the $\alg{sl}(2|1)$ R-matrix discussed in sections~\ref{surmat} and~\ref{altern}. The result reads
\begin{equation}
  \hat{H}^{(-1/2;1/2)}_{\tilde{m}_l \tilde{m}_{l+1}} = \sum_{j=0}^{\infty} 2 h(j) \hat{\Pi}^{j}_{\tilde{m} \tilde{m}+1} \,.
\end{equation}
Note that this Hamiltonian acts trivially on any site containing a singlet as well as on the groundstate of the $(-1/2;1/2)$ sub spin-chain.

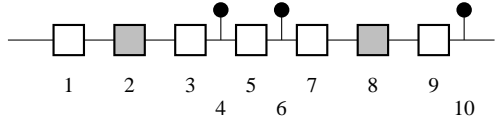
\begin{figure}

\centering
\begin{tikzpicture}
  \coordinate (start) at (-1\sitedist,0);
  \node (n1) at (0\sitedist,0) [odd site] {};
  \node (n2) at (1\sitedist,0) [even site] {};
  \node (n3) at (2\sitedist,0) [odd site] {};
  \coordinate (m4) at (2.5\sitedist,0);
  \node (n4) at (2.5\sitedist,0.5\sitedist) [null site] {};
  \node (n5) at (3\sitedist,0) [odd site] {};
  \coordinate (m6) at (3.5\sitedist,0);
  \node (n6) at (3.5\sitedist,0.5\sitedist) [null site] {};
  \node (n7) at (4\sitedist,0) [odd site] {};
  \node (n8) at (5\sitedist,0) [even site] {};
  \node (n9) at (6\sitedist,0) [odd site] {};
  \coordinate (m10) at (6.5\sitedist,0);
  \node (n10) at (6.5\sitedist,0.5\sitedist) [null site] {};
  \coordinate (end) at (7\sitedist,0) {};

  \draw [chain line] (start) -- (n1) -- (n2) -- (n3) -- (m4) -- (n5) -- (m6) -- (n7) -- (n8) -- (n9) -- (m10) -- (end);
  \draw [chain line] (m4) -- (n4);
  \draw [chain line] (m6) -- (n6);
  \draw [chain line] (m10) -- (n10);

  \node at ($(n1)-(0,\labeldist)$) {$\scriptstyle 1$};
  \node at ($(n2)-(0,\labeldist)$) {$\scriptstyle 2$};
  \node at ($(n3)-(0,\labeldist)$) {$\scriptstyle 3$};
  \node at ($(m4)-(0,1.5\labeldist)$) {$\scriptstyle 4$};
  \node at ($(n5)-(0,\labeldist)$) {$\scriptstyle 5$};
  \node at ($(m6)-(0,1.5\labeldist)$) {$\scriptstyle 6$};
  \node at ($(n7)-(0,\labeldist)$) {$\scriptstyle 7$};
  \node at ($(n8)-(0,\labeldist)$) {$\scriptstyle 8$};
  \node at ($(n9)-(0,\labeldist)$) {$\scriptstyle 9$};
  \node at ($(m10)-(0,1.5\labeldist)$) {$\scriptstyle 10$};
\end{tikzpicture}
  
  \caption{The reducible alternating spin-chain state from figure~\ref{fig:redaltchain} drawn in a way that reflects locality as dictated by the integrable structure. This is quite similar to figure~\ref{fig:redaltchainloc}, but the boson/fermion grading of the representations at sites 2 and 8 are now opposite to the grading of the representations at sites 1,3,5,7 or 9.
}
  \label{fig:redaltsuchainloc}

\end{figure}

Finally, the full spin-chain relevant to $AdS_3/CFT_2$ consists of a left-moving and a right- moving sector, together with a momentum conservation condition for physical states which combines both sectors. This construction was already discussed in~\cite{OhlssonSax:2011ms} and we refer the interested reader to it.

\section{The Lax connection and the algebraic Bethe ansatz}
\label{sec8}

In this section we discuss how reducible integrable spin-chains can be solved using the algebraic Bethe ansatz (ABA).\footnote{%
  For other unconventional settings generalising the ABA see~\cite{Fioravanti:2001bx,Fioravanti:2001pt}.%
} %
For any representation of $\alg{sl}(2)$ given by operators $T^3$, $T^\pm$ satisfying the conventional commutation relations and acting on a vector space $V$, one defines the Lax operator as
\begin{equation}
L_k(\mu)\equiv\left(\begin{array}{cc}
\mu - i T^3_k & i T^-_k \\
i T^+_k & \mu + i T^3_k \end{array}\right)\,,
\label{lax}
\end{equation}
where the subscript $k$ labels the site on the corresponding spin-chain: $L_k$ acts on $V\otimes v$ where $v$ is a two-dimensional auxiliary space. The above expression is slightly different from the conventional one in~\cite{Tarasov:1983cj}, because we take $T^-$ to be the ``creation'' operator. It is easy to check that the above Lax operator satisfies the fundamental commutation relation
\begin{equation}
R^{2\times 2}_{12}(\mu-\nu)L_{n,1}(\mu)L_{n,2}(\nu)=L_{n,2}(\nu)L_{n,1}(\mu)R^{2\times 2}_{12}(\mu-\nu)\,,
\label{fcr}
\end{equation}
where
\begin{equation}
  \begin{aligned}
    R^{2\times 2}_{12}(\mu) &=\mu \mbox{Id}_{4}+i \mbox{Perm}_{12}=(\mu+i/2)\mbox{Id}_{4}+i\sigma^i\otimes \sigma^i/2\,, \\
    L_{n,1} &= L_k\otimes\mbox{Id}_2\,, \\
    L_{n,2} &= \mbox{Id}_2\otimes L_k\,.
  \end{aligned}
\end{equation}
$R^{2\times 2}_{12}(\mu)$ takes the same form as the R-matrix used in~\cite{Faddeev:1996iy}. The $L_k$ can be used to construct an algebraic Bethe ansatz (ABA) for finding the eigenvectors and eigenvalues of the operator
\begin{equation}
\rho(\mu)\equiv
\tr_v\left(L_{1}(\mu)
L_{2}(\mu)\dots L_{J}(\mu)\right)\,.
\label{rhomat}
\end{equation}
where $J$ is the length of the spin-chain and the trace is taken over the auxiliary space $v$. Typically, one thinks of integrable spin-chains as coming equipped with an R-matrix $R_{0n}(\mu)$, which acts on $V\otimes V$. In such cases, the spin-chain Hamiltonian is one of the charges contained in the monodromy matrix
\begin{equation}
\tau(\mu)\equiv
\tr_0\left(R_{01}(\mu)
R_{02}(\mu)\dots R_{0J}(\mu)\right)\,,
\end{equation}
where the trace is taken over the auxiliary space $V$. In~\cite{Tarasov:1983cj}, the authors show that 
\begin{equation}
\label{fttcomm}
\left[\tau(\mu_1)\,,\,\rho(\mu_2)\right]=0\,,
\end{equation}
implying that, up to degeneracy, the eigenvectors of $\rho(\mu)$, found through the ABA procedure, are also eigenvectors of $\tau(\mu)$, and hence also of the Hamiltonian. In~\cite{Tarasov:1983cj}, a procedure is also given for extracting the eigenvalues of $\tau(\mu)$ from the ABA.

In this section we will follow the ABA construction for the toy model from section~\ref{sec:homogenous-toy-model}, the homogeneous R-module spin-chain from section~\ref{sec:homogenous-R-spin-chain} and the alternating R-module spin-chain from section~\ref{sec:alternating-R-spin-chain}.

\subsection{The ABA for the \texorpdfstring{$\mathbf{1} \oplus \mathbf{2}$}{1+2} spin-chain}
\label{sec:ABA-toy-model}

Before we study the full $\alg{sl}(2)$ spin-chain we will again consider the compact toy-model from section~\ref{sec:homogenous-toy-model}, with sites transforming in the $\mathbf{1} \oplus \mathbf{2}$ representation. As explained above, we can introduce the Lax operator $L_k(\mu)$ defined in~\eqref{lax} and satisfying~\eqref{fcr}. However, it will be convenient to consider a shifted operator
\begin{equation}
\label{Lshift}
  \tilde{L}_k(\mu) \equiv L_k(\mu) + \ell(\mu) \Pi_k^{(0)} \,,
\end{equation}
where $\Pi_k^{(0)}$ is a projector onto the singlet at site $k$. The operator $\tilde{L}_k$ satisfies the same fundamental commutation relation~\eqref{fcr} as $L_k$. We also introduce the corresponding operator
\begin{equation}
  \tilde{\rho}(\mu) \equiv \tr_v \left(
    \tilde{L}_1(\mu) \tilde{L}_2(\mu) \dotsb \tilde{L}_J(\mu) 
  \right) \,.
\end{equation}
As above, $\tilde{\rho}(\mu_1)$ commutes with $\tilde{\rho}(\mu_2)$ and $\tau(\mu_2)$.

The representation at each site is reducible. Hence there are two highest weight states $\ket{0}$ and $\ket{1}$. Let us denote these two possible states at site $k$ by
\begin{equation}
  \ket{\omega_k^0} \equiv \ket{0}_k \,, \qquad \text{and} \qquad
  \ket{\omega_k^1} \equiv \ket{1}_k \,.
\end{equation}
We can then construct a ``vacuum'' state of length $J$ of the form
\begin{equation}
  \ket{\Omega^{\mathbf{v}}}_J = \bigotimes_{k=1}^J \ket{\omega_k^{v_k}} \,,
\end{equation}
where $\mathbf{v} = (v_1,v_2,\dotsc,v_J)$ with $v_k \in \{0,1\}$ labels the choice of highest weight state at each site. Alternatively, we can use the notation of section~\ref{sec:homogenous-toy-model} and partition the sites $\mathbf{N} = \{1,\dotsc,J\}$ into two complementary sets
\begin{equation}
  \mathbf{n} = \{n_1, n_2, \dotsc , n_n\} \subset \mathbf{N} \,, \qquad
  \tilde{\mathbf{n}} = \mathbf{N} \setminus \mathbf{n} \,,
\end{equation}
with $\mathbf{n}$ specifying the sites transforming in the singlet submodule. In other words
\begin{equation}
  v_k =
  \begin{cases}
    0 , & \text{if $k \in \mathbf{n}$} \,, \\
    1 , & \text{if $k \not\in \mathbf{n}$}  \,.
  \end{cases}
\end{equation}
As before we denote the number of sites in the state $\ket{0}$ and $\ket{1}$ by $n$ and $\tilde{n}$, respectively. 

Labeling the entries in the $2 \times 2$ matrix defining $\rho(\mu)$ as
\begin{equation}\label{eq:toy-model-rho-ABCD}
  \rho(\mu) = \tr
  \begin{pmatrix}
    A(\mu) & B(\mu) \\
    C(\mu) & D(\mu)
  \end{pmatrix}
\end{equation}
we find
\begin{equation}
  A(\mu) \ket{\Omega^{\mathbf{v}}}_J = \alpha_J^{\mathbf{v}}(\mu) \ket{\Omega^{\mathbf{v}}}_J \,, \qquad
  D(\mu) \ket{\Omega^{\mathbf{v}}}_J = \delta_J^{\mathbf{v}}(\mu) \ket{\Omega^{\mathbf{v}}}_J \,, \qquad
  B(\mu) \ket{\Omega^{\mathbf{v}}}_J = 0 \,,
\end{equation}
where
\begin{equation}\label{eq:toy-model-a-d}
  \alpha_J^{\mathbf{v}}(\mu) = \bigg( \mu + \ell \bigg)^n \bigg( \mu + \frac{i}{2} \bigg)^{\tilde{n}} \,, \qquad
  \delta_J^{\mathbf{v}}(\mu) = \bigg( \mu + \ell \bigg)^n \bigg( \mu - \frac{i}{2} \bigg)^{\tilde{n}} \,.
\end{equation}

Following~\cite{Faddeev:1996iy}, the fundamental commutation relation~(\ref{fcr}) is
equivalent to certain relations between $A_J$, $B_J$, $C_J$ and $D_J$
including\footnote{%
  The expressions are exactly the same as in~\cite{Faddeev:1996iy}
  because our matrix $R^{2\times 2}$ is exactly the same as the one
  in~\cite{Faddeev:1996iy}.%
}%
\begin{equation}\label{eq:toy-model-ABCD-com-rel}
  \begin{aligned}
    B(\mu)B(\nu) &= B(\nu)B(\mu) \,, \\
    A(\mu)B(\nu) &= f(\mu-\nu) B(\nu)A(\mu) + g(\mu-\nu) B(\mu)A(\nu) \,, \\
    D(\mu)B(\nu) &= h(\mu-\nu) B(\nu)D(\mu) + k(\mu-\nu) B(\mu)D(\nu) \,,
  \end{aligned}
\end{equation}
where we have temporarily dropped the subscript $J$ for clarity and
\begin{equation}
    f(\mu) = \frac{\mu-i}{\mu} \,, \qquad
    g(\mu) = +\frac{i}{\mu} \,, \qquad
    h(\mu) = \frac{\mu+i}{\mu} \,, \qquad
    k(\mu) = -\frac{i}{\mu} \,.
\end{equation}

We saw above that the $2^J$ groundstates $\ket{\Omega^{\mathbf{v}}}_J$ are eigenvectors of $\rho(\mu)$. We now look for additional eigenstates by applying the ansatz
\begin{equation}
  \ket{\Phi_J^{\mathbf{v}}(\{\mu_i\})} \equiv B(\mu_1) \dotsb B(\mu_k) \ket{\Omega^{\mathbf{v}}}_J \,,
\end{equation}
where we have introduced $\{\mu_i\}$ to indicate the set $\{\mu_1,\dotsc,\mu_l\}$. Following the same arithmetic as in~\cite{Faddeev:1996iy} we find that the above are eigenvectors of $\rho(\mu)$ with eigenvalue
\begin{equation}
  \mu^{\mathbf{v}}_J(\mu,\{\mu_i\})=
  \alpha^{\mathbf{v}}_J(\mu)\prod_{k=1}^l f(\mu-\mu_k)
  +
  \delta^{\mathbf{v}}_J(\mu)\prod_{k=1}^l h(\mu-\mu_k)\,,
\end{equation}
as long as the $\mu_i$ satisfy
\begin{equation}\label{eq:toy-model-BE-af-dh}
  \alpha^{\mathbf{v}}_J(\mu_k) \prod_{j \neq k}^l f(\mu_k-\mu_j)
  =
  \delta^{\mathbf{v}}_J(\mu_k) \prod_{j \neq k}^l h(\mu_k-\mu_j)
\end{equation}
for $k=1,\dotsc,l$. Using the explicit expressions for $f$, $h$, $\alpha^{\mathbf{v}}$ and $\delta^{\mathbf{v}}$ we obtain the Bethe equations
\begin{equation}\label{eq:toy-model-BE}
  \bigg(\frac{\mu_k + \frac{i}{2}}{\mu_k - \frac{i}{2}}\bigg)^{\tilde{n}} = \prod_{j \neq k}^l \frac{\mu_k - \mu_j + i}{\mu_k - \mu_j - i} \,.
\end{equation}
Note that the Bethe equations only depend on the number of non-singlets sites $\tilde{n}$. Hence there are only $J+1$ distinct sets of Bethe equations even though the number of groundstates is $2^J$. The above equation coincides with the Bethe equation for an XXX$_{1/2}$ spin-chain of length $\tilde{n}$~\cite{Faddeev:1996iy}. Since the raising operator $B(\mu)$ does not change the position of the singlet states, the obtained spectrum consists of sectors labeled by the set of singlet sites $\mathbf{n}$. The highest weight states in such a sector matches the states of the XXX$_{1/2}$ spin-chain on the non-singlet sites.

When acting on a vacuum state containing singlets equation~\eqref{eq:toy-model-BE-af-dh} has a solution at $\mu_k = \ell(\mu_k)$. This is not a solution of the Bethe equations~\eqref{eq:toy-model-BE} and does not correspond to a new eigenstate of $\rho(\mu)$, but to a zero of the raising operator $B(\mu_k)$. If we had not introduced the additional shift in $\tilde{L}_n(\mu)$ this sporadic solution would sit at $\mu = 0$, and would hence hide a physical solution of the Bethe equations. In this regard, the shift  introduced in equation~(\ref{Lshift}) acts as a regulator of the raising operator. The function $\ell(\mu)$ is arbitrary. In particular we can choose, \emph{e.g.}, $\ell(\mu) = 1 - \mu$, which completely removes the sporadic solutions.

\subsection{The ABA for the \texorpdfstring{$R$}{R} module}
\label{sec:ABA-R-module}

We are now ready to apply the algebraic Bethe ansatz to a homogenous spin-chain with the sites transforming in the $R$ module. Like the toy model considered above, the $R$ module has two highest weight states. One of these is a singlet and the other state generates the rest of the module. Hence, the structure of the ABA is very similar to that of the toy model.
At each site we define
\begin{equation}
  \ket{\omega_n^0} \equiv \ket{0}_n \,, \qquad \text{and} \qquad
  \ket{\omega_n^1} \equiv \ad_n\ket{0}_n \,.
\end{equation}
As above a vacuum state for a length $J$ spin-chain can then be defined by
\begin{equation}
  \ket{\Omega^{\mathbf{v}}}_J \equiv \bigotimes_{n=1}^J\ket{\omega^{v_n}_n}
\end{equation}
where again $\mathbf{v}\equiv (v_1\,,\,v_2\,,\dots\,,v_J)$, with $v_i\in\{0\,,\,1\}$.

As in the previous section it is useful to consider the shifted  Lax operator
\begin{equation}
  \tilde{L}_k(\mu) = L_k(\mu) + \ell(\mu) \Pi_k^{0},
\end{equation}
and the corresponding transfer matrix $\tilde{\rho}(\mu)$. From $\tilde{\rho}(\mu)$ we construct operators $A(\mu)$, $B(\mu)$, $C(\mu)$ and $D(\mu)$ by the same prescription as in~\eqref{eq:toy-model-rho-ABCD}. These operators satisfy the relations in equation~\eqref{eq:toy-model-ABCD-com-rel}, with the only difference being their action on groundstates, which is now given by
\begin{equation}
 A_J\ket{\Omega^{\mathbf{v}}}_J =\alpha^{\mathbf{v}}_J(\mu)\ket{\Omega^{\mathbf{v}}}_J \,,
 \qquad
 D_J\ket{\Omega^{\mathbf{v}}}_J = \delta^{\mathbf{v}}_J(\mu)\ket{\Omega^{\mathbf{v}}}_J \,,
 \qquad 
 C_J\ket{\Omega^{\mathbf{v}}}_J =0 \,,
\end{equation}
where
\begin{equation}
  \alpha^{\mathbf{v}}_J(\mu) = (\mu + \ell)^{n}(\mu-i)^{\tilde{n}} \,,
  \qquad
  \delta^{\mathbf{v}}_J(\mu) = (\mu + \ell)^{n}(\mu+i)^{\tilde{n}}\,.
\end{equation}
Note that the extra parameter $\ell$ that we introduced in the shifted Lax operator $\tilde{L}(\mu)$ again acts as a regulator at $\mu=0$.

Following the derivation in the last section, we find that the state
\begin{equation}
  \ket{\Phi_J^{\mathbf{v}}(\{\mu_i\})} \equiv B(\mu_1) \dotsb B(\mu_K) \ket{\Omega^{\mathbf{v}}}_J
\end{equation}
is an eigenstate of $\tilde{\rho}(\mu)$, provided the parameters $\mu_i$ satisfy the Bethe equations
\begin{equation}
  \label{BER}
  \left(\frac{\mu_j - i}{\mu_j + i}\right)^{\tilde{n}}
  =
  \prod_{k\neq j}^K \frac{\mu_j - \mu_k + i}{\mu_j - \mu_k - i} \,.
\end{equation}
These equations take the same form as the Bethe equations of a homogenous Heisenberg spin-chain with $\tilde{n}$ sites in the $s=-1$ representation~\cite{Faddeev:1996iy}.

As in the toy model, there are $2^J$ different ``ground-states'' labeled by $\mathbf{v}$, but only $J+1$ distinct Bethe equations labeled by $\tilde{n}=0\,,\,\dots\,,J$.
Since the raising operator $B(\mu)$ is constructed out of $\alg{sl}(2)$ generators, any state obtained by acting with a set of raising operators on a groundstate which contains a singlet at a particular site will again have a singlet at that site. Hence we can use one solution to~\eqref{BER} to construct several different states by acting with the same set of raising operators on different groundstates of the same length and with the same number of singlet sites. Below we will check that we in this way obtain enough states to cover the full spectrum.

\paragraph{The counting of states.}

We will now show that the above Bethe equations give the correct number of highest weight states, \textit{i.e.}, that the obtained spectrum is complete. To do this we will asume that the ABA description of the $s=-1$ Heisenberg spin-chain is complete. The decomposition into highest weight states of the $J$-fold tensor product of
spin $s$ infinite dimensional irreducible $\alg{sl}(2)$ representations is given by
\begin{equation}
  s \otimes s \otimes \dotsm \otimes s = \oplus_{S=0}^{\infty} \, M_{J,S}  \times ( J\,s - S),
\end{equation}
where the multiplicities $M_{J,S}$ are given by\footnote{%
  To see this, we note that in an oscillator representation we can use a basis with states of the form
  \begin{equation*}
    \ket{n_1,n_2,\dotsc,n_J} = (a_1^{\dag})^{n_1} (a_1^{\dag})^{n_2} \dotsm (a_1^{\dag})^{n_J} \ket{0}.
  \end{equation*}
  The total number of states with $S = n_1 + n_2 + \dotsb n_J$ oscillators is
  \begin{equation*}
    N_{J,S} = \binom{J+S-1}{S}
  \end{equation*}
  However, $N_{J,S-1}$ of those states are descendants from the level below. This leaves us with
  \begin{equation*}
    N_{J,S} - N_{J,S-1} = \binom{J+S-1}{S} - \binom{J+S-2}{S-1} = \binom{J+S-2}{S} = M_{J,S}
  \end{equation*}
  highest weight states.%
}%
\begin{equation}
  M_{J,S} = \binom{J+S-2}{S} .
\end{equation}

We now note that the Bethe equation in the $R$ module for a state with total
weight $S$ above a groundstate with $n_1$ non-zero entries coincides with the
Bethe equation for a homogenous $s=-1$ spin-chain with $n_1$ sites and
$K_{n_1}=S-n_1$ Bethe roots. As we have just seen, this latter equation has
$M_{n_1,K_{n_1}}$ non-trivial solutions. For the $R$ module there are $2^J$
different groundstates, and to obtain all states we need to sum over them. In
doing this we note that there are $\binom{J}{n_1}$ groundstates with weight
$n_1$. Hence the total number of highest weight states of weight $S$ is
\begin{equation}
  \sum_{n_1=0}^{J} \binom{J}{n_1} \binom{n_1+K_{n_1}-2}{K_{n_1}} = \sum_{n_1=0}^{J} \binom{J}{n_1} \binom{S-2}{n_1-2} = \binom{J+S-2}{S},
\end{equation}
which agrees with the expect number of states, $M_{J,S}$.

\subsection{The ABA for an alternating \texorpdfstring{$R$}{R} module spin-chain}
\label{sec:ABA-alt-R-module}

Let us finish this section by considering the ABA for a spin-chain where the even sites transform in the $R$ module and the odd sites in a spin $-1$ representation. Since the $R$ module contains a $s=-1$ representation as a sub module, the structure of this alternating chain will be very similar to that of the homogenous chain considered in the previous section. In fact, the only difference between the two cases is that the groundstates of the alternating chain always has a non-singlet at the odd sites.

As in section~\ref{sec:alternating-R-spin-chain} we denote the set of singlet sites by $\mathbf{m}$ and the non-singlets by $\tilde{\mathbf{m}}$. In the alternating chain $\tilde{\mathbf{m}}$ in particular contains all the odd sites. For a chain of length $2L$ we can then construct $2^L$ groundstates of the form
\begin{equation}
  \ket{\Omega^{\mathbf{v}}}_J = \bigotimes_{k=1}^J \ket{\omega_k^{v_k}} \,, 
\end{equation}
where again
\begin{equation}
  \ket{\omega_n^0} \equiv \ket{0}_n \,, \qquad \text{and} \qquad
  \ket{\omega_n^1} \equiv \ad_n\ket{0}_n \,,
\end{equation}
and
\begin{equation}
  v_k =
  \begin{cases}
    0 , & \text{if $k \in \mathbf{m}$} \,, \\
    1 , & \text{if $k \not\in \mathbf{m}$} \,.
  \end{cases}
\end{equation}
Above such a groundstate we construct an excited state by acting with a raising operator $B(\mu)$,
\begin{equation}
  \ket{\Phi_J^{\mathbf{v}}(\{\mu_i\})} \equiv B(\mu_1) \dotsb B(\mu_K) \ket{\Omega^{\mathbf{v}}}_J .
\end{equation}
This is an eigenstate of the monodromy matrix $\tilde{\rho}(\mu)$ provided $\mu_i$ satisfy the Bethe equations
\begin{equation}
  \label{BER-alt}
  \left(\frac{\mu_j - i}{\mu_j + i}\right)^{\tilde{m}}
  =
  \prod_{k\neq j}^K \frac{\mu_j - \mu_k + i}{\mu_j - \mu_k - i} \,.
\end{equation}
As expected from the results of section~\ref{sec:alternating-R-spin-chain}, this is the Bethe equations of a homogenous $\mathbf{s}=-1$ spin-chain of length $\tilde{m}$.

\section{Missing massless modes}
\label{sec9}

Let us now turn to the relation between the reducible spin-chains we have investigated in the previous sections and the missing massless mode puzzle of the $AdS_3/CFT_2$ integrable system discussed in the introduction. In order to highlight the essential features of our discussion, throughout this section we will focus on the $\alg{sl}(2|1)$ subsector of the full $\alg{d}(2,1;\alpha=0)$ alternating reducible spin-chain. We remind the reader here that at $\alpha=0$ the odd sites of the spin chain are in the $(\frac{1}{2};\frac{1}{2})$ highest weight chiral irrep of $\alg{sl}(2|1)$ and the even sites are in the reducible 
${\bf 0}\oplus (\frac{1}{2};\frac{1}{2})$  representation. In particular, we recall that the reducible representation at the even sites has one singlet, and it's non-singlet highest weight state is a fermion. In analogy with the results in section~\ref{sec7}, a basis of states in the alternating reducible $\alg{sl}(2|1)$  spin-chain takes the form
\begin{equation}
\left|{\bf m}\right>\equiv {\cal P}\left(\bigotimes_{j=1}^m\ket{0}_{m_j}\bigotimes_{k=1}^{{\tilde m}}\left|v\right>_{{\tilde m}_k}\right)\,,
\end{equation}
where ${\bf m}$, $m$, ${\tilde m}$, $m_j$ and ${\tilde m}_k$ are defined in equations~(\ref{eqMdef})-(\ref{Raltbasis}) and the text around them. ${\cal P}$ orders the sites according to the ${\bf M}$-order (see also the text following equation~(\ref{rbasis4})).

\subsection{A glut of groundstates}

In section~\ref{sec76} we have constructed the Hamiltonian for this spin-chain. As a first exercise we may find all the groundstates of this chain of a given length. From the analysis in section~\ref{sec7} it is easy to see that any state of the form
\begin{equation}
 {\cal P}\left(\bigotimes_{j=1}^m\ket{0}_{m_j}\bigotimes_{k=1}^{{\tilde m}}\left|v=0\right>_{{\tilde m}_k}\right)\,,
\label{glutg}
\end{equation}
is a ground-state. Above $\left|v=0\right>$ is the highest weight state of the $(\frac{1}{2};\frac{1}{2})$ irrep. In other words, ground-states always have $\left|v=0\right>$ on the odd sites, but on the even sites one is free to chose the (bosonic) singlet state $\ket{0}$ or the (fermionic) highest weight state $\left|v=0\right>$. All such states must be groundstates since they sit in short multiplets of the overall $\alg{sl}(2|1)$. As a result the reducible alternating spin-chains have a very large degeneracy: given a spin-chain of length $2N$, there are $2^N$ ground-states, half of which are bosonic and the other half fermionic. These ground-states do not all carry the same $\alg{sl}(2|1)$ Cartan charges, since the singlet state
$\ket{0}$ has no charge under the $U(1)$ R-current, while the highest weight state $\left|v=0\right>$ has charge $1/2$. 

\subsection{Lifting the degeneracy}

This glut of groundstates is in fact unphysical. Perhaps the easiest way to see this is by considering what happens to certain magnon states in the $\alpha\rightarrow 0$ limit. For $\alpha\neq 0$ the alternating $\alg{sl}(2|1)$ spin-chain of length $2N$ has a unique vacuum
\begin{equation}
\ket{0}_{2N}\equiv\bigotimes_{i=1}^N \ket{0}_{2i-1}\otimes\ket{0}_{2i}
\label{unvac}
\end{equation}
where $\ket{0}_{2i-1}$ is the highest weight state of the $(\frac{1-\alpha}{2};\frac{1-\alpha}{2})$ irrep and $\ket{0}_{2i}$ is the highest weight state of the $(\frac{\alpha}{2};-\frac{\alpha}{2})$
irrep. On this groundstate we can build a fermionic magnon state
\begin{equation}
\left|Q^+_p;\alpha \neq 0\right>_{2N}\equiv\frac{1}{\sqrt{\alpha}}\sum_{m=1}^N e^{i p m} Q^+_{2m} \ket{0}_{2N}
\label{fermmag}
\end{equation}
where the momentum of the magnon is $p=2\pi m/N$ with $m=0,1,2,\dots,N-1$, and we have chosen the overall normalisation for later convenience. $Q^+_{2m}$ is the fermionic creation operator in $\alg{sl}(2|1)$ which does not annihilate the (chiral) highest weight state; the subscript ${}_{2m}$ indicates that the operator $Q^+$ is acting at site $2m$. 

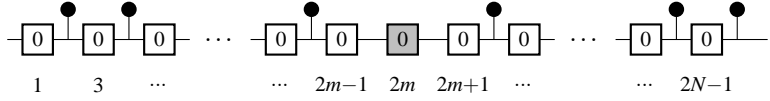
\begin{figure}
  \centering
\begin{tikzpicture}
  \coordinate (start) at (-0.5\sitedist,0);
  \node (n1) at (0\sitedist,0) [odd site] {$\scriptstyle 0$};
  \coordinate (n2) at (0.5\sitedist,0); \node (m2) at (0.5\sitedist,0.5\sitedist) [null site] {};
  \node (n3) at (1\sitedist,0) [odd site] {$\scriptstyle 0$};
  \coordinate (n4) at (1.5\sitedist,0); \node (m4) at (1.5\sitedist,0.5\sitedist) [null site] {};
  \node (n5) at (2\sitedist,0) [odd site] {$\scriptstyle 0$};

  \coordinate (n6) at (2.5\sitedist,0);
  \node at (3\sitedist,0) {\small $\dots$};
  \coordinate (n6p) at (3.5\sitedist,0);

  \node (n7) at (4\sitedist,0) [odd site] {$\scriptstyle 0$};  
  \coordinate (n8) at (4.5\sitedist,0); \node (m8) at (4.5\sitedist,0.5\sitedist) [null site] {};
  \node (n9) at (5\sitedist,0) [odd site] {$\scriptstyle 0$};
  \node (n10) at (6\sitedist,0) [even site] {$\scriptstyle 0$};
  \node (n11) at (7\sitedist,0) [odd site] {$\scriptstyle 0$};  
  \coordinate (n12) at (7.5\sitedist,0); \node (m12) at (7.5\sitedist,0.5\sitedist) [null site] {};
  \node (n13) at (8\sitedist,0) [odd site] {$\scriptstyle 0$};  

  \coordinate (n14) at (8.5\sitedist,0);
  \node at (9\sitedist,0) {\small $\dots$};
  \coordinate (n14p) at (9.5\sitedist,0);

  \node (n15) at (10\sitedist,0) [odd site] {$\scriptstyle 0$};
  \coordinate (n16) at (10.5\sitedist,0); \node (m16) at (10.5\sitedist,0.5\sitedist) [null site] {};
  \node (n17) at (11\sitedist,0) [odd site] {$\scriptstyle 0$};
  \coordinate (n18) at (11.5\sitedist,0); \node (m18) at (11.5\sitedist,0.5\sitedist) [null site] {};

  \coordinate (end) at (12\sitedist,0) {};

  \draw [chain line] (start) -- (n1) -- (n2) -- (n3) -- (n4) -- (n5) -- (n6);
  \draw [chain line] (n6p) -- (n7) -- (n8) -- (n9) -- (n10) -- (n11) -- (n12) -- (n13) -- (n14);
  \draw [chain line] (n14p) -- (n15) -- (n16) -- (n17) -- (end);

  \draw [chain line] (n2) -- (m2);
  \draw [chain line] (n4) -- (m4);
  \draw [chain line] (n8) -- (m8);
  \draw [chain line] (n12) -- (m12);
  \draw [chain line] (n16) -- (m16);
  \draw [chain line] (n18) -- (m18);

  \node at ($(n1)-(0,\labeldist)$) {$\scriptstyle 1$};
  \node at ($(n3)-(0,\labeldist)$) {$\scriptstyle 3$};
  \node at ($(n5)-(0,\labeldist)$) {$\scriptstyle \dots$};

  \node at ($(n7)-(0,\labeldist)$) {$\scriptstyle \dots$};
  \node at ($(n9)-(0,\labeldist)$) {$\scriptstyle 2m-1$};
  \node at ($(n10)-(0,\labeldist)$) {$\scriptstyle 2m\vphantom{+1}$};
  \node at ($(n11)-(0,\labeldist)$) {$\scriptstyle 2m+1$};
  \node at ($(n13)-(0,\labeldist)$) {$\scriptstyle \dots$};

  \node at ($(n15)-(0,\labeldist)$) {$\scriptstyle \dots$};
  \node at ($(n17)-(0,\labeldist)$) {$\scriptstyle 2N-1$};
\end{tikzpicture}

\caption{A groundstate of length $2m$ with a non-zero representation at site $2m$, and singlets on every other even site. As previously we indicate the odd sites by a box, the singlets by a dot and the non-singlet site at the even sites by a filled grey circle. The zeros on the non-singlet sites represents the highest weight state of the corresponding representations. In the notation of section~\ref{sec:alpha-0-limit}, the odd sites contain the state $\bar{\phi}_0$, and the even site at $2m$ contains the state $\psi_0$, while the singlets are given by the state $\phi_0$.}

  \label{fig:deg-mag-state-m}
\end{figure}

In the $\alpha\rightarrow 0$ limit the magnons 
$\left|Q^+_p;\alpha\neq0\right>_{2N}$ remain well-defined elements of the Hilbert space of states of the spin-chain, since $Q^+$ scales like $\sqrt{\alpha}$. However, they can no longer be written using the action of $\alg{sl}(2|1)$ generators, However, if we define an operator $\Psi^+$, which maps the singlet state $\ket{0}$ to the highest weight state $\left|v=0\right>$ in the $(\frac{1}{2};\frac{1}{2})$ module
\begin{equation}
\Psi^+\ket{0}\equiv \left|v=0\right>
\label{degmagcrop}
\end{equation}
then the magnon state in equation~(\ref{fermmag}) becomes at $\alpha=0$
\begin{equation}
\left|\Psi^+_p\right>_{2N}\equiv\sum_{m=1}^N e^{i p m} \, \Psi^+_{2m} \ket{0}_{2N}\,.
\label{degmag}
\end{equation}
It is easy to see that such magnon states  are in fact particular linear combinations of the groundstates of the reducible alternating spin-chain given in equation~(\ref{glutg}). We will refer to them as degenerate magnons. Any groundstate of the reducible alternating spin-chain is a linear combination of states given in equation~(\ref{glutg}) and so equivalently is given by a superposition of a number of degenerate magnons~(\ref{degmag}). Figure~\ref{fig:deg-mag-state-m} depicts one of the states $\Psi^+_{2m} \ket{0}_{2N}$ appearing in the construction of the degenerate magnon.

Now to see that these degenerate magnons are in fact not groundstates of the full $\AdS_3/\CFT_2$ integrable system, let us make the following observation. Recall that, using the non-perturbative dispersion relation, a magnon state~(\ref{fermmag}) has energy
\begin{equation}
E(\left|Q^+_p\right>_{2N}) = \sqrt{\alpha^2 + 4 h(\lambda)^2 \sin^2 \frac{p}{2}}\,,
\label{disprel}
\end{equation}
where $h(\lambda)$ is the ubiquitous and undetermined function of the 't Hooft coupling $\lambda$. In the $\alpha\rightarrow 0$ limit the energies of these magnons become
\begin{equation}
E(\left|\Psi^+_p\right>_{2N}) = 2\Bigl| h(\lambda) \sin \frac{p}{2} \Bigr|\,.
\label{degmagdisprel}
\end{equation}
So at $\alpha=0$, only the $p=0$ magnon is a ground-state, and all the other degenerate magnons are in fact excited states. The glut of groundstates is vastly reduced!

From the above, we see that the apparent glut of groundstates discussed in the previous subsection comes about because our R-matrix and resulting Hamiltonian are computed at small $h(\lambda)$. From the dispersion relation~(\ref{disprel}) it is clear that for the states that become massless in the $\alpha\rightarrow 0$ limit, we should really be re-summing all the $\lambda$ contributions to get the leading term in the dispersion relation at $\alpha=0$. Our preceding small $h(\lambda)$ analysis however has been very useful in determining the space-state of the spin-chain and how one should incorporate massless modes into it. We conclude that non-perturbatively in $\lambda$, for each  length $L$ our reducible alternating $\alg{sl}(2|1)$ spin-chain has two (degenerate) groundstates. One is given by the $\alpha\rightarrow 0$ limit of the state~(\ref{unvac}) - this is simply the state consisting of the singlet state at all even sites and the highest weight state at all the odd sites. The second groundstate is the $p=0$ degenerate magnon state~(\ref{degmag}).

 This degeneracy is in agreement with what one expects to see of the chiral ring in the plane wave limit~\cite{Gomis:2002qi}. The $CFT_2$ for the $\alpha=0$ theory is expected to be  the $(4,4)$ $Sym^N(T^4)$ sigma model or some deformation of it. Such $CFT_2$s have a chiral ring of chiral primary operators~\cite{Lerche:1989uy} whose dimensions are protected by supersymmetry - in particular the chiral ring is expected to be invariant under deformations, and can be studied at the orbifold point directly. A beautiful description of operators in chiral rings in a wide class of $CFT_2$s was given in~\cite{Vafa:1994tf} and more explicitly for the case of interest here in~\cite{Maldacena:1998bw} . A chiral ring operator can be thought of as a state in the Fock space of the left-moving sector of a sigma-model whose target space is the cohomology (in this case) of $T^4$. In particular, they are of the form
\begin{equation}
\prod_{i=1}^M \alpha^{A_i}_{-n_i}\ket{0}
\label{cpo}
\end{equation}
where $M$ is an integer and satisfies $1\le M\le N$, with $N$ defined as
\begin{equation}
\sum_{i=1}^Mn_i = N\,.
\end{equation}
The $A_i$ label the complex cohomology classes of $T^4$.\footnote{Recall that $T^4$ has, with multiplicity, the following Dolbeault cohomology classes 
\begin{equation*}
\oplus_{p,q} H^{p,q}(T^4) = 
(0,0)\oplus (1,0)^{\oplus 2}\oplus (0,1)^{\oplus 2}\oplus
(2,0)\oplus (0,2)\oplus (1,1)^{\oplus 4}\oplus(1,2)^{\oplus 2}\oplus
(2,1)^{\oplus 2}\oplus(2,2)\,.
\end{equation*}}
$A=0$ denotes the $(p,q)=(0,0)$ cohomology and operators $\alpha^A$ for which $p+q$ is even or odd are respectively bosonic or fermionic. The R-charges of the chiral-primaries~(\ref{cpo}) are 
\begin{equation}
(B_L,B_R)=(N-M+\sum_{i=1}^M p_i\,,\,N-M+\sum_{i=1}^M q_i)\,.
\end{equation}
Above $p_i$ and $q_i$ are the $(p,q)$-cohomology degree of $A_i$.

In the plane-wave limit the chiral primaries corresponding to single particle states were shown in~\cite{Gomis:2002qi} to be of the form\footnote{The analysis of~\cite{Gomis:2002qi} is mainly focused on the $AdS_3\times S^3\times K3$ background rather than the $AdS_3\times S^3\times T^4$ we are interested in, but it is straightforward to extend these results to the $T^4$ background.}
\begin{equation}
\left(\alpha^0_{-1}\right)^{N+\frac{p+q}{2}-J-1}\alpha^A_{-J+\frac{p+q}{2}-1}\ket{0}\,.
\label{chiralbmn}
\end{equation}
Picking $A=0$ gives an operator which in the spin-chain is the $\alpha=0$ limit of the groundstate~(\ref{unvac}). With $A$ chosen as a $(1,0)$ form, the operator in equation~(\ref{chiralbmn}) is precisely the $p=0$ degenerate magnon in equation~(\ref{degmag}) which we have shown is a genuine groundstate of the 
all-orders-in-$\lambda$ reducible spin-chain.

\subsection{Speculations on the degenerate magnon Hamiltonian}

In the previous subsection we have argued that the glut of groundstates present in the small $\lambda$ reducible spin-chain is lifted by a resummation of higher order in $\lambda$ terms. For $\alpha\neq 0$ these higher orders are suppressed, but for $\alpha=0$ they combine to give the degenerate magnons~(\ref{degmag}) a non-trivial dispersion relation and energy~(\ref{degmagdisprel}). 
This implies that the Hamiltonian of the full integrable system is not just given by $H$ the Hamiltonian of the one-loop reducible spin-chain of the form discussed in section~\ref{sec76}. Rather, we expect there to be an additional piece $H_d$, which will be responsible for giving energy to the degenerate magnons. In this subsection we will present an example of the sort of form that $H_d$ could take.

On general grounds the full Hamiltonian of the $\AdS_3/\CFT_2$ spin-chain should be given by
\begin{equation}
  H_{\text{\small total}} = h(\lambda)^2 H + h(\lambda) H_d \,,
\end{equation}
where $H$ is the Hamiltonian of the homogenous $\alg{sl}(2|1)$ spin-chain. $H_d$ should commute with the $\alg{sl}(2|1)$ generators and with $H$ as well as with the full transfer matrix $\tau(u)$. Acting on a single degenerate magnon $H_d$ should give
\begin{equation}
  H_d \ket{\Psi^+_p}_{2N} = \sqrt{ 4\sin^2 \frac{p}{2}} \ket{\Psi^+_p}_{2N} \,.
\end{equation}

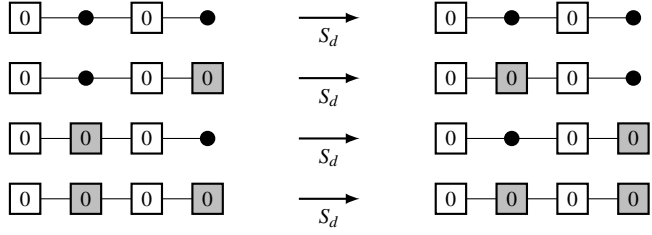
\begin{figure}
  \centering
  \begin{tikzpicture}
  \begin{scope}
    \begin{scope}[xshift=-5\sitedist]
      \node (n1) at (0\sitedist,0) [odd site] {$\scriptstyle 0$}; 
      \node (n2) at (1\sitedist,0) [null site] {}; 
      \node (n3) at (2\sitedist,0) [odd site] {$\scriptstyle 0$}; 
      \node (n4) at (3\sitedist,0) [null site] {};

      \draw [chain line] (n1) -- (n2) -- (n3) -- (n4);
    \end{scope}
    \draw [thick,-latex] (-0.4cm,0) -- (0.4cm,0); \node at (0,0) [anchor=north] {$\scriptstyle S_d$};
    \begin{scope}[xshift=+2\sitedist]
      \node (n1) at (0\sitedist,0) [odd site] {$\scriptstyle 0$}; 
      \node (n2) at (1\sitedist,0) [null site] {}; 
      \node (n3) at (2\sitedist,0) [odd site] {$\scriptstyle 0$}; 
      \node (n4) at (3\sitedist,0) [null site] {};

      \draw [chain line] (n1) -- (n2) -- (n3) -- (n4);
    \end{scope}
  \end{scope}
  \begin{scope}[yshift=-\sitedist]
    \begin{scope}[xshift=-5\sitedist]
      \node (n1) at (0\sitedist,0) [odd site] {$\scriptstyle 0$}; 
      \node (n2) at (1\sitedist,0) [null site] {}; 
      \node (n3) at (2\sitedist,0) [odd site] {$\scriptstyle 0$}; 
      \node (n4) at (3\sitedist,0) [even site] {$\scriptstyle 0$};

      \draw [chain line] (n1) -- (n2) -- (n3) -- (n4);
    \end{scope}
    \draw [thick,-latex] (-0.4cm,0) -- (0.4cm,0); \node at (0,0) [anchor=north] {$\scriptstyle S_d$};
    \begin{scope}[xshift=+2\sitedist]
      \node (n1) at (0\sitedist,0) [odd site] {$\scriptstyle 0$}; 
      \node (n2) at (1\sitedist,0) [even site] {$\scriptstyle 0$};
      \node (n3) at (2\sitedist,0) [odd site] {$\scriptstyle 0$}; 
      \node (n4) at (3\sitedist,0) [null site] {};

      \draw [chain line] (n1) -- (n2) -- (n3) -- (n4);
    \end{scope}
  \end{scope}
  \begin{scope}[yshift=-2\sitedist]
    \begin{scope}[xshift=-5\sitedist]
      \node (n1) at (0\sitedist,0) [odd site] {$\scriptstyle 0$}; 
      \node (n2) at (1\sitedist,0) [even site] {$\scriptstyle 0$};
      \node (n3) at (2\sitedist,0) [odd site] {$\scriptstyle 0$}; 
      \node (n4) at (3\sitedist,0) [null site] {};

      \draw [chain line] (n1) -- (n2) -- (n3) -- (n4);
    \end{scope}
    \draw [thick,-latex] (-0.4cm,0) -- (0.4cm,0); \node at (0,0) [anchor=north] {$\scriptstyle S_d$};
    \begin{scope}[xshift=+2\sitedist]
      \node (n1) at (0\sitedist,0) [odd site] {$\scriptstyle 0$}; 
      \node (n2) at (1\sitedist,0) [null site] {}; 
      \node (n3) at (2\sitedist,0) [odd site] {$\scriptstyle 0$}; 
      \node (n4) at (3\sitedist,0) [even site] {$\scriptstyle 0$};

      \draw [chain line] (n1) -- (n2) -- (n3) -- (n4);
    \end{scope}
  \end{scope}
  \begin{scope}[yshift=-3\sitedist]
    \begin{scope}[xshift=-5\sitedist]
      \node (n1) at (0\sitedist,0) [odd site] {$\scriptstyle 0$}; 
      \node (n2) at (1\sitedist,0) [even site] {$\scriptstyle 0$};
      \node (n3) at (2\sitedist,0) [odd site] {$\scriptstyle 0$}; 
      \node (n4) at (3\sitedist,0) [even site] {$\scriptstyle 0$};

      \draw [chain line] (n1) -- (n2) -- (n3) -- (n4);
    \end{scope}
    \draw [thick,-latex] (-0.4cm,0) -- (0.4cm,0); \node at (0,0) [anchor=north] {$\scriptstyle S_d$};
    \begin{scope}[xshift=+2\sitedist]
      \node (n1) at (0\sitedist,0) [odd site] {$\scriptstyle 0$}; 
      \node (n2) at (1\sitedist,0) [even site] {$\scriptstyle 0$};
      \node (n3) at (2\sitedist,0) [odd site] {$\scriptstyle 0$}; 
      \node (n4) at (3\sitedist,0) [even site] {$\scriptstyle 0$};

      \draw [chain line] (n1) -- (n2) -- (n3) -- (n4);
    \end{scope}
  \end{scope}
\end{tikzpicture}

  \caption{The action of the shift operator $S_d$ on four highest weight states.}
  \label{fig:Sd-ground-state}
\end{figure}

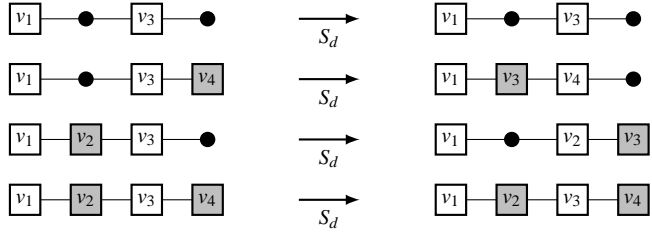
\begin{figure}
  \centering
\begin{tikzpicture}
  \begin{scope}
    \begin{scope}[xshift=-5\sitedist]
      \node (n1) at (0\sitedist,0) [odd site] {$\scriptstyle v_1$}; 
      \node (n2) at (1\sitedist,0) [null site] {}; 
      \node (n3) at (2\sitedist,0) [odd site] {$\scriptstyle v_3$}; 
      \node (n4) at (3\sitedist,0) [null site] {};

      \draw [chain line] (n1) -- (n2) -- (n3) -- (n4);
    \end{scope}
    \draw [thick,-latex] (-0.4cm,0) -- (0.4cm,0); \node at (0,0) [anchor=north] {$\scriptstyle S_d$};
    \begin{scope}[xshift=+2\sitedist]
      \node (n1) at (0\sitedist,0) [odd site] {$\scriptstyle v_1$}; 
      \node (n2) at (1\sitedist,0) [null site] {}; 
      \node (n3) at (2\sitedist,0) [odd site] {$\scriptstyle v_3$}; 
      \node (n4) at (3\sitedist,0) [null site] {};

      \draw [chain line] (n1) -- (n2) -- (n3) -- (n4);
    \end{scope}
  \end{scope}
  \begin{scope}[yshift=-\sitedist]
    \begin{scope}[xshift=-5\sitedist]
      \node (n1) at (0\sitedist,0) [odd site] {$\scriptstyle v_1$}; 
      \node (n2) at (1\sitedist,0) [null site] {}; 
      \node (n3) at (2\sitedist,0) [odd site] {$\scriptstyle v_3$}; 
      \node (n4) at (3\sitedist,0) [even site] {$\scriptstyle v_4$};

      \draw [chain line] (n1) -- (n2) -- (n3) -- (n4);
    \end{scope}
    \draw [thick,-latex] (-0.4cm,0) -- (0.4cm,0); \node at (0,0) [anchor=north] {$\scriptstyle S_d$};
    \begin{scope}[xshift=+2\sitedist]
      \node (n1) at (0\sitedist,0) [odd site] {$\scriptstyle v_1$}; 
      \node (n2) at (1\sitedist,0) [even site] {$\scriptstyle v_3$};
      \node (n3) at (2\sitedist,0) [odd site] {$\scriptstyle v_4$}; 
      \node (n4) at (3\sitedist,0) [null site] {};

      \draw [chain line] (n1) -- (n2) -- (n3) -- (n4);
    \end{scope}
  \end{scope}
  \begin{scope}[yshift=-2\sitedist]
    \begin{scope}[xshift=-5\sitedist]
      \node (n1) at (0\sitedist,0) [odd site] {$\scriptstyle v_1$}; 
      \node (n2) at (1\sitedist,0) [even site] {$\scriptstyle v_2$};
      \node (n3) at (2\sitedist,0) [odd site] {$\scriptstyle v_3$}; 
      \node (n4) at (3\sitedist,0) [null site] {};

      \draw [chain line] (n1) -- (n2) -- (n3) -- (n4);
    \end{scope}
    \draw [thick,-latex] (-0.4cm,0) -- (0.4cm,0); \node at (0,0) [anchor=north] {$\scriptstyle S_d$};
    \begin{scope}[xshift=+2\sitedist]
      \node (n1) at (0\sitedist,0) [odd site] {$\scriptstyle v_1$}; 
      \node (n2) at (1\sitedist,0) [null site] {}; 
      \node (n3) at (2\sitedist,0) [odd site] {$\scriptstyle v_2$}; 
      \node (n4) at (3\sitedist,0) [even site] {$\scriptstyle v_3$};

      \draw [chain line] (n1) -- (n2) -- (n3) -- (n4);
    \end{scope}
  \end{scope}
  \begin{scope}[yshift=-3\sitedist]
    \begin{scope}[xshift=-5\sitedist]
      \node (n1) at (0\sitedist,0) [odd site] {$\scriptstyle v_1$}; 
      \node (n2) at (1\sitedist,0) [even site] {$\scriptstyle v_2$};
      \node (n3) at (2\sitedist,0) [odd site] {$\scriptstyle v_3$}; 
      \node (n4) at (3\sitedist,0) [even site] {$\scriptstyle v_4$};

      \draw [chain line] (n1) -- (n2) -- (n3) -- (n4);
    \end{scope}
    \draw [thick,-latex] (-0.4cm,0) -- (0.4cm,0); \node at (0,0) [anchor=north] {$\scriptstyle S_d$};
    \begin{scope}[xshift=+2\sitedist]
      \node (n1) at (0\sitedist,0) [odd site] {$\scriptstyle v_1$}; 
      \node (n2) at (1\sitedist,0) [even site] {$\scriptstyle v_2$};
      \node (n3) at (2\sitedist,0) [odd site] {$\scriptstyle v_3$}; 
      \node (n4) at (3\sitedist,0) [even site] {$\scriptstyle v_4$};

      \draw [chain line] (n1) -- (n2) -- (n3) -- (n4);
    \end{scope}
  \end{scope}
\end{tikzpicture}

  \caption{The action of the shift operator $S_d$ on a generic state, with the states of the $\alg{sl}(2|1)$ representations labeled by the occupation numbers $v_1,\dotsc,v_4$.}
\label{fig:Sd-generic-state}
\end{figure}

Degenerate magnons considered in the previous sub-section carry a momentum, which is a conserved charge associated with an operator which we will call $P_d$. Let us construct this operator. In analogy with a conventional spin-chain, we will write $e^{i P_d}$ as a "shift" operator which acts as
\begin{equation}\label{shd}
  e^{i P_d}\Psi^+_{2k}\left|0\right>_{2N}=\Psi^+_{2k+2}\left|0\right>_{2N}\,.
\end{equation}
This ensures that degenerate magnons defined in equation~(\ref{degmag}) have $P_d=p$. In a conventional spin-chain the shift operator can be expressed as a product of permutation operators. For our spin-chain this can be generalized to
\begin{equation}
  e^{i P_d}=\prod_{k=1}^{N-1} (S_d)_{2k-1}\,,
\end{equation}
where $(S_d)_{2k-1}$ is an operator to be defined presently, which acts on the sites $2k-1$, $2k$, $2k+1$ and $2k+1$. In order to satisfy equation~(\ref{shd}) $S_d$
has to satisfy\footnote{%
  We remind the reader that the reducible representation that sits at the even sites of the spin chain has two highest weight states: the singlet denoted by $\left|0\right>$ and the highest weight of the infinite-dimensional irrep denoted by $\left|v=0\right>$.%
} %
\begin{equation}
  \begin{aligned}
    S_d \Bigl(\left|v_1=0\right>\left|0\right>\left|v_3=0\right>\left|0\right>\Bigr) &= \left|v_1=0\right>\left|0\right>\left|v_3=0\right>\left|0\right> \,, \\
    S_d \Bigl(\left|v_1=0\right>\left|v_2=0\right>\left|v_3=0\right>\left|0\right>\Bigr) &= \left|v_1=0\right>\left|0\right>\left|v_3=0\right>\left|v_4=0\right> \,, \\
    S_d \Bigl(\left|v_1=0\right>\left|0\right>\left|v_3=0\right>\left|v_4=0\right>\Bigr) &= \left|v_1=0\right>\left|v_2=0\right>\left|v_3=0\right>\left|0\right> \,,
  \end{aligned}
\end{equation}
and it is also natural to define
\begin{equation}
S_d \Bigl(\left|v_1=0\right>\left|v_2=0\right>\left|v_3=0\right>\left|v_4=0\right>\Bigr)
=\left|v_1=0\right>\left|v_2=0\right>\left|v_3=0\right>\left|v_4=0\right>
\,.
\end{equation}
In figure~\ref{fig:Sd-ground-state} the action of $S_d$ is shown pictorially.
We would like for $P_d$ to be defined on any state of the spin-chain, not just the degenerate magnon states. Since $P_d$ should be a conserved quantity that is independent of the $\alg{sl}(2|1)$ occupation numbers $v_i$ we will also take
\begin{equation}
  \begin{aligned}
    S_d \Bigl(\ket{v_1}\ket{\mathrlap{\,0}\phantom{v_1}}\ket{v_3}\ket{\mathrlap{\,0}\phantom{v_1}}\Bigr) &= \ket{v_1}\ket{\mathrlap{\,0}\phantom{v_1}}\ket{v_3}\ket{\mathrlap{\,0}\phantom{v_1}} \,, \\
    S_d \Bigl(\ket{v_1}\ket{v_2}\ket{v_3}\ket{\mathrlap{\,0}\phantom{v_1}}\Bigr) &= \ket{v_1}\ket{\mathrlap{\,0}\phantom{v_1}}\ket{v_2}\ket{v_3} \,, \\
    S_d \Bigl(\ket{v_1}\ket{\mathrlap{\,0}\phantom{v_1}}\ket{v_3}\ket{v_4}\Bigr) &= \ket{v_1}\ket{v_3}\ket{v_4}\ket{\mathrlap{\,0}\phantom{v_1}} \,, \\
    S_d \Bigl(\ket{v_1}\ket{v_2}\ket{v_3}\ket{v_4}\Bigr) &= \ket{v_1}\ket{v_2}\ket{v_3}\ket{v_4} \,.
  \end{aligned}
\end{equation}
This action is shown in figure~\ref{fig:Sd-generic-state}.
We may write $S_d$ as
\begin{equation}
(S_d)_{2k-1}=1-2R(0)_{2k,2k+1}\Pi^A_{2k,2k+2} R(0)_{2k,2k+1}\,.
\end{equation}
Above, $R(0)_{2k,2k+1}$ is the $\alpha\rightarrow 0$ limit of the $\alg{sl}(2|1)$ R-matrix~\eqref{eq:R-matrix-ob-ca-ac}. As discussed in section~\ref{sec:alpha-0-limit}, $R(0)_{2k,2k+1}$ acts as a permutation operator if the states at sites $2k$ and $2k+1$ are both in the $(\frac{1}{2};\frac{1}{2})$ representation, and as identity otherwise.

Let us now turn to the construction of a possible $H_d$ operator. In order to preserve the $\alg{sl}(2|1)$ structure of the spin-chain we would like for $H_d$ to commute with $H$. This condition, together with the dispersion relation~(\ref{degmagdisprel}) places restrictions on the form of $H_d$. To see this it is instructive to consider the action of $H_d$ and $H$ on a state in the spin-chain which is a superposition of a degenerate magnon and a conventional $\alg{sl}(2|1)$ magnon. Such a state can be constructed by acting with the supercharge $\gen{Q}^-$ on the sites $\tilde{m}_n$ of the homogenous part of the chain. The homogeneous part of the degenerate magnon states in equation~\eqref{degmag} has length $N+1$ -- the $N$ odd sites plus the single even site $2m$ containing a non-singlet representation. We thus find the state
\begin{equation}\label{eq:two-magnons-mq}
  \ket{Q^-_q;\Psi^+_m} = \sum_{n=1}^{N+1} e^{iqn} Q^-_{\tilde{m}_n} \Psi^+_{2m} \ket{0}_{2N}.
\end{equation}
On the odd sites, $\gen{Q}^-$ acts by replacing the boson $\bar{\phi}_0$ by the fermionic excitation $\bar{\psi}_0$. One such state appearing in the sum above is shown in figure~\ref{fig:deg-mag-state-excited-mn}. When $\gen{Q}^-$ acts on the site $2m$, the fermionic highest weight state $\psi_0$ turns into the boson $\varphi_0$, as shown in figure~\ref{fig:deg-mag-state-excited-mm}. From the above state, we can now construct a state where the degenerate magnon carries momentum $p$ under the operator $P_d$ by writing
\begin{equation}\label{eq:two-magnons}
  \ket{Q^-_q;\Psi^+_p} = \sum_{m=1}^{N} e^{ipm} \ket{Q^-_q;\Psi^+_m} = \sum_{m=1}^{N} \sum_{n=1}^{N+1} e^{i(pm+qn)} Q^-_{\tilde{m}_n} \Psi^+_{2m} \ket{0}_{2N}.
\end{equation}%
Note that the ordered set $\tilde{m}$ appearing on the right hand side contains the odd numbers as well as the even number $2m$, so that $\tilde{m} = \{1,3,\dotsc,2m-1,2m,2m+1,\dotsc,2N-1\}$.
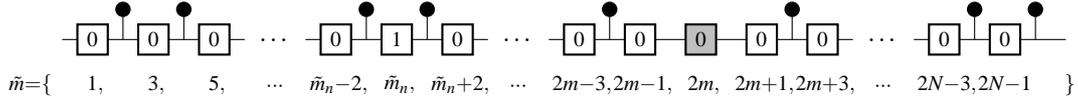
\begin{figure}
  \centering
\begin{tikzpicture}
  \coordinate (start) at (-0.5\sitedist,0);
  \node (n1) at (0\sitedist,0) [odd site] {$\scriptstyle 0$};
  \coordinate (n2) at (0.5\sitedist,0); \node (m2) at (0.5\sitedist,0.5\sitedist) [null site] {};
  \node (n3) at (1\sitedist,0) [odd site] {$\scriptstyle 0$};
  \coordinate (n4) at (1.5\sitedist,0); \node (m4) at (1.5\sitedist,0.5\sitedist) [null site] {};
  \node (n5) at (2\sitedist,0) [odd site] {$\scriptstyle 0$};

  \coordinate (n6) at (2.5\sitedist,0);
  \node (n6pp) at (3\sitedist,0) {\small $\dots$};
  \coordinate (n6p) at (3.5\sitedist,0);

  \node (n7) at (4\sitedist,0) [odd site] {$\scriptstyle 0$};
  \coordinate (n8) at (4.5\sitedist,0); \node (m8) at (4.5\sitedist,0.5\sitedist) [null site] {};
  \node (n9) at (5\sitedist,0) [odd site] {$\scriptstyle 1$};
  \coordinate (n10) at (5.5\sitedist,0); \node (m10) at (5.5\sitedist,0.5\sitedist) [null site] {};
  \node (n11) at (6\sitedist,0) [odd site] {$\scriptstyle 0$};

  \coordinate (n12) at (6.5\sitedist,0);
  \node (n12pp) at (7\sitedist,0) {\small $\dots$};
  \coordinate (n12p) at (7.5\sitedist,0);

  \node (n13) at (8\sitedist,0) [odd site] {$\scriptstyle 0$};  
  \coordinate (n14) at (8.5\sitedist,0); \node (m14) at (8.5\sitedist,0.5\sitedist) [null site] {};
  \node (n15) at (9\sitedist,0) [odd site] {$\scriptstyle 0$};
  \node (n16) at (10\sitedist,0) [even site] {$\scriptstyle 0$};
  \node (n17) at (11\sitedist,0) [odd site] {$\scriptstyle 0$};  
  \coordinate (n18) at (11.5\sitedist,0); \node (m18) at (11.5\sitedist,0.5\sitedist) [null site] {};
  \node (n19) at (12\sitedist,0) [odd site] {$\scriptstyle 0$};  

  \coordinate (n20) at (12.5\sitedist,0);
  \node (n20pp) at (13\sitedist,0) {\small $\dots$};
  \coordinate (n20p) at (13.5\sitedist,0);

  \node (n21) at (14\sitedist,0) [odd site] {$\scriptstyle 0$};
  \coordinate (n22) at (14.5\sitedist,0); \node (m22) at (14.5\sitedist,0.5\sitedist) [null site] {};
  \node (n23) at (15\sitedist,0) [odd site] {$\scriptstyle 0$};
  \coordinate (n24) at (15.5\sitedist,0); \node (m24) at (15.5\sitedist,0.5\sitedist) [null site] {};

  \coordinate (end) at (16\sitedist,0) {};

  \draw [chain line] (start) -- (n1) -- (n2) -- (n3) -- (n4) -- (n5) -- (n6);
  \draw [chain line] (n6p) -- (n7) -- (n8) -- (n9) -- (n10) -- (n11) -- (n12);
  \draw [chain line] (n12p) -- (n13) -- (n14) -- (n15) -- (n16) -- (n17) -- (n18) -- (n19) -- (n20);
  \draw [chain line] (n20p) -- (n21) -- (n22) -- (n23) -- (n24) -- (end);

  \draw [chain line] (n2) -- (m2);
  \draw [chain line] (n4) -- (m4);
  \draw [chain line] (n8) -- (m8);
  \draw [chain line] (n10) -- (m10);
  \draw [chain line] (n14) -- (m14);
  \draw [chain line] (n18) -- (m18);
  \draw [chain line] (n22) -- (m22);
  \draw [chain line] (n24) -- (m24);

  \node at ($(start) -(0,\labeldist)$) [anchor=east] {$\scriptstyle\tilde{m} = \{$};

  \node at ($(n1)-(0,\labeldist)$) {$\scriptstyle 1 \mathrlap{,}$};
  \node at ($(n3)-(0,\labeldist)$) {$\scriptstyle 3 \mathrlap{,}$};
  \node at ($(n5)-(0,\labeldist)$) {$\scriptstyle 5 \mathrlap{,}$};

  \node at ($(n6pp) -(0,\labeldist)$) {$\scriptstyle \dotsb\vphantom{+1}$};

  \node at ($(n7)-(0,\labeldist)$) {$\scriptstyle \tilde{m}_n-2 \mathrlap{,}$};
  \node at ($(n9)-(0,\labeldist)$) {$\scriptstyle \tilde{m}_n \mathrlap{,}$};
  \node at ($(n11)-(0,\labeldist)$) {$\scriptstyle \tilde{m}_n+2 \mathrlap{,}$};

  \node at ($(n12pp) -(0,\labeldist)$) {$\scriptstyle \dotsb\vphantom{+1}$};

  \node at ($(n13)-(0,\labeldist)$) {$\scriptstyle 2m-3 \mathrlap{,}$};
  \node at ($(n15)-(0,\labeldist)$) {$\scriptstyle 2m-1 \mathrlap{,}$};
  \node at ($(n16)-(0,\labeldist)$) {$\scriptstyle 2m\vphantom{+1}\mathrlap{,}$};
  \node at ($(n17)-(0,\labeldist)$) {$\scriptstyle 2m+1 \mathrlap{,}$};
  \node at ($(n19)-(0,\labeldist)$) {$\scriptstyle 2m+3 \mathrlap{,}$};

  \node at ($(n20pp) -(0,\labeldist)$) {$\scriptstyle \dotsb\vphantom{+1}$};

  \node at ($(n21)-(0,\labeldist)$) {$\scriptstyle 2N-3 \mathrlap{,}$};
  \node at ($(n23)-(0,\labeldist)$) {$\scriptstyle 2N-1 \vphantom{,}$};

  \node at ($(end)-(0.25\labeldist,\labeldist)$) [anchor=west] {$\scriptstyle \}$};
\end{tikzpicture}

\caption{An excitation at site $\tilde{m}_n$ of the homogenous spin-chain, above a groundstate with a non-singlet at site $2m$. The excitation is denoted by a $1$ inside a box, corresponding to the state $\bar{\psi}_0$ sitting at this place. In total there are the excitation $\bar{\psi_0}$ can sit in $N$ different positions $1,\dotsc,2N-1$. Below the spin-chain state drawn above we have written out explicitly the set $\tilde{m}$, using the notation introduced in equation~\eqref{eqmdef}.}
  \label{fig:deg-mag-state-excited-mn}
\end{figure}%
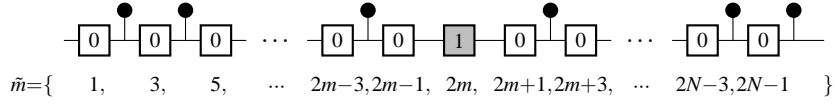
\begin{figure}
  \centering
\begin{tikzpicture}
  \coordinate (start) at (-0.5\sitedist,0);
  \node (n1) at (0\sitedist,0) [odd site] {$\scriptstyle 0$};
  \coordinate (n2) at (0.5\sitedist,0); \node (m2) at (0.5\sitedist,0.5\sitedist) [null site] {};
  \node (n3) at (1\sitedist,0) [odd site] {$\scriptstyle 0$};
  \coordinate (n4) at (1.5\sitedist,0); \node (m4) at (1.5\sitedist,0.5\sitedist) [null site] {};
  \node (n5) at (2\sitedist,0) [odd site] {$\scriptstyle 0$};

  \coordinate (n6) at (2.5\sitedist,0);
  \node (n6pp) at (3\sitedist,0) {\small $\dots$};
  \coordinate (n6p) at (3.5\sitedist,0);

  \node (n7) at (4\sitedist,0) [odd site] {$\scriptstyle 0$};  
  \coordinate (n8) at (4.5\sitedist,0); \node (m8) at (4.5\sitedist,0.5\sitedist) [null site] {};
  \node (n9) at (5\sitedist,0) [odd site] {$\scriptstyle 0$};
  \node (n10) at (6\sitedist,0) [even site] {$\scriptstyle 1$};
  \node (n11) at (7\sitedist,0) [odd site] {$\scriptstyle 0$};  
  \coordinate (n12) at (7.5\sitedist,0); \node (m12) at (7.5\sitedist,0.5\sitedist) [null site] {};
  \node (n13) at (8\sitedist,0) [odd site] {$\scriptstyle 0$};  

  \coordinate (n14) at (8.5\sitedist,0);
  \node (n14pp) at (9\sitedist,0) {\small $\dots$};
  \coordinate (n14p) at (9.5\sitedist,0);

  \node (n15) at (10\sitedist,0) [odd site] {$\scriptstyle 0$};
  \coordinate (n16) at (10.5\sitedist,0); \node (m16) at (10.5\sitedist,0.5\sitedist) [null site] {};
  \node (n17) at (11\sitedist,0) [odd site] {$\scriptstyle 0$};
  \coordinate (n18) at (11.5\sitedist,0); \node (m18) at (11.5\sitedist,0.5\sitedist) [null site] {};

  \coordinate (end) at (12\sitedist,0) {};

  \draw [chain line] (start) -- (n1) -- (n2) -- (n3) -- (n4) -- (n5) -- (n6);
  \draw [chain line] (n6p) -- (n7) -- (n8) -- (n9) -- (n10) -- (n11) -- (n12) -- (n13) -- (n14);
  \draw [chain line] (n14p) -- (n15) -- (n16) -- (n17) -- (end);

  \draw [chain line] (n2) -- (m2);
  \draw [chain line] (n4) -- (m4);
  \draw [chain line] (n8) -- (m8);
  \draw [chain line] (n12) -- (m12);
  \draw [chain line] (n16) -- (m16);
  \draw [chain line] (n18) -- (m18);

  \node at ($(start) -(0,\labeldist)$) [anchor=east] {$\scriptstyle\tilde{m} = \{$};

  \node at ($(n1)-(0,\labeldist)$) {$\scriptstyle 1\mathrlap{,}$};
  \node at ($(n3)-(0,\labeldist)$) {$\scriptstyle 3\mathrlap{,}$};
  \node at ($(n5)-(0,\labeldist)$) {$\scriptstyle 5\mathrlap{,}$};

  \node at ($(n6pp) -(0,\labeldist)$) {$\scriptstyle \dotsb\vphantom{+1}$};

  \node at ($(n7)-(0,\labeldist)$) {$\scriptstyle 2m-3\mathrlap{,}$};
  \node at ($(n9)-(0,\labeldist)$) {$\scriptstyle 2m-1\mathrlap{,}$};
  \node at ($(n10)-(0,\labeldist)$) {$\scriptstyle 2m\vphantom{+1}\mathrlap{,}$};
  \node at ($(n11)-(0,\labeldist)$) {$\scriptstyle 2m+1\mathrlap{,}$};
  \node at ($(n13)-(0,\labeldist)$) {$\scriptstyle 2m+3\mathrlap{,}$};

  \node at ($(n14pp) -(0,\labeldist)$) {$\scriptstyle \dotsb\vphantom{+1}$};

  \node at ($(n15)-(0,\labeldist)$) {$\scriptstyle 2N-3\mathrlap{,}$};
  \node at ($(n17)-(0,\labeldist)$) {$\scriptstyle 2N-1\vphantom{,}$};

  \node at ($(end)-(0.25\labeldist,\labeldist)$) [anchor=west] {$\scriptstyle \}$};
\end{tikzpicture}

  \caption{An excitation sitting on top of the non-singlet at site $2m$. The $1$ inside the circle denotes the excited state $\varphi_0$.}
  \label{fig:deg-mag-state-excited-mm}
\end{figure}%

Before discussing a Hamiltonian for the degenerate magnons that reproduce the dispersion relation~\ref{degmagdisprel}, let us write down a simpler Hamiltonian $H_d'$ with a nearest-neighbor type interactions, similar to the Hamiltonian of an XXX spin-chain. It is natural to construct such a Hamiltonian using the shift operator $S_d$ discussed above, by writing
 \begin{equation}
  H_d' = \sum_{m=1}^{L} (H_d')_{2m} \,, \qquad
  (H_d')_{2m} = 1 - (S_d)_{2m-1} \,.
\end{equation}
Since $H_d'$ is written in terms of the operator $S_d$, it will automatically preserve the ordering of any excitations in the homogenous $\alg{sl}(2|1)$ part of the spin-chain. Hence, $H_d'$ commutes with the $\alg{sl}(2|1)$ Hamiltonian $H$. Acting on the state $\ket{Q^-_q;\Psi^+_p}$ with $H_d'$ we find
\begin{equation}
  H_d'\ket{Q^-_q;\Psi^+_p} = 4\sin^2\frac{p}{2} \ket{Q^-_q;\Psi^+_p} \,.
\end{equation}
This is the normal form of the energy of an XXX spin-chain.

Above we have constructed a Hamiltonian $H_d'$ acting on the degenerate magnons and commuting with $H$. But this Hamiltonian does not reproduce the dispersion relation~\eqref{degmagdisprel}. To obtain the correct dispersion relation we can define $H_d$ as
\begin{equation}
  \label{eq:degmag-ham}
  H_d = \sqrt{\sum_{m=1}^L 1 - (S_d)_{2m-1}} ,
\end{equation}
we obtain an operator that does have the correct eigenvalue on states containing a single degenerate magnon. This operator again acts by changing the position of the non-singlet even sites, leaving the ordering of the excitations of the homogenous chain fixed. Therefore, it commutes not only with $H$ but with the whole transfer matrix $\tau(u)$, as well as with the momentum operator $P_d$.

When acting on a state where all the even sites are either $\left|0\right>$ or $\left| v=0\right>$, expression~\eqref{eq:degmag-ham} simplifies, since the R-matrices always gives the identity. Furthermore, the projector $\Pi^A$, when acting on such a state, can simply be written as $1 - P$, where $P$ exchanges the representations at two even sites. The square root can then be expanded
\begin{equation}
  H_d = \sqrt{L} \left[
    1 - \frac{1}{2L} \left(\sum_{m=0}^L P_{2m,2m+2}\right) - \frac{1}{8L^2} \left(\sum_{m=0}^L P_{2m,2m+2}\right) \left(\sum_{n=0}^L P_{2n,2n+2}\right) + \dotsb
  \right] \,.
\end{equation}
From this expansion, it is clear $H_d$ acts non-locally on the spin-chain. This argument also shows that $H_d$ has an expansion in powers of the degenerate magnon momenta.

Finally, we would like to note that all of the groundstates~(\ref{glutg}) continue to satisfy the $\alg{sl}(2|1)$ shortening condition -- indeed $H$ in equation~(\ref{halgred}) annihilates them. Nevertheless, as we have argued above, in the full reducible spin-chain the degeneracy of these groundstates is lifted: $H_{\text{\small total}}$ does not annihilate them.\footnote{Apart from the $p=0$ degenerate magnon state~(\ref{degmag}).} In other words, while these operators are short with respect to the $\alg{sl}(2|1)$  supercharges $Q$, they cannot be short with respect to the supercharges $Q_{\text{\small total}}$ of the full spin-chain. Hence, we expect the supercharges to also receive corrections, schematically written as\footnote{During the question time following the presentation of some of this work at the ETH Z\"urich meeting, Niklas Beisert suggested this possibility. We would like to thank him for bringing this to our attention and for a discussion of this point.}
\begin{equation}
Q_{\text{\small total}}\sim h(\lambda) Q+\sqrt{h(\lambda)} \, Q_d \,,
\end{equation}
where $Q$ is a $\alg{sl}(2|1)$  supercharge and $Q_d$ commutes with the $\alg{sl}(2|1)$ algebra as well as the transfer matrix $\tau(u)$. It would be interesting to construct the charges $Q_d$.

\subsection{Speculations on the degenerate magnon Bethe Ansatz}

We end this section with some comments on a possible ABA for the degenerate magnons. The Bethe equations for the $\alg{su}(1|1)$ spin-chain are given by
\begin{equation}
  \left( \frac{x_k^+}{x_k^-} \right)^L = \prod_{j \neq k}^K \frac{1 - \frac{1}{x_k^+ x_j^-}}{1 - \frac{1}{x_k^- x_j^+}} \sigma^2(p_k,p_j)\,,
\label{su11allba}
\end{equation}
where $\sigma^2$ indicates the dressing phase and the parameters $x_k^\pm$ satisfy
\begin{equation}
  \frac{x_k^+}{x_k^-} = e^{ip_k} \,, \qquad
  x_k^\pm + \frac{1}{x_k^\pm} = x_k + \frac{1}{x_k} \pm \frac{im}{h}.
\end{equation}
These relations can be solved by
\begin{equation}
    x_k^\pm = \frac{m + \sqrt{m^2 + 4h^2\sin^2\frac{p_k}{2}}}{2h\sin\frac{p_k}{2}} e^{\pm\frac{ip_k}{2}} \,.
\end{equation}
From the above expressions we see that $m$ and $h$ enter only in the combination $h/m$; for the above expressions taking $m\rightarrow 0$ is equivalent to the $h\rightarrow\infty$ limit. In this latter limit the dressing phase reduces to the AFS phase~\cite{Arutyunov:2004vx}
\begin{equation}
\sigma^2(p_k,p_j)=\left(\frac{1 - \frac{1}{x_k^- x_j^+}}{1 - \frac{1}{x_k^+ x_j^-}}\right)^2
\left(\frac{\left(1 - \frac{1}{x_k^- x_j^+}\right)\left(1 - \frac{1}{x_k^+ x_j^-}\right)}{\left(1 - \frac{1}{x_k^+ x_j^+}\right)\left(1 - \frac{1}{x_k^- x_j^-}\right)}\right)^{i\frac{h}{m}(x_k+1/x_k-x_j-1/x_j)}
\end{equation}
Inserting this expression into equation~(\ref{su11allba}) we obtain
\begin{equation}
1= \left( \frac{x_k^-}{x_k^+} \right)^{L\frac{m}{h}}  \prod_{j \neq k}^K \left(\frac{1 - \frac{1}{x_k^- x_j^+}}{1 - \frac{1}{x_k^+ x_j^-}}\right)^{\frac{m}{h}}
\left(\frac{\left(1 - \frac{1}{x_k^- x_j^+}\right)\left(1 - \frac{1}{x_k^+ x_j^-}\right)}{\left(1 - \frac{1}{x_k^+ x_j^+}\right)\left(1 - \frac{1}{x_k^- x_j^-}\right)}\right)^{i(x_k+1/x_k-x_j-1/x_j)}\,.
\end{equation}
We can now send $\frac{m}{h}\rightarrow 0$, however, in order to keep the $L$ dependence of the equation we also take $L\to \frac{h}{m} \tilde{L}$. The above equation then becomes
\begin{equation}
1= \left( \frac{x_k^-}{x_k^+} \right)^{{\tilde L}}  \prod_{j \neq k}^K 
\left(\frac{\left(1 - \frac{1}{x_k^- x_j^+}\right)\left(1 - \frac{1}{x_k^+ x_j^-}\right)}{\left(1 - \frac{1}{x_k^+ x_j^+}\right)\left(1 - \frac{1}{x_k^- x_j^-}\right)}\right)^{i(x_k+1/x_k-x_j-1/x_j)}\,.
\label{intermba}
\end{equation}
In order to keep the momenta $p_k$ and $p_j$ non-zero, we will also take in this limit
\begin{equation}
x^{\pm}_k=e^{\pm ip_k/2}\,,\qquad x^{\pm}_j=e^{\pm ip_j/2}
\,.
\end{equation}
Equation~(\ref{intermba}) then becomes
\begin{equation}
e^{ip_k{\tilde L}}=\prod_{j \neq k}^K \left(\frac{\sin\frac{p_j-p_k}{4}}{\sin\frac{p_j+p_k}{4}}\right)^{8i(\cos\frac{p_k}{2}-\cos\frac{p_j}{2})}\,.
\end{equation}
We propose this equation as a possible Bethe Ansatz for the degenerate magnons~(\ref{fermmag}). We hope to investigate this further in the future.

\section*{Acknowledgments}

We would like to thank Gleb Arutyunov, Joe Chuang, Davide Fioravanti, Andreas Fring, Matthias Gaberdiel, Joe Minahan, Vidas Regelskis, Evgeny Sklyanin, Arkady Tseytlin and Kostya Zarembo for stimulating discussions and sharing their insights with us. 
BS is grateful to the Tata Institute of Fundamental Research and the Kavli Institute of Theoretical Physics for support during extended stays at these institutions while this work was ongoing. BS also thanks  Caltech, Harvard, Institute of Advanced Studies and  SUNY Stony Brook for hospitality during parts of this work. We would like to thank the organisers of the ``Exact Results in Gauge-String Dualities 2012'' workshop held at NORDITA for hospitality which allowed us to work together during a part of this project while participating in an inspirational research programme.
 We would like to thank the organizers of the ``Integrability in Gauge and String Theory 2012''  held at ETH, Zurich for a stimulating atmosphere and a great meeting,  where some of the results were first presented. The work of BS is funded by an Advanced Research Fellowship from the EPSRC and by an STFC Consolidated Grant "Theoretical Physics at City University" ST/J00037X/1. AT thanks the EPSRC for funding under grant EP/H000054/1 at the initial stage of this project.
OOS acknowledges support from the Netherlands Organization for Scientific Reasearch (NWO) under the VICI grant 680-47-602.

\appendix

\section{The \texorpdfstring{$\alg{d}(2,1;\alpha)$}{d(2,1;a)} algebra}
\label{sec:d21a-algebra}

In this section we give the full commutation relations of the $\alg{d}(2,1;\alpha)$ algebra, as well as the action of the generators on the representations of the even and odd spin-chain sites. The generators of $\alg{d}(2,1;\alpha)$ satisfy the algebra
\begin{equation}
  \begin{gathered}
    \begin{aligned}
      \comm{\gen{S}_0}{\gen{S}_\pm} &= \pm \gen{S}_\pm , &
      \comm{\gen{S}_+}{\gen{S}_-} &= 2 \gen{S}_0 , &
      \comm{\gen{S}_0}{\gen{Q}_{\pm\beta\dot{\beta}}} &= \pm\frac{1}{2} \gen{Q}_{\pm\beta\dot{\beta}} , &
      \comm{\gen{S}_\pm}{\gen{Q}_{\mp\beta\dot{\beta}}} &= \gen{Q}_{\pm\beta\dot{\beta}} , \\
      \comm{\gen{L}_5}{\gen{L}_\pm} &= \pm \gen{L}_\pm , &
      \comm{\gen{L}_+}{\gen{L}_-} &= 2 \gen{L}_5 , &
      \comm{\gen{L}_5}{\gen{Q}_{b\pm\dot{\beta}}} &= \pm\frac{1}{2} \gen{Q}_{b\pm\dot{\beta}} , &
      \comm{\gen{L}_\pm}{\gen{Q}_{b\mp\dot{\beta}}} &= \gen{Q}_{b\pm\dot{\beta}} , \\
      \comm{\gen{R}_8}{\gen{R}_\pm} &= \pm \gen{R}_\pm , &
      \comm{\gen{R}_+}{\gen{R}_-} &= 2 \gen{R}_8 , &
      \comm{\gen{R}_8}{\gen{Q}_{b\beta\pm}} &= \pm\frac{1}{2} \gen{Q}_{b\beta\pm} , &
      \comm{\gen{R}_\pm}{\gen{Q}_{b\beta\mp}} &= \gen{Q}_{b\beta\pm} ,
    \end{aligned} \\
    \begin{aligned}
      \acomm{\gen{Q}_{\pm++}}{\gen{Q}_{\pm--}} = &\pm \gen{S}_{\pm} , &
      \acomm{\gen{Q}_{\pm+-}}{\gen{Q}_{\pm-+}} = &\mp \gen{S}_{\pm} , \\
      \acomm{\gen{Q}_{+\pm+}}{\gen{Q}_{-\pm-}} = &\mp \alpha \gen{L}_{\pm} , &
      \acomm{\gen{Q}_{+\pm-}}{\gen{Q}_{-\pm+}} = &\pm \alpha \gen{L}_{\pm} , \\
      \acomm{\gen{Q}_{++\pm}}{\gen{Q}_{--\pm}} = &\mp (1-\alpha) \gen{R}_{\pm} , &
      \acomm{\gen{Q}_{+-\pm}}{\gen{Q}_{-+\pm}} = &\pm (1-\alpha) \gen{R}_{\pm} , \\
      \acomm{\gen{Q}_{+\pm\pm}}{\gen{Q}_{-\mp\mp}} = - \gen{S}_0 \pm \alpha\gen{L}_5 &\pm (1-\alpha)\gen{R}_8 , &
      \acomm{\gen{Q}_{+\pm\mp}}{\gen{Q}_{-\mp\pm}} = &+ \gen{S}_0 \mp \alpha\gen{L}_5 \pm (1-\alpha)\gen{R}_8 .
    \end{aligned}
  \end{gathered}
\end{equation}
The non-vanishing action of the $\alg{d}(2,1;\alpha)$ generators on the states of the $(-\tfrac{\alpha}{2};\tfrac{1}{2};0)$ representation is given by
\begin{equation}
  \begin{gathered}
    \begin{aligned}
      L_5 \ket{\phi_{\pm}^{(n)}} &= \pm \frac{1}{2} \ket{\phi_{\pm}^{(n)}} \,, &
      L_+ \ket{\phi_{-}^{(n)}} &= \ket{\phi_+^{(n)}} \,, &
      L_- \ket{\phi_{+}^{(n)}} &= \ket{\phi_-^{(n)}} \,, \\
      R_8 \ket{\psi_{\pm}^{(n)}} &= \pm \frac{1}{2} \ket{\psi_{\pm}^{(n)}} \,, &
      R_+ \ket{\psi_{-}^{(n)}} &= \ket{\psi_+^{(n)}} \,, &
      R_- \ket{\psi_{+}^{(n)}} &= \ket{\psi_-^{(n)}} \,,
    \end{aligned} \\
    \begin{aligned}
      S_0 \ket{\phi_{\beta}^{(n)}} &= - \left( \tfrac{\alpha}{2} + n \right) \ket{\phi_{\beta}^{(n)}} \,, &
      S_0 \ket{\psi_{\dot\beta}^{(n)}} &= - \left( \tfrac{\alpha}{2} + \tfrac{1}{2} + n \right) \ket{\psi_{\dot\beta}^{(n)}} \,, \\
      S_- \ket{\phi_{\beta}^{(n)}} &= -\sqrt{(n + \alpha)(n + 1)} \ket{\phi_{\beta}^{(n+1)}} \,, &
      S_- \ket{\psi_{\dot\beta}^{(n)}} &= -\sqrt{(n + 1+ \alpha) (n+1)} \ket{\psi_{\dot\beta}^{(n+1)}} \,, \\
      S_+ \ket{\phi_{\beta}^{(n)}} &= +\sqrt{(n - 1 + \alpha) n} \ket{\phi_{\beta}^{(n-1)}} \,, &
      S_+ \ket{\psi_{\dot\beta}^{(n)}} &= +\sqrt{(n + \alpha)n} \ket{\psi_{\dot\beta}^{(n-1)}} \,,
    \end{aligned} \\
    \begin{aligned}
      Q_{-\pm\dot\beta} \ket{\phi_{\mp}^{(n)}} &= \pm \sqrt{n+\alpha} \ket{\psi_{\dot\beta}^{(n)}} \,, &
      Q_{+\pm\dot\beta} \ket{\phi_{\mp}^{(n)}} &= \pm \sqrt{n} \ket{\psi_{\dot\beta}^{(n-1)}} \,, \\
      Q_{-\beta\pm} \ket{\psi_{\mp}^{(n)}} &= \mp \sqrt{n+1} \ket{\phi_{\beta}^{(n+1)}} \,, &
      Q_{+\beta\pm} \ket{\psi_{\mp}^{(n)}} &= \mp \sqrt{n+\alpha} \ket{\phi_{\beta}^{(n)}} \,.
    \end{aligned}
  \end{gathered}
\end{equation}
On the $(-\tfrac{1-\alpha}{2};0;\tfrac{1}{2})$ representation the generators act as
\begin{equation}
  \begin{gathered}
    \begin{aligned}
      L_5 \ket{\bar{\psi}_{\pm}^{(n)}} &= \pm \frac{1}{2} \ket{\bar{\psi}_{\pm}^{(n)}} \,, &
      L_+ \ket{\bar{\psi}_{-}^{(n)}} &= \ket{\bar{\psi}_+^{(n)}} \,, &
      L_- \ket{\bar{\psi}_{+}^{(n)}} &= \ket{\bar{\psi}_-^{(n)}} \,, \\
      R_8 \ket{\bar{\phi}_{\pm}^{(n)}} &= \pm \frac{1}{2} \ket{\bar{\phi}_{\pm}^{(n)}} \,, &
      R_+ \ket{\bar{\phi}_{-}^{(n)}} &= \ket{\bar{\phi}_+^{(n)}} \,, &
      R_- \ket{\bar{\phi}_{+}^{(n)}} &= \ket{\bar{\phi}_-^{(n)}} \,,
    \end{aligned} \\
    \begin{aligned}
      S_0 \ket{\bar{\phi}_{\dot\gamma}^{(n)}} &= - \left( \tfrac{1-\alpha}{2} + n \right) \ket{\bar{\phi}_{\dot\gamma}^{(n)}} \,, &
      S_0 \ket{\bar{\psi}_{\gamma}^{(n)}} &= - \left( \tfrac{1 - \alpha}{2} + \tfrac{1}{2} + n \right) \ket{\bar{\psi}_{\gamma}^{(n)}} \,, \\
      S_- \ket{\bar{\phi}_{\dot\gamma}^{(n)}} &= -\sqrt{(n + 1 - \alpha)(n + 1)} \ket{\bar{\phi}_{\dot\gamma}^{(n+1)}} \,, &
      S_- \ket{\bar{\psi}_{\gamma}^{(n)}} &= -\sqrt{(n + 2 - \alpha) (n+1)} \ket{\bar{\psi}_{\gamma}^{(n+1)}} \,, \\
      S_+ \ket{\bar{\phi}_{\dot\gamma}^{(n)}} &= \sqrt{(n - \alpha) n} \ket{\bar{\phi}_{\dot\gamma}^{(n-1)}} \,, & 
      S_+ \ket{\bar{\psi}_{\gamma}^{(n)}} &= \sqrt{(n + 1 - \alpha) n} \ket{\bar{\psi}_{\gamma}^{(n-1)}} \,,
    \end{aligned} \\
    \begin{aligned}
      Q_{-\gamma\pm} \ket{\bar{\phi}_{\mp}^{(n)}} &= \pm \sqrt{n+1-\alpha} \ket{\bar{\psi}_{\gamma}^{(n)}} \,, &
      Q_{+\gamma\pm} \ket{\bar{\phi}_{\mp}^{(n)}} &= \pm \sqrt{n} \ket{\bar{\psi}_{\gamma}^{(n-1)}} \,, \\
      Q_{-\pm\dot\gamma} \ket{\bar{\psi}_{\mp}^{(n)}} &= \mp \sqrt{n+1} \ket{\bar{\phi}_{\dot\gamma}^{(n+1)}} \,, &
      Q_{+\pm\dot\gamma} \ket{\bar{\psi}_{\mp}^{(n)}} &= \mp \sqrt{n+1-\alpha} \ket{\bar{\phi}_{\dot\gamma}^{(n)}} \,.
    \end{aligned}
  \end{gathered}
\end{equation}

\section{Universal R-matrix calculations for \texorpdfstring{$\alg{sl}(2)$}{sl(2)}}\label{appA}

Let us follows \cite{Arutyunov:2009ce} and evaluate the universal R-matrix starting with the $P$ module of sections \ref{Pmodulo} and \ref{moduli}, in order to display the computational details. All the other $\alg{sl}(2)$ modules will be dealt with analogously. 

\subsection{\texorpdfstring{$P$}{P} module}

We will compute the various factors of the universal R-matrix separately, and assemble the results in the main text.

\subsubsection{The factor $R_H$}
Let us start by computing the factor $R_H$, and let us consider the logarithmic derivative of the Drinfeld current, $\frac{\mathrm{d}}{\mathrm{d}\,t} \log H^+ (t)$ in the first factor of the tensor product, acting on a state $|n_1\rangle$. The strategy is to first act on single-particle states in each factor of the tensor product, and then combine the results found to obtain the action of $R_H$ on two-particle states. The way to combine these two single-particle results is provided by the residue formula (\ref{eqn;Res}). 

The action in the first factor of the tensor product is diagonal, and the eigenvalue can be expanded in power series in $t$ as follows:
\begin{eqnarray}
\label{iduo11}
\frac{\mathrm{d}}{\mathrm{d}\,t} \log H_1^+ (t) \, |n_1\rangle = \sum_{m=1}^\infty \, \big( \a_1^m + \a_2^m - \a_3^m - \a_4^m\big) \, t^{-m-1} \, |n_1\rangle
\end{eqnarray}
with
\begin{equation}
\a_1 = u_1+\frac{1}{2}, \qquad \a_2 = u_1-\frac{1}{2}, \qquad\
\a_3 = u_1+n_1+\frac{1}{2}, \qquad \a_4 = u_1+n_1-\frac{1}{2}.
\end{equation}
where $u_1$ is the Yangian evaluation parameter appearing in (\ref{d2P}).

To be able to use formula (\ref{eqn;Res}), we need now the contribution from the second single-particle factor in the tensor product. In the second factor we have the Drinfeld current $\log H^- (v + 2n +1)$ acting diagonally on a generic state $|n_2\rangle$. This can be expanded as
\begin{eqnarray}
\label{tay}
\log H^- (v+2q +1) \, |n_2\rangle = K(q) + \sum_{m=1}^\infty \, \bigg( \beta_1(q)^{-m} + \beta_2(q)^{-m} - \beta_3(q)^{-m} - \beta_4(q)^{-m}\bigg) \, \frac{v^m}{m} \, |n_2\rangle
\end{eqnarray}
with
\begin{equation}
  \begin{gathered}
    \begin{aligned}
      \beta_1(q) &= u_2+n_2-2q-\frac{1}{2}, & \beta_2(q) = u_2+n_2-2q-\frac{3}{2} \\
      \beta_3(q) &= u_2-2q-\frac{1}{2}, & \beta_4(q) = u_2-2q-\frac{3}{2}, \\
    \end{aligned} \\
    K(q) = \log \, \bigg[1+\frac{2 n_2(1+n_2)}{(1 - 2 n_2 + 4 q - 2 u_2)} + \frac{2 n_2(- 1+n_2)}{(- 3 + 2 n_2 - 4 q + 2 u_2)}\bigg].
  \end{gathered}
\end{equation}
The term $K(q)$ corresponds to the constant term in the Taylor expansion (\ref{tay}). Because of the structure of the residue formula (\ref{eqn;Res}) and of (\ref{iduo11}), $K(q)$ does not play any role in the final result (this will be true in all subsequent calculations of this appendix and of appendix \ref{appB}.

By combining these contributions together into (\ref{eqn;Res}), we get a factor 
\begin{eqnarray}
\label{cartaf}
&&\exp \left\{ \Res_{t=v}\left[
\frac{\mathrm{d}}{\mathrm{d}\,t}(\log H^+(t))\otimes { 
}  \log H^-(v+ 2n+1)\right]\right\} |n_1\rangle \otimes |n_2\rangle= \\
&&\exp \left\{\sum_{q=0}^{\infty} \, \sum_{m=1}^{\infty} \big( \a_1^m + \a_2^m - \a_3^m - \a_4^m\big)\frac{\bigg( \beta_1(q)^{-m} + \beta_2(q)^{-m} - \beta_3(q)^{-m} - \beta_4(q)^{-m}\bigg)}{m} \right\}|n_1\rangle \otimes |n_2\rangle \nonumber\\
&&= \, \frac{\Gamma(n_1 + u_1 - u_2)\Gamma(1 + n_1 + u_1 - u_2)\Gamma(- n_2 + u_1 - u_2)\Gamma(1 - n_2 + u_1 - u_2)}{\Gamma(u_1 - u_2)\Gamma(1 + u_1 - u_2)\Gamma(n_1 - n_2 + u_1 - u_2)\Gamma(1 + n_1 - n_2 + u_1 - u_2)} \, |n_1,n_2\rangle\nonumber\\
&&\qquad \qquad \qquad \qquad \qquad \qquad \qquad \qquad \qquad \qquad \qquad \qquad \qquad \qquad \qquad \equiv R_H (n_1, n_2) \, |n_1,n_2\rangle,\nonumber
\end{eqnarray} 
where we denote $|m,n\rangle \equiv |m\rangle \otimes |n\rangle$.

\subsubsection{The root factors $R_E$ and $R_F$}

Let us focus our attention on $R_E$ acting on two states $|m_1 \rangle \otimes |m_2 \rangle$. We closely follow appendix A.3 of \cite{Arutyunov:2009ce}.
One has, using (\ref{d2P}),
\begin{equation}
\label{eqn;RF21}
\prod_{n\ge 0}^{\rightarrow}\exp(- e_n \otimes f_{-n-1}) |m_1\rangle \otimes |m_2\rangle \equiv \sum_{m=0}^{m_2} B_m (m_1,m_2)\, \,  |m_1+m\rangle \otimes |m_2-m\rangle.
\end{equation}
We will suppress the dependence of $B_m$ (and of $A_m$ to be defined in (\ref{defA}) below) on $m_1$ and $m_2$ whenever it is not ambiguous. Let us define 
\begin{eqnarray}
\label{tildi}
\tilde{d} = (\frac{1}{2} + m_1 + u_1),\qquad \qquad
\tilde{c} = (- \frac{1}{2} + m_2 + u_2).
\end{eqnarray}
The term $B_m$ is built up out of $m$ copies of
$-e \otimes f$ acting on $|m_1\rangle \otimes |m_2\rangle$. On the left factor
\begin{eqnarray}
\label{string11}
e_{k_m}\ldots e_{k_2} e_{k_1} |m_1\rangle 
= \frac{(n_1+m-1)!}{(n_1-1)!} \, \, \tilde{d}^{k_1}\ldots (\tilde{d}+m-1)^{k_m}|m_1+m\rangle.
\end{eqnarray}
On the right factor
\begin{eqnarray}
\label{string1111}
f_{-k_m-1}\ldots f_{-k_2-1} f_{-k_1-1} |m_2\rangle 
= (-)^m \, \frac{(m_2+m-1)!}{(m_2-1)!} \, \, \, \, \tilde{c {}{}}{}^{-k_1-1}\ldots (\tilde{c}-m+1)^{-k_m-1}|m_2-m\rangle.\nonumber
\end{eqnarray}
From the ordered exponential
(\ref{eqn;RF21}) we have $k_i\leq k_{i-1}$. In case $k_i =
k_{i+1}$, we pick up a combinatorial
factor coming from the series of the exponential. One finds
\begin{equation}
\label{combcase}
B_m = \, \frac{(m_1+m-1)!(m_2+m-1)!}{(m_1-1)!(m_2-1)!} \sum_{k_1\geq\ldots\geq k_m} \frac{1}{N(\{k_1,\ldots,k_m\})}\frac{\tilde{d}^{k_1}}{\tilde{c}^{{}\, k_1+1}}\ldots \frac{(\tilde{d}+m-1)^{k_m}}{(\tilde{c}-m+1)^{{}\, k_m+1}}.
\end{equation}
$N$ is a combinatorial factor which is defined as 
the order of the permutation group of the set
$\{k_1,\ldots,k_m\}$. For example, $N(\{2,1,1\})=\frac{1}{2}$ and
$N(\{5,4,3,3,2,1,1,1\})=\frac{1}{3!}\frac{1}{2!}=\frac{1}{12}$. The sum
evaluates to~\cite{Arutyunov:2009ce}
\begin{equation}
B_m = (-)^m \, \frac{(m_1+m-1)!(m_2+m-1)!}{m! \, (m_1-1)!(m_2-1)!} \, \prod_{p=0}^{m-1}\frac{1}{\tilde{d}-\tilde{c}-p+m-1}.
\end{equation}

A similar computations allows to determine the contribution of $R_F$. One obtains correspondingly
\begin{gather}
\label{defA}
R_F \, |m_1, m_2\rangle \, \equiv \, \sum_{m=0}^{m_1} \, A_m (m_1,m_2)\, |m_1 - m, m_2 + m\rangle, \\
\begin{aligned}
  A_m  &= \frac{(-)^m}{m!} \, [m_1 (m_1 - 1) \dotsb (m_1 - m + 1)][(m_2 + m - 1)\dotsb(m_2 + 1) \, m_2] \times \\
  & \qquad \qquad \qquad \qquad \qquad \qquad \qquad \qquad \qquad  \times \prod_{i=0}^{m-1} \, \frac{1}{u_1 - u_2 + m_1 - m_2 - m - i}.
\end{aligned}
\end{gather}

\subsubsection{Final expressions for the $P$ module}

In summary, let us re-write the above expressions for the R-matrix in the following way. One has 
\begin{eqnarray}
R^{(P)}_F \, |m_1, m_2\rangle = \sum_{m=0}^{m_1} \, A^{(P)}_m \, |m_1 - m, m_2 + m\rangle,
\end{eqnarray}
where the superscript $(P)$ stands for $P$ module, and 
\begin{eqnarray}
&&A^{(P)}_m = \frac{(-)^m}{m!} \, [m_1 (m_1 - 1) \dotsb (m_1 - m + 1)][(m_2 + m - 1)\dotsb(m_2 + 1) \, m_2] \times\nonumber\\
&&\qquad \qquad \qquad \qquad \qquad \qquad \qquad  \, \, \, \, \times \prod_{p=0}^{m-1} \, \frac{1}{u_1 - u_2 + m_1 - m_2 - m - p} \qquad \forall \, \, m>0,\nonumber\\
&&A^{(P)}_0 = 1.
\end{eqnarray}
Then, we obtain for the Cartan part
\begin{eqnarray}
R^{(P)}_H \, |n_1, n_2\rangle = \frac{\Gamma(n_1 + u_1 - u_2)\Gamma(1 + n_1 + u_1 - u_2)\Gamma(- n_2 + u_1 - u_2)\Gamma(1 - n_2 + u_1 - u_2)}{\Gamma(u_1 - u_2)\Gamma(1 + u_1 - u_2)\Gamma(n_1 - n_2 + u_1 - u_2)\Gamma(1 + n_1 - n_2 + u_1 - u_2)} \, |n_1,n_2\rangle.\nonumber
\end{eqnarray} 
Finally, we obtain
\begin{eqnarray}
R^{(P)}_E \, |k_1, k_2\rangle = \sum_{k=0}^{k_2} \, B^{(P)}_k \, |k_1 + k, k_2 - k\rangle,
\end{eqnarray}
with
\begin{eqnarray}
&&B^{(P)}_k = \frac{(-)^k}{k!}  \, [k_2 (k_2 - 1) \dotsb (k_2 - k + 1)][(k_1 + k - 1)\dotsb(k_1 + 1) k_1] \prod_{j=0}^{k-1} \, \frac{1}{u_1 - u_2 + k_1 - k_2 + k - j}\nonumber\\
&&\qquad \qquad \qquad \qquad \qquad \qquad \qquad \qquad \qquad \qquad \qquad \qquad \qquad \qquad \qquad \qquad \forall \, \, k>0,\nonumber\\
&&B^{(P)}_0 = 1.
\end{eqnarray}
In this re-writing all terms are explicitly well defined, in particular for small $m$.

\subsection{\texorpdfstring{$S$}{S} module}
If we repeat the entire procedure for the $S$ module of section \ref{Smodulo} and \ref{moduli}, we find now 
\begin{eqnarray}
R^{(S)}_F \, |m_1, m_2\rangle = \sum_{m=0}^{m_1} \, A^{(S)}_m \, |m_1 - m, m_2 + m\rangle,
\end{eqnarray}
where the superscript $(S)$ stands for $S$ module, and 
\begin{eqnarray}
&&A^{(S)}_m = \frac{(-)^m}{m!}  \, [m_1 (m_1 - 1) \dotsb (m_1 - m + 1)][(m_1 - 1) (m_1 - 2)\dotsb(m_1 - m)] \times\nonumber\\
&&\qquad \qquad \qquad \qquad \qquad \qquad \qquad  \, \, \, \,\prod_{p=0}^{m-1} \, \frac{1}{u_1 - u_2 + m_1 - m_2 - m - p} \qquad \, \, \, \forall \, \, m>0,\nonumber\\
&&A^{(S)}_0 = 1.
\end{eqnarray}
The Cartan part turns out to be the same as for the $P$ module, since not only the level zero Cartan generator $h$ is the same across all three modules (see (\ref{azin1}), (\ref{azin2}) and (\ref{azin3})), but also all higher level Yangian partners $h_n$ (\ref{exte}) happens to be the same for the $P$,  $S$ and $R$ module by explicit computation. One has therefore  
\begin{eqnarray}
\label{eqo}
R^{(S)}_H \, |n_1, n_2\rangle = R^{(P)}_H \, |n_1, n_2\rangle .\nonumber
\end{eqnarray} 
Finally, one obtains
\begin{eqnarray}
R^{(S)}_E \, |k_1, k_2\rangle = \sum_{k=0}^{k_2} \, B^{(S)}_k \, |k_1 + k, k_2 - k\rangle,
\end{eqnarray}
with
\begin{eqnarray}
&&B^{(S)}_k = \frac{(-)^k}{k!} \, [k_2 (k_2 - 1) \dotsb (k_2 - k + 1)][(k_2 - 1) (k_2 - 2)\dotsb(k_2 - k)]  \prod_{j=0}^{k-1} \, \frac{1}{u_1 - u_2 + k_1 - k_2 + k - j}\nonumber\\
&&\qquad \qquad \qquad \qquad \qquad \qquad \qquad \qquad \qquad \qquad \qquad \qquad \qquad \qquad \qquad \qquad \forall \, \, k>0,\nonumber\\
&&B^{(S)}_0 = 1.
\end{eqnarray}

\subsection{\texorpdfstring{$R$}{R} module}
Since the $R$ module of section \ref{Rmodulo} and \ref{moduli} is the physically relevant one, et us give a few more details of the corresponding computation. 
The term $B_m$ is built up out of $m$ copies of
$-e \otimes f$ acting on $|m_1\rangle \otimes |m_2\rangle$. On the left factor
\begin{eqnarray}
\label{string11r}
&&e_{k_m}\ldots e_{k_2} e_{k_1} |m_1\rangle 
= \sqrt{\frac{(n_1+m-1)!}{(n_1-1)!}} \, \, \tilde{d}^{k_1}\ldots (\tilde{d}+m-1)^{k_m}|m_1+m\rangle,
\end{eqnarray}
where $\tilde{d}$ (and $\tilde{c}$ below) are still given by (\ref{tildi}).
On the right factor
\begin{eqnarray}
\label{string1111r}
&&f_{-k_m-1}\ldots f_{-k_2-1} f_{-k_1-1} |m_2\rangle 
= (-)^m \, \frac{(m_2+m-1)!}{(m_2-1)!} \, \sqrt{\frac{(m_2+m-2)!}{(m_2-2)!}} \times\, \nonumber\\
&&\qquad \qquad \qquad \qquad \qquad \qquad \qquad \qquad \qquad \times \, \, \tilde{c {}{}}{}^{-k_1-1}\ldots (\tilde{c}-m+1)^{-k_m-1}|m_2-m\rangle.\nonumber
\end{eqnarray}
From the ordered exponential
(\ref{eqn;RF21}) we have $k_i\leq k_{i-1}$. In case $k_i =
k_{i+1}$, we pick up a combinatorial
factor coming from the series of the exponential. One finds
\begin{eqnarray}
\label{combcaser}
&&B_m = \frac{(m_2+m-1)!}{(m_2-1)!}  \, \sqrt{\frac{(m_1+m-1)!(m_2+m-2)!}{(m_1-1)!(m_2-2)!}} \times \nonumber\\
&&\qquad \qquad \qquad \qquad \qquad \times \sum_{k_1\geq\ldots\geq k_m} \frac{1}{N(\{k_1,\ldots,k_m\})}\frac{\tilde{d}^{k_1}}{\tilde{c}^{{}\, k_1+1}}\ldots \frac{(\tilde{d}+m-1)^{k_m}}{(\tilde{c}-m+1)^{{}\, k_m+1}}.\nonumber
\end{eqnarray}
$N(\{k_1,\ldots,k_m\})$ is the same combinatorial factor discussed below (\ref{combcase}). One obtains
\begin{eqnarray}
R^{(R)}_E \, |k_1, k_2\rangle = \sum_{k=0}^{k_2} \, B^{(R)}_k \, |k_1 + k, k_2 - k\rangle,
\end{eqnarray}
with
\begin{align*}
B^{(R)}_k &= \frac{(-)^k}{k!} \, [k_2 (k_2 - 1) \dotsb (k_2 - k + 1)] \sqrt{(k_2 - 1) (k_2 - 2)\dotsb(k_2 - k)}\nonumber\\
&\qquad \qquad \qquad \times \, \sqrt{(k_1 + k - 1)\dotsb(k_1 + 1) \, k_1} \, \, \prod_{j=0}^{k-1} \, \frac{1}{u_1 - u_2 + k_1 - k_2 + k - j} \qquad \, \, \, \forall \, \, k>0,\nonumber\\
B^{(R)}_0 &= 1.
\end{align*}
Analogously, one has for the other root factor
\begin{eqnarray}
R^{(R)}_F \, |m_1, m_2\rangle = \sum_{m=0}^{m_1} \, A^{(R)}_m \, |m_1 - m, m_2 + m\rangle,
\end{eqnarray}
where the superscript $(R)$ stands for $R$ module, and 
\begin{align*}
    A^{(R)}_m  &= \frac{(-)^m}{m!}  \, [m_1 (m_1 - 1) \dotsb (m_1 - m + 1)] \sqrt{(m_1 - 1) (m_1 - 2)\dotsb(m_1 - m)} \\
    &\qquad \qquad \qquad  \times \, \sqrt{(m_2 + m - 1)\dotsb(m_2 + 1) \, m_2} \, \, \prod_{p=0}^{m-1} \, \frac{1}{u_1 - u_2 + m_1 - m_2 - m - p} \qquad \, \, \, \forall \, \, m>0, \\
  A^{(R)}_0 &= 1.
\end{align*}
The Cartan part is again the same as for the $P$ module for the reasons explained before (\ref{eqo}):
\begin{eqnarray}
R^{(R)}_H \, |n_1, n_2\rangle = R^{(P)}_H \, |n_1, n_2\rangle .\nonumber
\end{eqnarray} 

\section{Unitarity in the \texorpdfstring{$\alg{sl}(2)$}{sl(2)} case}\label{appU}

We can check unitarity of the R-matrices we have derived in section \ref{ddddd} for the $P$, $S$ and $R$ modules. The unitarity condition on states reads
\begin{equation}
R_{21} (u_2 - u_1) \, \, R (u_1 - u_2) = \mathds{1},
\end{equation}
where
\begin{equation}
R_{21} (x) = P \, R (x) \, P
\end{equation}
and $P$ is the permutation matrix implementing the permutation on states: $P \ket{a} \otimes \ket{b} = \ket{b} \otimes \ket{a}$.

It is straightforward to see that, for all three modules $P$, $S$ and $R$, the permutation on highest weight states simply acts
\begin{equation}
P \, |hw\rangle_j = (-)^j \, |hw\rangle_j 
\end{equation}
(one can simply change variable $q \to j-q$ in the summation in all three cases). Since the R-matrix acts diagonally on highest weight states, namely
\begin{equation}
R \, |hw\rangle_j = R_j \, |hw\rangle_j 
\end{equation}
one deduces that unitarity implies
\begin{equation}
R_j (u_1 - u_2) \, R_j (u_2 - u_1) = 1 \qquad \forall j,
\end{equation}
which is indeed satisfied in all three modules, see (\ref{mssv}), (\ref{eqPS}) and (\ref{eqPR}). 

We now assume that all other states in the tensor product of two modules can be obtained by acting with the $\alg{sl}(2)$ tensor product raising generator $\Delta(e) = e \otimes \mathds{1} + \mathds{1} \otimes e$ on highest weight states. If this is the case, then, since one simultaneously has that
\begin{equation}
\Delta^{op} (e) \, R =   \Delta (e) \, R = R \, \Delta (e),  \qquad 
\Delta (e) \, P = P \, \Delta (e),
\end{equation}
one concludes that unitarity on highest weight states implies unitarity on all states. In fact (denoting again $\delta u=u_1 - u_2$)
\begin{equation}
P R(-\delta u) P R(\delta u) \Delta (e) \, |hw\rangle_j = \Delta (e) \, P R(-\delta u) P R(\delta u) \, |hw\rangle_j \, = \Delta(e) |hw\rangle_j,
\end{equation}
since having $-\delta u$ instead of $\delta u$ makes no difference when it comes to $\Delta (e) \, R = R \, \Delta (e)$.

Let us also comments on the \emph{physical unitarity} of the R-matrices of section~\ref{ddddd}. This is a reality condition taking the form~\cite{Arutyunov:2009ga}
\begin{equation}\label{eq:physical-unitarity}
  R^\dag (u_1 - u_2) \, R(u_1 - u_2) = \mathds{1}.
\end{equation}
The R-matrix in the $P$ module is given in equation~\eqref{mssv} by
\begin{equation}
  R(u_1 - u_2) = \sum_{j=0}^{\infty} R_j(u_1 - u_2) \Pi_j , \qquad
  R_j(u_1 - u_2) = \prod_{k=1}^{j-1} \frac{u_1 - u_2 + k}{u_1 - u_2 - k} ,
\end{equation}
where $\Pi_j$ is a projector onto the $j$:th representation in the decomposition of $P \otimes P$ in equation~\eqref{PPirep}. As shown in section~\ref{ddddd}, the same expression holds in the $R$ and $S$ modules provided we substitute in the relevant projectors. Since the projectors are self-adjoint,~\footnote{%
The operation $\dagger$ is the adjoint operator in the $R$ module. The $P$ and $S$ modules are not unitary and there is no adoint operator compatible with the Lie algebra. In these cases $\dagger$, which sends $a\rightarrow a^\dagger$ and $a\rightarrow a^\dagger$  is still an operator on the vector space on which the Lie algebra acts; it ensures positivity of the inner product and is the $s\rightarrow 0$ limit of the adjoint operator for a ${\bf s}$ module. These properties make it a sensible choice for an adjoint operator for $S$ and $P$ modules.%
} %
the physical unitarity condition~\eqref{eq:physical-unitarity} requires the coefficients of the R-matrices to satisfy
\begin{equation}
  1 = R_j (u_1 - u_2)^\dag \, R_j(u_1 - u_2) = \prod_{k=1}^{j-1} \frac{u_1 - u_2 + k}{u_1 - u_2 - k} \frac{u_1^* - u_2^* + k}{u_1^* - u_2^* - k} .
\end{equation}
This is true provided the spectral parameters $u_1$ and $u_2$ are purely imaginary, which is consistent with $u_i$ being relate to the parameter $\mu_i$ appearing in the Bethe equations of section~\ref{sec8} by $u_i = -i \mu_i$.

\section{Universal R-matrix calculations for \texorpdfstring{$\alg{sl}(2|1)$}{sl(2|1)}}\label{appB}
In this appendix, we calculate the action of the universal R-matrix (\ref{univ1}) on the chiral $\alg{sl}(2|1)$ module (\ref{represl21-yangian}). 

\subsection{The factor \texorpdfstring{$R_H$}{RH}}
Let us start by computing the factor $R_H$, following \cite{Arutyunov:2009ce}. For example, consider the Drinfeld current $H_1^+ (t)$, taken in the first factor of the tensor product, acting on a bosonic state $|\phi_{n_1}\rangle$. It will act diagonally, with an eigenvalue which can be re-expressed as a power series in $t$ as follows:
\begin{eqnarray}
\label{iduo1}
\frac{\mathrm{d}}{\mathrm{d}\,t} \log H_1^+ (t) \, |\phi_{n_1}\rangle = \sum_{m=1}^\infty \, \big(- \a_1^m - \a_2^m + \a_3^m + \a_4^m\big) \, t^{-m-1} \, |\phi_{n_1}\rangle
\end{eqnarray}
with
\begin{eqnarray}
&&\a_1 = u_1+{n_1}+\frac{1}{4}+s_1, \qquad \a_2 = u_1+{n_1}-\frac{3}{4}+s_1,\nonumber\\
&&\a_3 = u_1-\frac{3}{4}+s_1, \qquad \a_4 = u_1+\frac{1}{4}-s_1.
\end{eqnarray}
where $u_1$ is the Yangian evaluation parameter appearing in (\ref{represl21-yangian}).
There are four of these actions, corresponding to $H_i^+ (t)$, $i=1,2$ acting on bosons and fermions, respectively. 

In the second factor of the tensor product we have for instance to compute the Drinfeld current $H_1^- (v + q)$, for a shift $q$ to be determined shortly. This can be re-expressed as
\begin{eqnarray}
\log H_1^- (v+q) \, |\phi_{n_2}\rangle = K(q) + \sum_{m=1}^\infty \, \bigg(\beta_1(q)^{-m} + \beta_2(q)^{-m} - \beta_3(q)^{-m} - \beta_4(q)^{-m}\bigg) \, \frac{v^m}{m} \, |\phi_{n_2}\rangle
\end{eqnarray}
with
\begin{eqnarray}
&&\beta_1(q) = u_2+{n_2}-q-\frac{3}{4}+s_2, \qquad \beta_2(q) = u_2+{n_2}-q+\frac{1}{4}+s_2\nonumber\\
&&\beta_3(q) = u_2-q+\frac{1}{4}-s_2, \qquad \beta_4(q) = u_2-q-\frac{3}{4}+s_2,\nonumber\\
&&K(q) = \log \, \frac{(3+4q-4s_2-4u_2)(-1+4q+4s_2-4u_2)}{(-3+4n_2-4q+4s_2+4u_2)(1+4n_2-4q+4s_2+4u_2)}.
\end{eqnarray}
We have again four combinations in the second factor of the tensor product, corresponding to $H_j^- (v+q)$, $j=1,2$ acting on bosons and fermions, respectively. 
Let us compute another one of these contributions for future convenience, for instance the Drinfeld current $H_2^- (v + q)$ on a boson, for a corresponding $q$ to be fixed below. This is re-expressed as
\begin{eqnarray}
\label{iduo2}
\log H_2 (v+q) \, |\phi_{n_2}\rangle = \tilde{K}(q) + \sum_{m=1}^\infty \, \big(- \gamma_1(q)^{-m} + \gamma_2(q)^{-m}\big) \, \frac{v^m}{m} \, |\phi_{n_2}\rangle
\end{eqnarray}
with
\begin{eqnarray}
&&\gamma_1(q) = u_2-q-\frac{1}{4}+s_2, \qquad \gamma_2(q) = u_2+{n_2}-q-\frac{1}{4}+s_2\nonumber\\
&&\tilde{K}(q) = \log \, \bigg[ 1 + \frac{4 n_2}{(-1-4q+4s_2+4u_2)} \bigg].
\end{eqnarray}

The shift $q$ depends on how these contributions are paired up in $R_H$ with the first factor via $D^{-1}_{ij}$, which reads explicitly
\begin{equation}
\label{dinverso}
D^{-1}_{ij} = 
\begin{pmatrix}
0&\frac{1}{T^{\frac{1}{2}} - T^{-\frac{1}{2}}}\\
\frac{1}{T^{\frac{1}{2}} - T^{-\frac{1}{2}}}&\frac{T^{\frac{1}{2}} + T^{-\frac{1}{2}}}{T^{\frac{1}{2}} - T^{-\frac{1}{2}}}
\end{pmatrix}.
\end{equation}
We remind that $T$ is the shift operator $T f(v) = f(v+1)$ defined in section \ref{chch}. 

We adopt the prescription of \cite{Rej:2010mu} and everywhere interpret 
\begin{eqnarray}
\frac{1}{T^{\frac{1}{2}} - T^{-\frac{1}{2}}} = - \sum_{p=0}^\infty T^{p + \frac{1}{2}}.
\end{eqnarray}
This means that, for instance, 
$D^{-1}_{12} = - \sum_{p=0}^\infty T^{p + \frac{1}{2}}$, therefore there will be an additional sum over $p$ in the exponent of $R_H$ for the term $i=1,j=2$. In each term of the sum over $p$ the factor $H_j^- (v+q)$ will have $q = p+ \frac{1}{2}$. Considering that the exponential in $R_H$ factorizes on individual contributions from the sum on $i,j$, and taking into account (\ref{eqn;Res}), (\ref{iduo1}) and (\ref{iduo2}), we get a factor 
\begin{eqnarray}
&&\exp \left\{ \Res_{t=v}\left[
\frac{\mathrm{d}}{\mathrm{d}\,t}(\log H_1^+(t))\otimes { 
} D^{-1}_{12} \log H_2^-(v)\right]\right\} |\phi_{n_1}\rangle \otimes |\phi_{n_2}\rangle= \nonumber\\
&&\exp \left\{ - \sum_{p=0}^{\infty} \, \sum_{m=1}^{\infty} \big(- \a_1^m - \a_2^m + \a_3^m + \a_4^m\big)\frac{\big(- \gamma_1(p+\frac{1}{2})^{-m} + \gamma_2(p+\frac{1}{2})^{-m}\big)}{m} \right\}|\phi_{n_1}\rangle \otimes |\phi_{n_2}\rangle = \nonumber\\
&&\frac{\Gamma[1 - n_2 - s_1 - s_2 + u_1 - u_2] \Gamma[
    n_1 + s_1 - s_2 + u_1 - u_2] }{\Gamma[1 - s_1 - s_2 + u_1 - u_2] \Gamma[
    s_1 - s_2 + u_1 - u_2] } \times \nonumber\\
&&\qquad \qquad \frac{\Gamma[
    1 + n_1 + s_1 - s_2 + u_1 - u_2] \Gamma[-n_2 + s_1 - s_2 + u_1 - 
     u_2]}{\Gamma[n_1 - n_2 + s_1 - s_2 + u_1 - u_2] \Gamma[
    1 + n_1 - n_2 + s_1 - s_2 + u_1 - u_2]} |\phi_{n_1}\rangle \otimes |\phi_{n_2}\rangle.
\end{eqnarray}
Since $D^{-1}_{11}=0$, we have only two more such factors ($i=2,j=1$ and $i=2,j=2$) acting on two bosons, to be computed similarly to the above\footnote{In (\ref{dinverso}), we consistently interpret $D^{-1}_{22}= - \sum_{p=0}^\infty T^{p + 1} - \sum_{p=0}^\infty T^p$.}. The overall product is 
\begin{eqnarray}
&&\prod_{i,j} \exp \left\{ \Res_{t=v}\left[
\frac{\mathrm{d}}{\mathrm{d}\,t} (\log H_i^+(t))\otimes { 
} D^{-1}_{ij} \log H_j^-(v)\right]\right\} |\phi_{n_1}\rangle \otimes |\phi_{n_2}\rangle= \nonumber\\
&&\frac{\Gamma[1 - n_2 - s_1 - s_2 + u_1 - u_2] \Gamma[
    1 + n_1 + s_1 - s_2 + u_1 - u_2]}{\Gamma[1 - s_1 - s_2 + u_1 - u_2] \Gamma[
    n_1 - n_2 + s_1 - s_2 + u_1 - u_2]}\times \nonumber\\
&&\qquad \qquad \frac{\Gamma[-n_2 + s_1 - s_2 + u_1 - u_2] \Gamma[
    n_1 + s_1 + s_2 + u_1 - u_2]}{\Gamma[
    1 + n_1 - n_2 + s_1 - s_2 + u_1 - u_2] \Gamma[s_1 + s_2 + u_1 - u_2])} |\phi_{n_1}\rangle \otimes |\phi_{n_2}\rangle.
\end{eqnarray}
Similarly, one calculates the contribution of $R_H$ on states with one or two fermions. 

\subsection{The root factors}
In this section, we describe the computation of one of the root factors, the other ones being dealt with in a similar fashion. Let us focus our attention on $R_1$ acting on two bosons. We again closely follow appendix A.3 of \cite{Arutyunov:2009ce}.
One has, using (\ref{represl21-yangian}),
\begin{eqnarray}\label{eqn;RF2s}
\prod_{n\ge 0}^{\rightarrow}\exp(- \xi_{1,n}^+ \otimes \xi_{1,-n-1}^-) |\phi_{n_1}\rangle \otimes |\phi_{n_2}\rangle &\equiv& \sum_{m=0}^{n_2} B_m \,  |\phi_{n_1+m}\rangle \otimes |\phi_{n_2-m}\rangle.
\end{eqnarray}
Let us define 
\begin{eqnarray}
\tilde{d} = (\frac{1}{4} + n_1 + s_1 + u_1),\qquad \qquad
\tilde{c} = (- \frac{3}{4} + n_2 + s_2 + u_2).
\end{eqnarray}
The term $B_m$ is built up out of $m$ copies of
$-\xi_1^+\otimes \xi^-_1$ acting on the state $|\phi_{n_1}\rangle \otimes |\phi_{n_2}\rangle$. One has
\begin{eqnarray}
\label{string}
\xi^+_{k_m}\ldots \xi^+_{k_2} \xi^+_{k_1} |\phi_{n_1}\rangle 
= (-)^m \, \sqrt{\frac{(n_1+m)!(n_1+2s_1+m-1)!}{(n_1)!(n_1+2s_1-1)!}} \, \, \tilde{d}^{k_1}\ldots (\tilde{d}+m-1)^{k_m}|\phi_{n_1+m}\rangle.
\end{eqnarray}
Similar expressions hold for the string of $m$ generators $\xi^-_{-n-1}$ acting correspondingly to (\ref{string}) on $|\phi_{n_2}\rangle$,
with $\tilde{c}$ instead of $\tilde{d}$, producing the
state $|\phi_{n_2-m}\rangle$. Specifically, we have
\begin{equation*}
\xi^-_{-k_m-1}\ldots \xi^-_{-k_2-1} \xi^-_{-k_1-1} |\phi_{n_2}\rangle 
= \, \sqrt{\frac{n_2!(n_2+2s_2-1)!}{(n_2-m)!(n_2+2s_2-m-1)!}} \, \, \tilde{c {}{}}{}^{-k_1-1}\ldots (\tilde{c}-m+1)^{-k_m-1}|\phi_{n_1-m}\rangle \,.
\end{equation*}
From the ordered exponential
(\ref{eqn;RF2s}) we have $k_i\leq k_{i-1}$. In case $k_i =
k_{i+1}$, we pick up a combinatorial
factor coming from the series of the exponential. We find, similarly to the $\alg{sl}(2)$ case (\ref{combcase}), 
\begin{eqnarray}
&&B_m = \sqrt{\frac{(n_1+m)! (n_1+2s_1+m-1)! \, \, n_2! (n_2+2s_2-1)!}{(n_1)! (n_1+2s_1-1)! (n_2-m)! (n_2+2s_2-m-1)!}}\nonumber\\
&&\qquad \qquad \qquad \qquad \qquad \qquad \times \sum_{k_1\geq\ldots\geq k_m} \frac{1}{N(\{k_1,\ldots,k_m\})}\frac{\tilde{d}^{k_1}}{\tilde{c}^{{}\, k_1+1}}\ldots \frac{(\tilde{d}+m-1)^{k_m}}{(\tilde{c}-m+1)^{{}\, k_m+1}}.\nonumber
\end{eqnarray}
$N$ is again a combinatorial factor which is defined as the order of the permutation group of the set
$\{k_1,\ldots,k_m\}$. For example, $N(\{2,1,1\})=\frac{1}{2}$ and
$N(\{5,4,3,3,2,1,1,1\})=\frac{1}{3!}\frac{1}{2!}=\frac{1}{12}$. The sum
evaluates at
\begin{eqnarray}
B_m  = (-)^m \, m! \sqrt{ \binom{n_1+m}{m} \binom{n_1+2s_1+m-1}{m} \binom{n_2}{m} \binom{n_2+2s_2-1}{m}}
\, \, \prod_{p=0}^{m-1}\frac{1}{\tilde{d}-\tilde{c}-p+m-1}.\nonumber
\end{eqnarray}

In a similar fashion, one needs to compute the action of all six root factors in the universal R-matrix, and then repeat again the computation for the other three combinations of states (boson-fermion, fermion-boson and fermion-fermion). For the fermionic exponentials (\textit{i.e.}, involving fermionic roots) $R_2$ and $R_{1+2}$ and barred version, the calculation is made easier by the fact that fermionic generators square to zero, therefore only the linear term survives. For instance,
\begin{eqnarray}
{ R}_2=\prod_{n\ge 0}^{\rightarrow}\exp(- \xi_{2,n}^+ \otimes \xi_{2,-n-1}^-) = \mathds{1} \otimes \mathds{1} \, - \, \sum_{n\ge 0} \, \xi_{2,n}^+ \otimes \xi_{2,-n-1}^-. 
\end{eqnarray}
The sum is then performed easily by recalling that the $n$-dependence in the generators 
(\ref{represl21-yangian}) is always of the form $a_n = \omega^n \, a$, for an appropriate $\omega$ linear function of $u$ and $s$. Therefore, one is systematically reduced to combinations like for instance the following:
\begin{eqnarray}
\sum_{n\ge 0} \, \xi_{2,n}^+ \otimes \xi_{2,-n-1}^- \, |v_1\rangle \otimes |v_2\rangle =  \frac{1}{\omega_2 - \omega_1} \, \, \xi_{2,0}^+ \otimes \xi_{2,0}^- \, |v_1\rangle \otimes |v_2\rangle .
\end{eqnarray}

When all these 24 blocks are ready, taking into account the corresponding formulas for $R_H$, one can compute the action of the universal R-matrix on arbitrary states. In the main text (see section \ref{sl21hwe}) we focus on the highest weight states (\ref{chirale}). 

\section{Jordan blocks in \texorpdfstring{$S$}{S} and \texorpdfstring{$P$}{P} module spin chains}
\label{sec:jordan-blocks}

In this appendix we demonstrate explicitly the presence of non-trivial Jordan blocks in the operators $\rho(u)$ and $\tau(u)$ in homogeneous $P$ and $S$ module spin-chains. Since these operators preserve length and excitation number we will consider their action mostly on states of length $L=3$ with three excitations. We have checked that longer states or states with more excitations have similar properties. In total there are ten states of length three with three excitations. Six of these states are descendants of states with fewer excitations, and the other four states are highest weight states. We will include in our discussion $R$ module spin chains as well since this will contrast the reducible and indecomposable cases.

The eigenvalues of the operator $\rho(u)$ in this sector are the same in all three $s=0$ modules, namely $2u^3$, $2(u^3-u)$ and $2(u^3-3u)$. The first two of these eigenvalues both have multiplicity three and correspond to the descendant states. The third eigenvalue has \emph{algebraic} multiplicity four. However, as we will see below, in the $P$ and $S$ modules there only exists three corresponding eigenvectors. Hence the eigenvalues has \emph{geometric} multiplicity three and the Jordan normal form of the operator is non-diagonal.

\subsection{\texorpdfstring{$P$}{P} module}

We introduce the following base of states with $L=3$ and containing three excitations:
\begin{equation}
  \begin{aligned}
    \ket{\psi_1}_P &= \ket{300} + \ket{030} + \ket{003} , \\
    \ket{\psi_2}_P &= \ket{300} - \ket{003} , \\
    \ket{\psi_3}_P &= \ket{030} - \ket{003} , \\
    \ket{\psi_4}_P &= 2 (\ket{003} + \ket{030} + \ket{300}) - (\ket{012} + \ket{120} + \ket{201}) - (\ket{210} + \ket{102} + \ket{021}) , \\
    \ket{\psi_5}_P &= (\ket{003} - 2 \ket{030} + \ket{300}) + (\ket{012} - 2 \ket{201} + \ket{120}) + (\ket{210} - 2 \ket{102} + \ket{021}) , \\
    \ket{\psi_6}_P &= (\ket{003} - \ket{300}) + (\ket{120} + \ket{210}) - (\ket{012} + \ket{021}) , \\
    \ket{\psi_7}_P &= ((\ket{012} + \ket{120} + \ket{201}) - (\ket{210} + \ket{102} + \ket{021})) , \\
    \ket{\psi_8}_P &= 3 (\ket{012} - \ket{021}) - 3 (\ket{120} - \ket{210}) - (\ket{300} - 2 \ket{030} + \ket{003}) , \\
    \ket{\psi_9}_P &= (\ket{012} - \ket{021}) + (\ket{120} - \ket{210}) - 2 (\ket{201} - \ket{102}) + \ket{300} - \ket{003} , \\ 
    \ket{\psi_{10}}_P &= 3 (\ket{012} + \ket{120} + \ket{201}) + 3 (\ket{210} + \ket{102} + \ket{021}) - 12 \ket{111} 
      \\&\qquad 
      - 2 (\ket{300} + \ket{030} + \ket{003}) , \\
  \end{aligned}
\end{equation}
The first nine state satisfies the equations
\begin{equation}
  \begin{aligned}
    \big(\rho_P-2u^3\big) \ket{\psi_k}_P &= 0 , & k &= 1,2,3, \\
    \big(\rho_P-2(u^3-u)\big) \ket{\psi_k}_P &= 0 , & k &= 4,5,6, \\
    \big(\rho_P-2(u^3-3u)\big) \ket{\psi_k}_P &= 0 , & k &= 7,8,9.
  \end{aligned}
\end{equation}
The state $\ket{\psi_{10}}_P$ satisfies
\begin{equation}
  \big(\rho_P-2(u^3-3u)\big) \ket{\psi_{10}}_P = 12i\,\ket{\psi_7}_P .
\end{equation}
Hence, $\ket{\psi_{10}}_P$ is not an eigenstate of $\rho_P(u)$. However, it does satisfy
\begin{equation}
  \big(\rho_P-2(u^3-3u)\big) \big(\rho_P-2(u^3-3u)\big) \ket{\psi_{10}}_P = 12i \big(\rho_P-2(u^3-3u)\big) \ket{\psi_7}_P = 0.
\end{equation}
This shows that the operator $\rho_P(u)$ only has nine eigenvectors, while
$\big(\rho_P-2(u^3-3u)\big)^2$ has ten and hence is completely
diagonalisable in the sector we consider here. In other words, $\rho_P(u)$
has a non-trivial Jordan normal form.

When we add more excitations we find that more eigenvectors are missing. For
example, for $L=3$ and five excitations there is a new eigenvalue $2(u^3 -
10u)$ with arithmetic multiplicity four, but with only three
excitations. The same is true for larger $L$ -- with three excitations at $L=4$
we find that the Jordan normal form of $\rho_P(u)$ has three non-zero
off-diagonal entries.

\paragraph{The $\tau(0)$ operator.}

In the $P$ module the operator $\tau(0)$ acts on any state except $\ket{000}$
as\footnote{%
  When acting with $\tau(0)$ on $\ket{000}$ there is a contribution from all
  states appearing in the trace, which leads to an infinite sum that needs to be
  regularized.%
  }%
\begin{equation}
  \begin{split}
    \tau(0) \ket{mnk} = \ket{mnk} - \ket{nkm} & + \ket{0k(m+n)} + \ket{(n+k)0m} + \ket{n(k+m)0} \\&
    - \ket{(m+n+k)00} - \ket{0(m+n+k)0} - \ket{00(m+n+k)} .
  \end{split}
\end{equation}
In particular we have
\begin{equation}
  \begin{aligned}
    \tau(0) \ket{\psi_k}_P &= 0 , & k&= 1,2,3 , \\
    \tau(0) \ket{\psi_k}_P &= 2\ket{\psi_k}_P , & k&= 4,5,6 , \\
    \tau(0) \ket{\psi_k}_P &= 0 , & k&= 7,8,9,
  \end{aligned}
\end{equation}
and
\begin{equation}
  \tau(0) \ket{\psi_{10}}_P = -6 \ket{\psi_7}_P .
\end{equation}
Hence $\tau(0)$ has the same non-trivial Jordan normal form as $\rho_P(u)$.

\subsection{\texorpdfstring{$S$}{S} module}

The action of $\rho_S(u)$ on simple state of the $S$ module is very similar
to the case of the $P$ module above. In this case we find the
eigenvectors\footnote{%
  The states $\ket{\psi_k}_S$ are chosen so that ${}_S\braket{\psi_k|\psi_m}_P =
  \delta_{km}$, where the conjugate states are given by the adjoint action $a
  \to a^\dag$. Under this action the $P$ and $S$ modules are exchanged. Hence the
  conjugate of the generalized eigenvectors of the operator $\rho_S$ when acting
  to the right are generalized eigenvectors of the left action of $\rho_P$.%
}%
\begin{equation}
  \begin{aligned}
    \ket{\psi_1}_S &= +\frac{1}{18} \big((\ket{300} + \ket{030} + \ket{003}) + 3 (\ket{012} + \ket{120} + \ket{201}) \\ & \qquad\qquad
    + 3(\ket{021} + \ket{210} + \ket{102}) + 6 \ket{111}\big) , \\
    \ket{\psi_2}_S &= +\frac{1}{18} \big((2\ket{300} - \ket{030} - \ket{003}) - 3 (\ket{012} - \ket{210} + \ket{021} - \ket{201})\big) , \\
    \ket{\psi_3}_S &= +\frac{1}{18} \big((\ket{300} + \ket{030} - 2\ket{003}) + 3 (\ket{120} - \ket{102} + \ket{210} - \ket{120})\big) , \\
    \ket{\psi_4}_S &= -\frac{1}{12} \big(\ket{012} + \ket{021} + \ket{102} + 3 \ket{111} + \ket{120} + \ket{201} + \ket{210}\big) , \\
    \ket{\psi_5}_S &= +\frac{1}{24} \big(\ket{012} + \ket{021} - 2 \ket{201} - 2 \ket{102} + \ket{120} + \ket{210}\big) , \\
    \ket{\psi_6}_S &= -\frac{1}{8} \big((\ket{012} + \ket{021}) - (\ket{120} + \ket{210})\big) , \\
    \ket{\psi_7}_S &= +\frac{1}{12} \big((\ket{012} - \ket{021}) - (\ket{120} - \ket{210})\big) , \\
    \ket{\psi_8}_S &= +\frac{1}{24} \big((\ket{012} - \ket{021}) + (\ket{120} - \ket{210}) - 2 (\ket{201} - \ket{102})\big) , \\
    \ket{\psi_9}_S &= +\frac{1}{24} \big((\ket{012} + \ket{120} + \ket{201}) - (\ket{210} + \ket{102} + \ket{021})\big) , \\
    \ket{\psi_{10}}_S &= -\frac{1}{12} \ket{111} , 
  \end{aligned}
\end{equation}
which satisfy
\begin{equation}
  \begin{aligned}
    \big(\rho_S-2u^3\big) \ket{\psi_k}_S &= 0 , & k &= 1,2,3, \\
    \big(\rho_S-2(u^3-u)\big) \ket{\psi_k}_S &= 0 , & k &= 4,5,6, \\
    \big(\rho_S-2(u^3-3u)\big) \ket{\psi_k}_S &= 0 , & k &= 8,9,10, \\
    \big(\rho_S-2(u^3-3u)\big) \ket{\psi_7}_S &= 12i\,\ket{\psi_{10}}_S .
  \end{aligned}
\end{equation}
Again we find that $\rho_S(u)$ cannot be fully diagonalized.

The action of $\tau(0)$ in the $S$ module is a bit more complicated. However, from
the relation between the $P$~and $S$ modules under the adjoint operation $a \to
a^\dag$, we see the $\tau(0)$ in the $S$ module has a non-trivial Jordan normal
form when acting to the left -- the corresponding generalized eigenstates are
the states ${}_P\bra{\psi_k}$ whose coefficients were given above.

\subsection{\texorpdfstring{$R$}{R} module}

As a base for the three-excitation $L=3$ states of the $R$ module we use
\begin{equation}
  \begin{aligned}
    \ket{\psi_1}_R &= \ket{003} , \\
    \ket{\psi_2}_R &= \ket{030} , \\
    \ket{\psi_3}_R &= \ket{300} , \\
    \ket{\psi_4}_R &= \ket{012} + \ket{021} , \\
    \ket{\psi_5}_R &= \ket{102} + \ket{201} , \\
    \ket{\psi_6}_R &= \ket{120} + \ket{210} , \\
    \ket{\psi_7}_R &= \ket{111} , \\
    \ket{\psi_8}_R &= \ket{012} - \ket{021} , \\
    \ket{\psi_9}_R &= \ket{102} - \ket{201} , \\
    \ket{\psi_{10}}_R &= \ket{120} - \ket{210} .
  \end{aligned}
\end{equation}
These states satisfy
\begin{equation}
  \begin{aligned}
    \big(\rho_R - 2u^3\big) \ket{\psi_k}_R &= 0 , & k &= 1,2,3, \\
    \big(\rho_R - 2(u^3-u)\big) \ket{\psi_k}_R &= 0 , & k &= 4,5,6, \\
    \big(\rho_R - 2(u^3-3u)\big) \ket{\psi_k}_R &= 0 , & k &= 7,8,9,10,
  \end{aligned}
\end{equation}
which shows that $\rho_R(u)$ in this sector is completely diagonalizable. In
the $R$ module, $\rho_R(u)$ is Hermitian under the adjoint action acting on
the oscillators $a \to a^\dag$, and hence has a complete basis of eigenvectors.

\section{Some comments on \texorpdfstring{$P$}{P} and \texorpdfstring{$S$}{S} module integrable spin-chains}

In this appendix we collect some formulae relating to homogeneous spin-chains with sites in the $P$ and $S$ representations. Despite having some peculiar properties discussed in section~\ref{sec71}, one may calculate many quantities in these chains, precisely because the R-matrix is explicitly known. Indeed the $P$ module homogeneous spin-chain is well known in the QCD literature~\cite{Faddeev:1994zg,Korchemsky:1994um}. In the first subsection below we discuss the application of the ABA to the $S$ module spin-chain. In the second subsection we present an example of an explicit calculation of the transfer matrix on specific states in the $P$ module. In this appendix we wanted to illustrate that $P$ and $S$ module spin-chains can be investigated quite explicitly using for example the projection operators we presented in appendix~\ref{sec:sl2-projectors} or ABA techniques, but that these spin-chains do have some unusual properties which deserve further study.

\subsection{The ABA for the \texorpdfstring{$S$}{S} module}
\label{sec:ABA-S-module}

The $S$ module is similar to the $R$ module, in that it has two highest weight states, the lower one of which generates the whole module. We can therefore try to apply the ABA directly, like we did in section~\ref{sec:ABA-R-module} for the $R$ module case. Since the basic formulas are expressed in terms of the $\alg{sl}(2)$ generators and their eigenvalues, they take the same form for the $S$ and $R$ modules. In particular we consider a Lax operator $\tilde{L}(\mu)$ that is given by
\begin{equation}
  \tilde{L}_k(\mu) \equiv L_k(\mu) + \ell(\mu) \Pi_k^0 \,,
\end{equation}
and a corresponding transfer matrix $\tilde{\rho}(\mu) = A(\mu) + D(\mu)$, as well as raising and lowering operators $B(\mu)$ and $C(\mu)$.

The resulting Bethe equations take the same for as in~\eqref{BER},
\begin{equation}\label{BES}
\left(\frac{\mu-i}{\mu+i}\right)^{\tilde{n}}=
\prod_{k\neq j}^l\frac{\mu_j-\mu_k+i}{\mu_j-\mu_k-i}\,.
\end{equation}

However, there is an important difference in the spectrum obtained in the $R$ and $S$ modules. To see this, let us consider the case of a spin-chain with three sites, $J=3$, and a groundstate containing one singlet, so that $n=1$ and $\tilde{n}=2$. For concreteness we will take this groundstate to be $\ket{011}$. We now add a single excitation to this site. The Bethe equation~\eqref{BES} have a single solution for $\tilde{n}=2$ and $K=1$, sitting at $\mu = 0$. To obtain the corresponding eigenstates we act with the raising operator $B(\mu=0)$. In the $R$ module this gives
\begin{equation}
  B_{R}(0) \ket{011} = \ell(\mu) ( \ket{012} - \ket{021} ) \,,
\end{equation}
while we for the $S$ module get
\begin{equation}
  B_{S}(0) \ket{011} = i \ket{111} - \ell(\mu) ( \ket{012} - \ket{021} ) \,.
\end{equation}
Hence the regularizing factor $\ell(\mu)$ only appears as an overall normalization in the $R$ module. After adjusting the normalization of the excited state, we can therefore set $\ell(\mu)=0$ to obtain the spectrum of the original Lax operator.

However, in the $S$ module, the obtained spectrum actually depends on $\ell(\mu)$. In particular, if we set $\ell(\mu) = 0$ in the above state we obtain the state $\ket{111}$. But this is one of the groundstates of the model, and not an excited state. Hence, the spectrum of the regularized Lax operator does not fully describe the spectrum of the original operator in the $S$ module.

\subsection{The monodromy matrix for the \texorpdfstring{$P$}{P} module}

In this subsection we will show that, using the projection operators defined in appendix~\ref{sec:sl2-projectors}, one can compute the monodromy matrix for $P$ module spin-chain. Since the monodromy matrix is one of the key ingredients of an integrable spin-chain, this calculation demonstrates that the $P$ module spin chain is a sensible integrable system, though with some unconventional properties. We consider the action of the monodromy matrix on simple states of length 2 with n excitations at either the first or second site
\begin{equation}
\left|n,0\right>_{12}\,,\qquad\mbox{and}\qquad
\left|0,n\right>_{12}\,,
\end{equation}
and we will evaluate the monodromy matrix $\tau(u)$ on these to establish whether they are degenerate. Throughout this subsection we will always take $n>0$.
From a representation-theory point of view it is natural to expect the following linear combinations to be eigenstates of $\tau(u)$
\begin{equation}
\left|n^\pm\right>_{12}\equiv\left|n,0\right>_{12}\pm\left|0,n\right>_{12}\,.
\label{rep}
\end{equation}
It is easy to see, using, for example, the projectors we have constructed previously, that $\left|n^+\right>_{12}$ belongs to the $P$ module and $\left|n^-\right>_{12}$ belongs to the ${\bf -1}$ module in the decomposition of $P\otimes P$ given in equation~(\ref{PPirep}).

\subsubsection{The $\left|n,0\right>_{12}$ state}

We wish to compute
\begin{equation}
\tau(u)\left|n,0\right>_{12}=
\tr_0(R_{01}(u)R_{02}(u))\left|n,0\right>_{12}
=\sum_{m=0}^\infty\frac{1}{m!}{}_0\!\left<m\right|
R_{01}(u)R_{02}(u)\left|m,n,0\right>_{012}\,.
\end{equation}
The R-matrix can be expressed as a sum of projectors
\begin{equation}
R_{ij}(u)=R^P(u)\sum_{k=0}^\infty \left|e^P_k\right>_{ij}{}_{ij}\!\left<e^P_k\right|+
\sum_{l=1}^\infty R^l(u)\sum_{k=0}^\infty \left|e^{{\bf -l}}_k\right>_{ij}{}_{ij}\!\left<e^{{\bf -l}}_k\right|\,,
\end{equation}
with explicit expressions for the projectors given in section~\ref{PPproj}, and the $u$ dependent coefficients can be read off from the universal $\alg{sl}(2)$ R-matrix evaluated on highest weight states with $J=0,-1,-2,\dots$
\begin{equation}
R^l(u)=
\frac{\Gamma(iu)\Gamma(iu+1)}{\Gamma(iu+l)\Gamma(iu-l+1)}\,,\qquad\mbox{and}\qquad R^P(u)=1\,,
\label{PRmatcoeffs}
\end{equation}
where $l=1,2,\dots$ and we have set the overall normalisation of the R-matrix $f(u)=1$, since it will not play a role in the discussion here.
Notice in particular that after some rescalings of $u$ this is the same as the $\alpha\rightarrow 0$ limit of the homogeneous part of the R-matrix proposed in~\cite{OhlssonSax:2011ms}, and also that $R^P(u)\equiv 1\equiv R^1(u)$, as is the case in~\cite{OhlssonSax:2011ms}.

For notational convenience let us define
\begin{equation}
\tau^{(n,0)}_m(u)\equiv
\frac{1}{m!}{}_0\!\left<m\right|
R_{01}(u)R_{02}(u)\left|m,n,0\right>_{012}\,.
\end{equation}
For $m=0$ we have
\begin{eqnarray}
\tau^{(n,0)}_0(u)&\equiv&
{}_0\!\left<0\right|
R_{01}(u)R_{02}(u)\left|0,n,0\right>_{012}
\nonumber \\
&=&\left|n,0\right>_{12}\,.
\end{eqnarray}
In doing this computation it is useful to note that states of the form $\left|k,0\right>_{ij}$ (for any $k>0$) are only found in the $P$ and ${\bf -1}$ modules in the decomposition~(\ref{PPirep}) and since $1=R^P(u)=R^{{\bf 1}}(u)$ we have for example
\begin{equation}
R_{ij}(u)\left|k,0\right>_{ij}=\left|k,0\right>_{ij}\,.
\label{simpract}
\end{equation}
Let us now evaluate the $m>0$ contributions
\begin{eqnarray}
m!\tau^{(n,0)}_m(u)&\equiv&
{}_0\!\left<m\right|
R_{01}(u)R_{02}(u)\left|m,n,0\right>_{012}
\nonumber \\
&=&
{}_0\!\left<m\right|R_{01}(u)
\left[R^P(u) \sum_{k=0}^\infty\left|e^P_k\right>_{02}{}_{02}\!\left<e^P_k\right|
+
\sum_{l=1}^\infty R^l(u)\sum_{k=0}^\infty \left|e^{{\bf -l}}_k\right>_{02}{}_{02}\!\left<e^{{\bf -l}}_k\right|
\right]\left|m,n,0\right>_{012}\nonumber \\
&=&
{}_0\!\left<m\right|R_{01}(u)
\left[R^P(u) \sum_{k=1}^\infty v_{P,k}(-1)^k(
\left|k,0\right>_{02}+\left|0,k\right>_{02}){}_{02}\!\left<k,0\right|
\right.\nonumber \\
&&\qquad\qquad\left.
+
R^1(u)\sum_{k=0}^\infty k!v_{1,k}(-1)^k
(\left|k+1,0\right>_{02}-\left|0,k+1\right>_{02}{}_{02}\!\left<k+1,0\right|
\right]\left|m,n,0\right>_{012}\nonumber \\
&=& 
\sum_{l=2}^\infty R^l(u)\sum_{k=0}^\infty
{}_0\!\left<m
 \left|e^{{\bf -l}}_k\right.\right>_{01}{}_{01}\!\left<\left.e^{{\bf -l}}_k\right|m,n,0\right>_{012}\,.
\label{taum}
\end{eqnarray}
Let us evaluate the two matrix elements on the line above. Remembering that $l=2,3,\dots$, one of the matrix element gives
\begin{eqnarray}
{}_{01}\!\left<\left.e^{{\bf -l}}_k\right|m,n,0\right>_{012}&=&
v_{l,k}\sum_{u=1}^{l-1}\sum_{t=0}^k\frac{u(l-u)}{l!}(-1)^{k+l-u}
\left(\begin{matrix}l \\ u \end{matrix}\right)^2
\left(\begin{matrix}k \\ t \end{matrix}\right)
{}_{01}\!\left<u+t,l+k-u-t\right|
\left.m,n,0\right>_{012}
\nonumber \\
&=&v_{l,k}m!n!\delta_{n+m,l+k}\sum_{u=1}^{l-1}\frac{u(l-u)}{l!}(-1)^{k+l-u}
\left(\begin{matrix}l \\ u \end{matrix}\right)^2
\left(\begin{matrix}k \\ m-u \end{matrix}\right)
\ket{0}_{2}\,.
\label{ma1}
\end{eqnarray}
This sum can be expressed in terms of regularized hypergeometric functions ${}_3FR_2$. Above, we use the generalized convention for binomial coefficients which extends the conventional to negative and zero values as follows
\begin{eqnarray}
\left(\begin{matrix}0 \\ k \end{matrix}\right)&=&\delta_{0,u}\,,\qquad\mbox{for }k\in{\bf Z}
\nonumber \\
\left(\begin{matrix} r \\ u \end{matrix}\right)&=&0\,,\qquad
\mbox{for }r\in{\bf N}\mbox{ and }u\in{\bf Z}\,,\,u<0\,.
\end{eqnarray}
Next,  we turn to the other matrix element. We will evaluate the other matrix element in the case where $m+n=k+l$, as required by the delta function on the last line of equation~(\ref{ma1}). After some algebra one finds
\begin{eqnarray}
{}_0\!\left<m
 \left|e^{{\bf -l}}_{m+n-l}\right.\right>_{01}
&=&m!\left[\sum_{s=1}^l
\left(\begin{matrix}m+n-l \\ m-s \end{matrix}\right)
\left(\begin{matrix}l \\ s \end{matrix}\right)(-1)^{l+s}
(s)^{(m-s)}(l-s)^{(n-l+s)}\right]
\left|n\right>_{1}
\nonumber \\
\label{ma2}
\end{eqnarray}
Above we have used the Pochhammer symbol
\begin{equation}
(s)^{(r)}\equiv s(s+1)\dots(s+r-1)=\frac{\Gamma(r+s)}{\Gamma(s)}\,.
\end{equation}
We can now insert the expressions we found for the two matrix elements back into equation~(\ref{taum})
\begin{eqnarray}
\tau^{(n,0)}_m(u)&=& 
\sum_{l=2}^\infty \frac{R^l(u)}{m!}\sum_{k=0}^\infty
{}_0\!\left<m
 \left|e^{{\bf -l}}_k\right.\right>_{01}
{}_{01}\!\left<\left.e^{{\bf -l}}_k\right|m,n,0\right>_{012}
\nonumber \\
&=& 
m!n!\left(\sum_{l=2}^{m+n} R^l(u)v_{l,m+n-l}
\left[\sum_{s=1}^l
\left(\begin{matrix}m+n-l \\ m-s \end{matrix}\right)
\left(\begin{matrix}l \\ s \end{matrix}\right)(-1)^{l+s}
(s)^{(m-s)}(l-s)^{(n-l+s)}\right]
\right.\nonumber \\ &&\qquad\qquad\qquad\times\left.
\left[
\sum_{u=1}^{l-1}\frac{u(l-u)}{l!}(-1)^{m+n-u}
\left(\begin{matrix}l \\ u \end{matrix}\right)^2
\left(\begin{matrix}m+n-l \\ m-u \end{matrix}\right)\right]\right)
\left|n,0\right>_{12}\,.
\nonumber \\
\end{eqnarray}
Finally, we can then write the complete action of the monodromy matrix on the state $\left|n,0\right>_{12}$
\begin{eqnarray}
\tau(u)\left|n,0\right>_{12}&=&
\sum_{m=0}^\infty\tau^{(n,0)}_m(u)
\nonumber \\
&=&
\left(1+n!\sum_{m=1}^\infty m!\sum_{l=2}^{m+n} R^l(u)v_{l,m+n-l}
\right.\nonumber \\
&&\qquad\qquad\left.\times
\left[\sum_{s=1}^l
\left(\begin{matrix}m+n-l \\ m-s \end{matrix}\right)
\left(\begin{matrix}l \\ s \end{matrix}\right)(-1)^{l+s}
(s)^{(m-s)}(l-s)^{(n-l+s)}\right]
\right.\nonumber \\ &&\qquad\qquad\qquad\times\left.
\left[
\sum_{u=1}^{l-1}\frac{u(l-u)}{l!}(-1)^{m+n-u}
\left(\begin{matrix}l \\ u \end{matrix}\right)^2
\left(\begin{matrix}m+n-l \\ m-u \end{matrix}\right)\right]\right)
\left|n,0\right>_{12}\,.
\nonumber \\
&=&
\Bigl(1+n!(n-1)!\sum_{m=1}^\infty m\sum_{l=2}^{m+n} R^l(u)\frac{l(2l-1)\Gamma(1-l+m+n)}{\Gamma(l+m+n)}
\nonumber \\ 
&&\qquad\qquad
\left[{}_3FR_2(1-l,1-l,1-m;1,2-l+n;1)
\right.\nonumber \\  && \qquad\qquad\qquad\left.
-l{}_3FR_2(1-l,1-l,1-m;2,2-l+n;1)\right]
\nonumber \\ 
&&\qquad\qquad
\times
{}_3FR_2(1-l,2-l,1-m;2,2-l+n;1)
\Bigr)
\left|n,0\right>_{12}\,.
\nonumber \\
\end{eqnarray}
Above, the regularized hypergeometric function $FR$ is generically defined (typically) in terms of the ordinary hypergeometric function $F$ as
\begin{equation}
{}_3FR_2(a_1,a_2,a_3;b_1,b_2;z)=
\frac{{}_3F_2(a_1,a_2,a_3;b_1,b_2;z)}{\Gamma(b_1)\Gamma(b_2)}\,,
\end{equation}
but is better behaved than the right-hand side when some of the arguments are zero or negative integers. The sum over $m$ above is finite, showing that, at least on such states, the monodromy matrix is a well-behaved object. Secondly, we see that  the state $\left|n,0\right>_{12}$ does not mix with other states and is an eigenvector of the monodromy matrix $\tau(u)$. A similar calculation yields
\begin{eqnarray}
\tau(u)\left|0,n\right>_{12}&=&
\sum_{m=0}^\infty\tau^{(0,n)}_m(u)
\nonumber \\
&=&
\left(1+n!\sum_{m=1}^\infty m!\sum_{l=2}^{m+n} R^l(u)v_{l,m+n-l}\right.
\nonumber \\ &&
\qquad\qquad \times\left.
\left[\sum_{s=1}^l
\left(\begin{matrix}m+n-l \\ m-s \end{matrix}\right)
\left(\begin{matrix}l \\ s \end{matrix}\right)(-1)^{l+s}
(s)^{(m-s)}(l-s)^{(n-l+s)}\right]
\right.\nonumber \\ &&\qquad\qquad\qquad\times\left.
\left[
\sum_{u=1}^{l-1}\frac{u(l-u)}{l!}(-1)^{m+n-u}
\left(\begin{matrix}l \\ u \end{matrix}\right)^2
\left(\begin{matrix}m+n-l \\ m-u \end{matrix}\right)\right]\right)
\left|0,n\right>_{12}\,.
\nonumber \\
\end{eqnarray}

\section{Projectors for tensor products of \texorpdfstring{$s=0$ $\alg{sl}(2)$}{s=0 sl(2)} representations}
\label{sec:sl2-projectors}

In this section we collect some results on projection operators relevant to our spin-chains.
As we have discussed in section~\ref{inof}, there are three inequivalent $\alg{sl}(2)$ representations at $s=0$. We will construct them using an oscillator basis
\begin{equation}
\left[ a\,,\,\ad\right]=1\,,
\end{equation}
and the Hilbert space whose orthogonal basis is given by
\begin{equation}
\left| n\right> \equiv \ad{}^n\ket{0}\,,\qquad n=0,1,2,3\dots\,,
\end{equation}
together with the inner product
\begin{equation}
\left<n\right|\left|m\right>\equiv \delta_{n,m}n!\,.
\label{hinner}
\end{equation}
The $\alg{sl}(2)$ representations are given explicitly in equations~(\ref{Pmod}),~(\ref{gens2}) and~(\ref{uno}). The $R$ representation is a reducible representation which is a direct sum of an $s=0$ and $s=-1$ representation, there are two highest weight states: $\ket{0}$ and $\left|1\right>$, which respectively generate the $s=0$ (trivial) and $s=-1$ modules by the action of $R^-$
\begin{equation}
R^-\left|{0}\right>=0\,,\qquad
(R^-)^n\left|1\right>= \sqrt{n!}\left|n+1\right>\,.
\end{equation}
 The $P$ representation contains an indecomposable $s=0$ (trivial) sub-representation; the $P$ representation has a single highest weight state $\ket{0}$ which generates the trivial sub-representation; all other states in the $P$ representation are generated by acting with $P^-$ on the state $\left|1\right>$
\begin{equation}
P^-\left|{0}\right>=0\,,\qquad
(P^-)^n\left|1\right>= n!\left|n+1\right>\,.
\end{equation}
The $S$ representation has two highest weight states $\ket{0}$ and $\left|1\right>$, and the whole module can be generated by acting on $\ket{0}$ with $S^-$
\begin{equation}
(S^-)^n \left|{0}\right>=(-1)^n \left|n\right>\,.
\end{equation}

\subsection{Projectors onto irreps of \texorpdfstring{$P\otimes P$}{P x P}, \texorpdfstring{$S\otimes S$}{S x S} and \texorpdfstring{$R\otimes R$}{R x R} representations}

A central ingredient of the integrability machinery is played by projectors onto irreps in the tensor product of the $s=0$ irreps $P$, $R$ and $S$. The decomposition of these tensor products is given in equations~(\ref{PPirep}),~(\ref{SSirep}) and~(\ref{RRirep}).
The highest weight states of the irreps in these decompositions are mostly given in equations~(\ref{pphw}),~(\ref{shw}) and~(\ref{rhw}), as well as equations near these for some simple special cases. In this subsection we construct the projectors onto these irreps. As discussed in section~\ref{sec23}, the tensor product representations decompose into direct sums, as a result finding orthonormal projectors is straightforward given that the vector-space underlying the modules has a positive-definite inner product given in equation~(\ref{hinner}). Recall that for a vector space 
$V\cong V_1\oplus V_2$, an orthonormal basis of $V$ can be found 
which is of the form
\begin{equation}
V\cong\mbox{span}\left\{\left|e_i^{V_1}\right>\right\}\oplus \mbox{span}\left\{\left|f_j^{V_2}\right>\right\}\,,
\end{equation}
where the $e^{V_1}_i$ and $f^{V_2}_j$ form a basis of $V_1$ and $V_2$, respectively. In that case the projector onto  $V_1$ takes the form
\begin{equation}
P^{V_1}\equiv \sum_i\left|e^{V_1}_i\right>
\left<e^{V_1}_i\right|\,.
\label{genprojform}
\end{equation}
The bra $\left<e^{V_1}_i\right|$ is defined as the state in the dual vector space which satisfies
\begin{equation}
\left<e^{V_1}_{i_1}\right|\left|e^{V_1}_{i_2}\right>=
\delta_{i_1,i_2}\,,\qquad\mbox{and}\qquad
\left<e^{V_1}_i\right|\left|f^{V_2}_j\right>=0\,.
\end{equation}
From this we see that we can use the results of the previous subsection to write down the states $\left|e^{V_1}_i\right>$. To find the  projectors onto the various irreps we will then need to find the states $\left<e^{V_1}_i\right|$ which satisfy the above orthonormality relations. We proceed to do this for our different $s=0$ modules presently. This procedure amounts to working out the relevant Clebsch-Gordan coefficients.

\subsubsection{Projectors onto irreps of $P\otimes P$}
\label{PPproj}

Let us construct the basis vectors $\left|e^{V_1}_i\right>$ for the irreps on the right-hand side of equation~(\ref{PPirep}). For the $P$ module we have
\begin{equation}
\left\{\left|e^{P}_i\right>\right\}\equiv 
\left\{\ket{0}_1\otimes \ket{0}_2, 
\left|k\right>_1\otimes \ket{0}_2+\ket{0}_1\otimes \left|k\right>_2
\right\}\,,
\end{equation}
where $k=1,2,\dots$. Above we have used 
\begin{equation}
(P^-_1+P^-_2)^k(\left|1\right>_1\otimes \ket{0}_2+\ket{0}_1\otimes \left|1\right>_2)=k!(\left|k+1\right>_1\otimes \ket{0}_2+\ket{0}_1\otimes \left|k+1\right>_2)\,.
\end{equation}
For the ${\bf -l}$ modules on the right-hand side of equation~(\ref{PPirep})
the basis vectors are
\begin{equation}
\left\{\left|e^{{\bf -l}}_i\right>\right\}\equiv \left\{
(P^-_1+P^-_2)^k(\ad_1-\ad_2)^l\ket{0}_1\otimes \ket{0}_2
\right\}\,,
\end{equation}
where $k=0,1,2,\dots$ and $l=1,2,3,\dots$. It is in principle possible to obtain an explicit expression for the states $\left|e^{{\bf -l}}\right>$ for $l>1$
\begin{equation}
 \sum_{r=0}^k\sum_{s=0}^l
\frac{k!l!(r+s-1)!(k+l-r-s-1)!s(l-s)}{r!s!^2(k-r)!(l-s)!^2}
\ad_1{}^{r+s}\ad_2{}^{k+l-r-s}\ket{0}_1\otimes
\ket{0}_2\,.
\end{equation}
When $l=1$ the expression simplifies dramatically to
\begin{equation}
\left\{\left|e^{{\bf -1}}_k\right>\right\}\equiv \left\{k!
(\ad_1{}^{k+1}-\ad_2{}^{k+1})\ket{0}_1\otimes \ket{0}_2
\right\}\,,
\end{equation}
where $k=0,1,2,\dots$. In any case we have now found all the $\left|e^{V_1}_i\right>$ states.

To define the projection operator we now need to find the $\left<e^{V_1}_i\right|$ states. Up to normalisation, these are given by
\begin{equation}
\left\{\left<e^{P}_i\right|\right\}\equiv 
\left\{{}_1\!\!\left<0\right|\otimes {}_2\!\!\left<0\right|
v_{P,k}
(P_1^++P_2^+)^k\right\}\,,
\end{equation}
where $k=0,1,2,\dots$,
\begin{equation}
\left\{\left<e^{{\bf -1}}_i\right|\right\}\equiv 
\left\{{}_1\!\!\left<0\right|\otimes {}_2\!\!\left<0\right|v_{1,k}
(a_1-a_2)
(P_1^++P_2^+)^k\right\}\,,
\end{equation}
where $k=0,1,2,\dots$, and
\begin{equation}
\left\{\left<e^{{\bf -n}}_i\right|\right\}\equiv 
\left\{{}_1\!\!\left<0\right|\otimes {}_2\!\!\left<0\right|\left[v_{n,k}
\sum_{l=1}^{n-1}\frac{n!l(n-l)}{l!^2(n-l)!^2}a_1^l(-a_2)^{n-l}\right]
(P_1^++P_2^+)^k\right\}\,,
\end{equation}
where $k=0,1,2,\dots$ and $n=2,3,\dots$. Above, the $v_{n,k}$ and $v_{P,k}$ are normalisation constants.
One can check that for $k=0$ all of the above states are annihilated by $P_1^-+P_2^-$ and so are highest weight ``bra'' states\footnote{These highest weight states are basically the same as the highest weight ``ket'' states for the $S$ module, after swapping all $\ad$ operators with $a$ operators, and the ket vacuum for the bra vacuum. This really just follows from the fact that if we define $(\ad)^\dagger=a$ the $S$ and $P$ modules are conjugate to each other.}. This observation ensures the orthogonality relations, and all that remains to be done is find the normalisations $v_{n,k}$. These are given by
\begin{eqnarray}
v_{P,0}&=&1\,,
\\
v_{P,k}&=&\frac{(-1)^k}{2\,k!}\qquad\mbox{for }k=1,2,\dots\,,
\label{vpk}
\\
v_{1,k}&=&\frac{(-1)^k}{2\,(k+1)!k!}\,,
\\
v_{n,k}&=&\frac{(-1)^k(n-1)!^2(2n-1)}{n(n-1)k!(k+2n-1)!}\,,
\qquad\mbox{for }n=2,3,\dots\,.
\end{eqnarray}

\subsubsection{Projectors onto irreps of $S\otimes S$}

Let us construct the basis vectors $\left|e^{V_1}_i\right>$ for the irreps on the right-hand side of equation~(\ref{SSirep}). For the $S$ module we have
\begin{equation}
\left\{\left|e^{S}_i\right>\right\}\equiv 
\left\{(S_1^-+S^-_2)^k
\ket{0}_1\otimes \ket{0}_2
\right\}\,,
\end{equation}
where $k=0,1,2,\dots$. The ${\bf -1}$ module gives
\begin{equation}
\left\{\left|e^{{\bf -1}}_i\right>\right\}\equiv 
\left\{(S_1^-+S^-_2)^k
(\ad_1-\ad_2)\ket{0}_1\otimes \ket{0}_2
\right\}\,,
\end{equation}
where $k=0,1,2,\dots$. The remaining modules are given by
\begin{equation}
\left\{\left|e^{{\bf -n}}_i\right>\right\}\equiv 
\left\{(S_1^-+S^-_2)^k
\sum_{l=1}^{n-1}
\frac{n!(\ad_1)^l(-\ad_2)^{n-l}}{l!(l-1)!(n-l)!(n-l-1)!} 
\ket{0}_1\otimes\ket{0}_2
\right\}
\,,
\end{equation}
for $n=2,3,\dots$ and $k=0,1,2,\dots$. At this stage we will not bother writing out explicit expressions in terms of creation operators for the above; this can be done easily, though the expressions are again somewhat lengthy.

To define the projection operator we now need to find the $\left<e^{V_1}_i\right|$ states. Up to normalisation, these are given by
\begin{equation}
\left\{\left<e^{S}_i\right|\right\}\equiv 
\left\{{}_1\!\!\left<0\right|\otimes {}_2\!\!\left<0\right|
v_{S,0},({}_1\!\!\left<k\right|\otimes {}_2\!\!\left<0\right|
+{}_1\!\!\left<0\right|\otimes {}_2\!\!\left<k\right|)
v_{S,k}\right\}\,,
\end{equation}
where $k=1,2,\dots$, and
\begin{equation}
\left\{\left<e^{{\bf -l}}_i\right|\right\}\equiv 
\left\{{}_1\!\!\left<0\right|\otimes {}_2\!\!\left<0\right|v_{l,k}
(a_1-a_2)^l
(S_1^++S_2^+)^k\right\}\,,
\end{equation}
where $l=1,2,\dots$ and $k=0,1,2,\dots$

\subsubsection{Projectors onto irreps of $R\otimes R$}\label{RRproj}

The projector onto the singlet representation $\Pi_{{\bf 0}}$ is
\begin{equation}
\Pi_{{\bf 0}}=\ket{0,0}\bra{0,0}\,,
\end{equation}
while the projectors onto the two ${\bf -1}$ representations are
\begin{align}
  \Pi_{\mathbf{-1}_S} &= \sum_{k=1}^{\infty} \frac{1}{k!} \frac{1}{2} \left( \ket{k,0} + \ket{0,k} \right) \left( \bra{k,0} + \bra{0,k} \right) \,, \\
  \Pi_{\mathbf{-1}_A} &= \sum_{k=1}^{\infty} \frac{1}{k!} \frac{1}{2} \left( \ket{k,0} - \ket{0,k} \right) \left( \bra{k,0} - \bra{0,k} \right)
\end{align}
To find the projector onto a ${\bf s}<-1$ irrep.\@ from the right-hand side of equation~(\ref{RRirep}) we first write down the highest weight bra states dual to the highest weight ket state $\left|n\right>_{12}$ given in equation~(\ref{rhw})
\begin{equation}
\left<n\right|_{12}\equiv 
\left<0,0\right|
\sum_{l=1}^{n-1}v_{n,0}
\frac{(a_1)^l(-a_2)^{n-l}}{l!\sqrt{(l-1)!}(n-l)!\sqrt{(n-l-1)!}} 
\,,
\end{equation}
where the normalisation is
\begin{equation}
v_{n,0}=\frac{(n-1)!^2(-1)^n\sqrt{(n-2)!}}{(2n-2)!}\,.
\end{equation}
The projectors then take the form given in equation~(\ref{genprojform}) with
\begin{equation}
\left\{\left|e^n_k\right>\right\}\equiv \left\{
(R^-_1+R^-_2)^k\left|n\right>\right\}\,,
\qquad\mbox{and}\qquad
\left\{\left<e^n_k\right|\right\}\equiv \left\{ v_{n,k}\left<n\right|
(R^+_1+R^+_2)^k\right\}\,,
\end{equation}
where $k=0,1,2,\dots$ and ${\bf s}=-2,-3,\dots$ and the normalisation for $k>0$ is given by
\begin{equation}
v_{n,k}=v_{n,0}\frac{(-1)^k(2n-1)!}{k!(k+2n-1)!}\,.
\end{equation}

\bibliographystyle{nb}
\bibliography{InfDimMod-refs}

\end{document}